\documentclass[reprint,amsmath,amssymb,floatfix,aip,apl,superscriptaddress]{revtex4-1}

\usepackage{color}
\usepackage{graphicx}
\graphicspath{ {./_Figures/} }
\usepackage{epstopdf}
\usepackage{color}
\definecolor{red}{rgb}{1.00,0.00,0.00}
\usepackage[cp1251]{inputenc} 
\usepackage{multirow}

\begin{document}

\title{Magnetic Order in 3D Topological Insulators\\ Wishful Thinking or Gateway to Emergent Quantum Effects?}

\author{A. I. Figueroa}
\affiliation{Catalan Institute of Nanoscience and Nanotechnology
(ICN2), CSIC, Campus UAB, Barcelona, 08193, Spain}
\email{Authors to whom correspondence should be addressed: adriana.figueroa@icn2.cat, Thorsten.Hesjedal@physics.ox.ac.uk, steinkenj@ill.eu}

\author{T. Hesjedal}
\affiliation{Clarendon Laboratory, Department of Physics, University of Oxford, Parks Road, Oxford, OX1~3PU, United Kingdom}
\email{Thorsten.Hesjedal@physics.ox.ac.uk}

\author{N.-J. Steinke}
\affiliation{Institut Laue-Langevin, 71 Avenue des Martyrs, 38000 Grenoble, France}
\email{steinkenj@ill.eu}

\begin{abstract}
Three-dimensional topological insulators (TIs) are a perfectly tuned quantum-mechanical machinery in which counter-propagating and oppositely spin-polarized conduction channels balance each other on the surface of the material. This topological surface state crosses the bandgap of the TI, and lives at the interface between the topological and a trivial material, such as vacuum.
Despite its balanced perfection, it is rather useless for any practical applications.
Instead, it takes the breaking of time-reversal symmetry (TRS), and the appearance of an exchange gap to unlock hidden quantum states. The quantum anomalous Hall effect, which has first been observed in Cr-doped (Sb,Bi)$_2$Te$_3$, is an example of such a state in which two edge channels are formed at zero field, crossing the magnetic exchange gap.
The breaking of TRS can be achieved by magnetic doping of the TI with transition metal or rare earth ions, modulation doping to keep the electronically active channel impurity free, or by proximity coupling to a magnetically ordered layer or substrate, in heterostructures or superlattices.
We review the challenges these approaches are facing in the famous 3D TI (Sb,Bi)$_2$(Se,Te)$_3$ family, and try to answer the question whether these materials can live up to the hype surrounding them.
\end{abstract}

\date{\today}
\maketitle

\section{Introduction\label{sec:intro}}

The prototypical three-dimensional (3D) topological insulators (TIs)\cite{Fu2007} of the (Bi,Sb)$_2$(Se,Te)$_3$ family of solid solutions, most notably Bi$_2$Se$_3$, Bi$_2$Te$_3$, and Sb$_2$Te$_3$, had a successful career as efficient thermoelectric materials\cite{Goldsmid1954,Goldsmid_1958} before the theoretical prediction of their topological surface states (TSSs) in 2009.\cite{Zhang2009}
The TSS results from their large spin-orbit coupling and is made up of spin-momentum locked, counter-propagating streams of oppositely spin-polarized electrons.
Elastic backscattering by nonmagnetic impurities is forbidden by time-reversal symmetry (TRS), in principle resulting in high carrier mobilities.
The existence of the gapless TSS in 3D TIs was first experimentally demonstrated using angle-resolved photoemission spectroscopy,\cite{Chen2009} instead of, as one might have expected, in transport measurements. 
The reason lies in the rather poor electronic properties of (Bi,Sb)$_2$(Se,Te)$_3$-based materials, which are in fact narrow-gap semiconductors with strong, unintentional charge doping due to chalcogen vacancies.
These high levels of bulk carriers can be most efficiently overcome by counter-doping,\cite{Hor2009} in particular in very thin films in which the relative bulk carrier concentration is naturally suppressed.

In order to observe the novel physical phenomena TIs are synonymous for, such as the quantum anomalous Hall effect (QAHE),\cite{Chang2013} the topological magnetoelectric
effect,\cite{QiPRL2009} the physics related to chiral edge states,\cite{Qi2011} and spintronic effects,\cite{Pesin:2012kx} TRS has to be broken and an exchange gap introduced in the TI.\cite{Niu2011}
The exchange gap was initially achieved by direct magnetic doping of the TI, by both transition-metals and rare-earth ions, in bulk crystals as well as thin films.\cite{Elkholdi1994, Song2012,Harrison2014_Gd,Harrison2015-Dy,Harrison2015-Ho, Harrison_SciRep2015, Collins-McIntyre2014, Liam2014_MnBS_AIP,Duffy2017} The QAHE was recently observed in the intrinsic magnetic TI MnBi$_2$Te$_4$.\cite{Deng2020}
Proximity-coupling to magnetically ordered substrates or layers, i.e., ferromagnets, ferrimagnets, or antiferromagnets, is another way to break TRS. One inherent advantage of this approach is that there are no dopants interfering with the electrically active part of the TI, which could be compromised by the added impurities.
Finally, the combination of the different approaches in the form of heterostructures and superlattices opens up new ways to achieve efficient TRS breaking without, in principle, compromising the electronic properties of the TI too much.
Nevertheless, neither approach is yielding an observable QAHE at decent (liquid He and above) temperatures, despite the fact the TSS is observable at room temperature. On the other hand, the magnetic transition temperature $T_\mathrm{C}$ itself is not setting the limit either; in fact, it can reach very high values. 
For instance,  Cr$_{0.15}$(Bi$_{0.1}$Sb$_{0.9}$)$_{1.85}$Te$_3$, which has been the most successful QAHE system so far, has a  $T_\mathrm{C}$ of 16~K whereas very low temperatures ($<$300~mK) are required to observe the QAHE.\cite{Chang2013,Chang2011,Chang:2015aa}
 
There are several outstanding recent reviews of magnetic TIs, their exotic phenomena, and applications, which can be found in Refs.\ \onlinecite{Tokura2019,He_Rev2019,Fei2020}.
In this review, we focus on the discussion of the various growth and doping approaches for achieving an exchange gap in TI thin films, with a particular emphasis on heterostructures, and we finally try to answer the question whether there is hope for these materials to fulfill their star potential.

\section{Growth}

In the context of thermoelectric applications and fundamental studies of the (Bi,Sb)$_2$(Se,Te)$_3$ family of solid solutions, a number of single crystal,\cite{Koehler1973,Watson2013} nanostructure,\cite{Echem_2006, Alegria_NS_MOCVD_2012, Harrison_nanowire_2014, Schoenherr_2014, Schoenherr_2014b, Cecchini2019} and thin film growth techniques have been employed,\cite{Ferhat1996, Ferhat2000, Cao_MOCVD_films_2012, Wang2016, Rusek_ALD_2017, Liao2019} however, for brevity and also given the versatility of the method, our review focuses solely on results obtained using molecular beam epitaxy (MBE)\cite{Charles1988,Iwata1999} (for reviews on TI thin film growth see, e.g., Refs.\ \onlinecite{Wang_Review_2011,He2013,Ginley2016}).
High-quality single crystalline TI thin film growth by MBE has been reported on a wide variety of single-crystalline substrates, e.g. Si,\cite{ZhangAPL2009,Li2010,He2011,Krumrain2011,Liu2013,Liu2012JVST}
Al$_2$O$_3$,\cite{TaskinPRL2012,Lee2012APL,BansalPRL2012,Harrison2013,Zhao2013} GaAs,\cite{RichardellaAPL2010,Chen2014,Eddrief2014,Liu2011}
Ge,\cite{Guillet2018, Kim2018}
CdS,\cite{Kou2011}
SrTiO$_3$,\cite{Zhang_STO_2011}
graphene (on 6H-SiC),\cite{Zhang2010NatPhys}
lattice-matched InP,\cite{Schreyeck2013,Guo2013}
and BaF$_2$,\cite{Bonell_2017} as well as amorphous SiO$_2$/Si for back-gated electrical transport measurements\cite{Jerng2013,Bansal2014,Jeon2014} and fused silica,\cite{Collins_silica} indicating that these materials may grow on virtually anything.\footnote{For achieving in-plane ordered films, the substrates should have hexagonal symmetry.}

\begin{figure*}[ht!]
	\begin{center}
		\includegraphics[width=12.6cm]{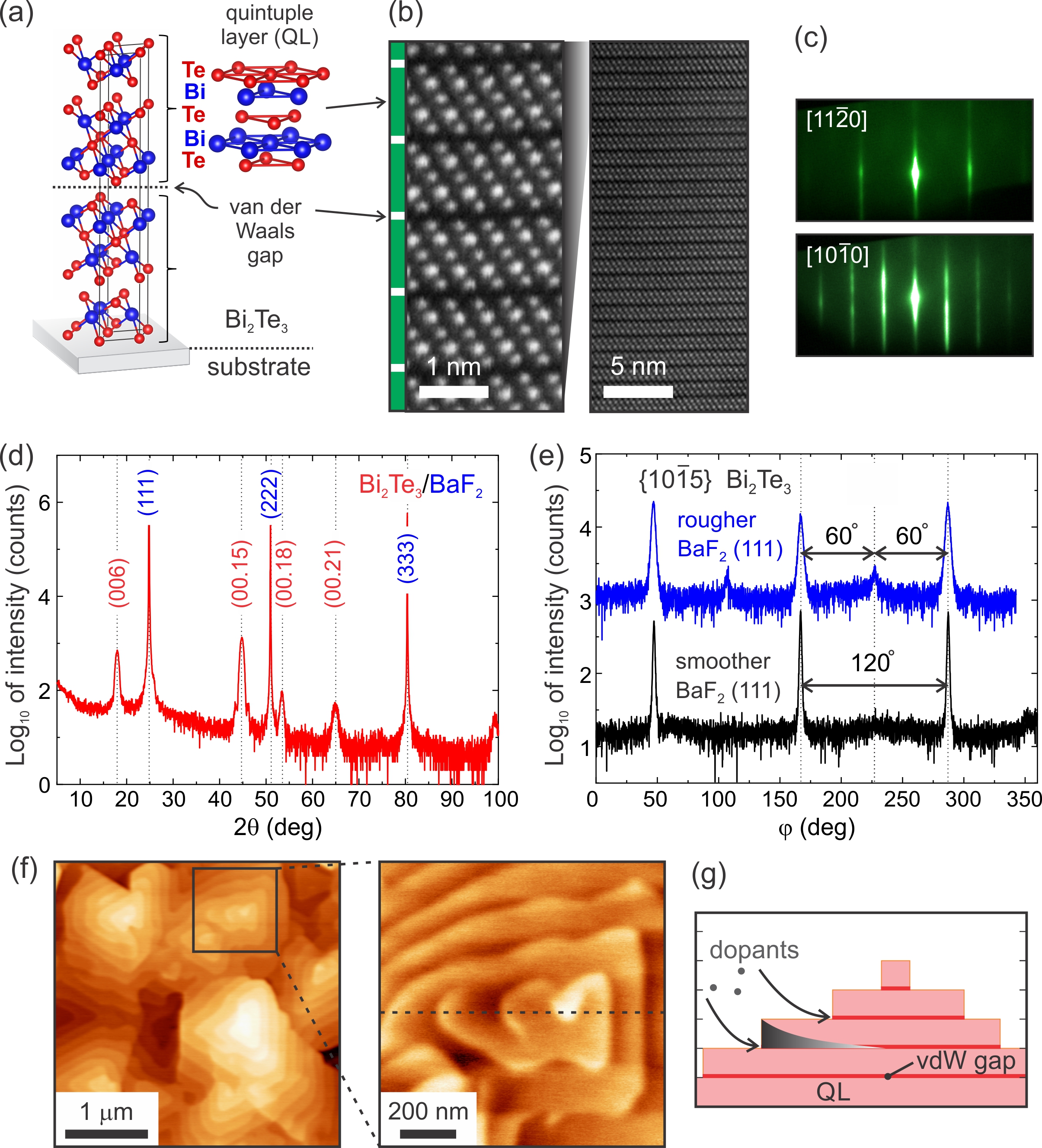}
		\caption{Structural properties and growth of Bi$_2$Te$_3$ thin films by MBE.
		(a) Crystal structure. On the left, the unit cell consisting of three  Te-Bi-Te-Bi-Te quintuple layer (QL) blocks is shown. The QL blocks (Bi and Te are shown in blue and red, respectively) are separated by the weakly bonding Te-Te van der Waals gap.
		(b) Cross-sectional high-angle annular dark field scanning transmission electron microscopy (HAADF-STEM).
		The high-resolution image one the left shows the fine structure of the QL blocks and the van der Waals gap (no intensity).
		The overview scan on the right shows the perfectly ordered QL stack.
		(c) RHEED images of Bi$_2$Te$_3$ on $c$-plane sapphire along the $[11\bar{2}0]$ and $[10\bar{1}0]$ azimuths, showing 
		streaky patterns, indicative of flat, crystalline surfaces. 
		(d) Out-of-plane $\theta$-$2 \theta$ XRD spectrum of Bi$_2$Te$_3$ on BaF$_2$ (111). The film peaks (red) and substrate peaks (blue) are indicated.
		(e) In-plane XRD $\varphi$ scan of a $\{10\bar{1}5\}$ Bi$_2$Te$_3$ peak. On smooth substrates (black line), only one domain is found (120$^\circ$ apart), consistent with the three-fold crystal symmetry.
		On rough substrates (blue line), two in-plane domains are found (60$^\circ$ apart).
		(f) The surface morphology of Bi$_2$Te$_3$ on $c$-plane sapphire, as obtained by AFM, is dominated by two domains of triangular islands, measuring $\sim$1~$\mu$m across.
		On the right, the close-up reveals details of the typical 1-QL-step terrace structure as well as the spiral-like growth center. 
		(g) Schematic illustration of a Bi$_2$Te$_3$ island. The stepped structure exposes the van der Waals gap, which can act as an entry point for the efficient, but uncontrolled, diffusion of dopants into the material.
		}
		\vspace*{-0.5cm}
		\label{fig:growth}
	\end{center}
\end{figure*}

In fact, the in-plane lattice mismatch spans a remarkable range in the context of conventional thin film growth (see Table 1 in Ref.\ \onlinecite{He2013}), reaching from 0.2\% for Bi$_2$Se$_3$ on InP all the way up to 43.8\% for Bi$_2$Te$_3$ on graphene.\cite{He2013}
The reason for this enormous tolerance of lattice mismatches lies in the layered nature of the rhombohedral crystal structure [space group $D^5_{3d}$ ($R \bar{3} m$)] of the (Bi,Sb)$_2$(Se,Te)$_3$ compounds, which is characterized by the van der Waals (vdW) gap.
Figure \ref{fig:growth}(a) illustrates a Bi$_2$Te$_3$ unit cell, in which the Te-Bi-Te-Bi-Te quintuple layer (QL) building blocks are illustrated, which are separated from one another by the vdW gap across which the Te-Te bonding is weak. In consequence, the bonding between the first QL and the substrate is also characterized by weak vdW forces.
The vdW gap can be clearly seen in transmission electron microscopy images shown in Fig.\ \ref{fig:growth}(b).
The vdW epitaxy growth mechanism\cite{Koma1985,Koma1992,Koma1999} does not require lattice matching between film and substrate, but is nevertheless characterized by in-plane (rotational) alignment between film and substrate.
Van der Waals epitaxy has seen a renaissance with the advent of 2D materials, for which TIs provide a suitable substrate for the direct 2D materials growth, or for integration with TIs in heterostructures for future electronic devices.\cite{Walsh_vdWepi_2017}
The flip-side of the lack of strong ionic or covalent bonding across the film-substrate interface is a lack of control over the TI film growth.

The growth of (Bi,Sb)$_2$(Se,Te)$_3$ thin films by MBE is usually carried out with a considerable chalcogen overpressure on the order of $>$5:1 due to the high rate of re-evaporation.
The growth rate of typically less than 1~nm/min is controlled by the group-V element flux and varies as a function of substrate temperature.
The substrate temperature, which typically lies in the range between 200$^\circ$C and 300$^\circ$C, has been shown to be the main parameter determining the quality of the films.\cite{Chen2011,He2013}
It has proven advantageous to first grow a seed layer at a temperature of $\sim$50$^\circ$C lower than the final growth temperature, followed by an anneal under chalcogen flux.\cite{Taskin2012, Harrison2013}
The growth can be monitored \emph{in-situ} by reflection high-energy electron diffraction
(RHEED). Figure \ref{fig:growth}(c) shows streaky RHEED patterns on Bi$_2$Te$_3$ on $c$-plane sapphire obtained along the $[11\bar{2}0]$ and $[10\bar{1}0]$ azimuths, which repeat every 60$^\circ$. 
The streaky patterns are indicative of flat, crystalline surfaces and they remain intact for doped films as well, up to the critical concentration above which the streaks become diffuse (and the films rough), indicative of their non-substitutional incorporation.\cite{Harrison2014_Gd,Harrison2015-Dy,Harrison2015-Ho}
Note that RHEED can not be continuously used for (Bi,Sb)$_2$(Se,Te)$_3$ growth as the electron beam visibly damages the films, most likely due to local heating.
For the evaporation of Bi and Sb, standard effusion cells are commonly used, whereas Se and Te are often (but not necessarily) evaporated out of special cracker cells.
For the evaporation of dopants, high temperature effusion cells have proven advantageous since they are thermally better shielded, thereby reducing the unintentional evaporation of the high vapor-pressure chalcogens from surrounding areas. 
In fact, Se is very hard to contain and some unintentional Se `co-evaporation' is unavoidable, as shown in high-resolution compositional analysis using Rutherford backscattering spectroscopy with 2.3~MeV He ions and particle-induced x-ray emission with 1~MeV protons.\cite{Harrison2015-Ho}

The structural properties of the films are usually further investigated \emph{ex-situ} using x-ray diffraction (XRD) with Cu K$_{\alpha 1}$ radiation. Figure 
\ref{fig:growth}(d) shows the spectrum for a Bi$_2$Te$_3$ film on BaF$_2$ (111), which is characterized by the (1 1 1) substrate peaks and the (allowed) (0 0 $3l$) film peaks, representative of $c$-axis oriented, rhombohedral single crystalline material. 
Upon doping, the analysis of the resulting peak shifts and peak broadening can be used to quantify the undesired degradation in crystalline quality, as well as giving clues on the doping scenario.\cite{Haazen2012,Li2013,Collins-McIntyre2014}
Further, the in-plane lattice constants, as well as the rotational symmetry of the system, can be obtained from asymmetric 2D reciprocal space mapping or $\varphi$-scans, as shown in Fig.\ \ref{fig:growth}(e).
Whereas the in-plane orientation of vdW epitaxially grown films in principle lock to the symmetry of the underlying substrate, this locking may not be perfect.
In particular, (Bi,Sb)$_2$(Se,Te)$_3$ films on $c$-plane sapphire are known to exhibit rotational twins, as can be seen in the $60^\circ$ repeat of the streaky patterns in RHEED [cf.\ Fig.\ \ref{fig:growth}(c)] --- instead of the three-fold symmetry expected from the $R\bar{3}m$ crystal structure.
In contrast, films grown on BaF$_2$ (111) are twin-free,\cite{Bonell_2017} which alters the electronic properties of the film as well.\cite{Kim2016}
Figure \ref{fig:growth}(e) shows a comparison of the three-fold symmetric $\{10\bar{1}5\}$ Bi$_2$Te$_3$ peaks on a smooth BaF$_2$ surface (120$^\circ$ apart), and the additional peaks due to twinning (60$^\circ$ apart) on a rougher BaF$_2$ surface (which dominates the growth on, e.g., $c$-plane sapphire).

The film morphology, and the distribution of rotational domains, can be conveniently visualized using atomic force microscopy (AFM) owing to the dominating QL-terrace structure with its 1-nm-high steps. 
Figure \ref{fig:growth}(f) shows the characteristic, triangular terraced islands, which are rotated with respect to each other by 60$^\circ$ in a Bi$_2$Te$_3$ film on $c$-plane sapphire.
The triangular shape in case of a hexagonal system can be understood in the framework of the Burton-Cabrera-Frank theory, when for particular crystallographic orientations, the distance between corners of the structure is larger than the diffusion length of the surface adatoms.\cite{Ferhat2000}
In general, the growth is of the spiral surface growth type.\cite{Karma1998}
The growth spirals are believed to be due to the pinning of 2D growth fronts at irregular substrate steps.\cite{Liu2012}
The formation of screw dislocations in Bi$_2$Te$_3$ has been linked to variations in stoichiometry of the deposited nucleation centers due to Te re-evaporation during the initial stages of film growth.\cite{Ferhat2000}

This characteristic morphology of (Bi,Sb)$_2$(Se,Te)$_3$ thin films poses a major challenge for their controlled doping [see illustration in Fig.\  \ref{fig:growth}(g)]. While during standard layer-by-layer growth the dopants can be incorporated in a deterministic way, allowing for the so-called $\delta$-doping of individual layers,\cite{Dingle1978,Stoermer1981} the exposed vdW gap of the terraced islands with its relatively weak electrostatic bonding forces and large layer spacings provides an entry portal for the uncontrolled incorporation of dopants.\cite{Carlson1960}
This is particularly concerning since the diffusion constants in these layered hosts are highly anisotropic, with the diffusion along the planes (in the vdW gap) being particularly fast.\cite{Carlson1960, Korzhuev1991, Koski2012_IntercalBS}
The diffusing dopants are then incorporated into the (Bi,Sb)$_2$(Se,Te)$_3$ host, illustrated by the concentration gradient in Fig.\ \ref{fig:growth}(g), as the solubility limit for interstitial metal incorporation is usually low.
This is certainly no surprise and a long-standing problem\cite{Carlson1960} in the thermoelectrics community where the electric contacting with the obvious choices of low electrical resistivity and high thermal conductivity metals has been challenging (with a few exceptions such as Ta\cite{Music2018}), mostly resulting in the degeneration (or even complete dissolution\cite{Shaughnessy2014}) of the electrodes in case of, e.g., Sn,\cite{Lan2008} Ag,\cite{Keys1963} and Au.\cite{Shaughnessy2014}
For instance, Cu is notorious for its high diffusivity in Bi$_2$Te$_3$ of 
$10^{-6}$~cm$^2$~s$^{-1}$ parallel to the plane and 3$\cdot$$10^{-15}$~cm$^2$~s$^{-1}$ along the $c$-axis at room temperature,\cite{Carlson1960} and the formation of binary chalcogenides.
For electrodes, specific diffusion barriers have been developed (e.g., Ni barriers for Sn electrodes), however, some of them rely on the formation of secondary chalcogen compounds at the interface,\cite{Lan2008} so do not provide a solution for achieving controlled, substitutional doping of TIs.

\section{Magnetic ordering in doped TI\lowercase{s}}
\newcommand{\musr}{$\mu^+\mathrm{SR}$}


The need to break TRS in TIs led to a very active quest for suitable dopants to introduce magnetic order. Magnetic TIs (MTI) were successfully achieved with magnetic dopants, both with transition metals (TM) and rare earths (RE). 

With TMs, magnetic order has been observed in {Fe-}, Mn-, Cr- and V-doped compounds. 
For Fe-doping in Sb$_2$Te$_3$ and Bi$_2$Se$_3$, the compounds remained paramagnetic while the electron concentration was increased, despite successful substitutional doping\cite{Zhou2006_FeSb2Te3, Sugama2001} and $pd$-hybridization predicted by spin density calculations.\cite{Rodriguez2019} 
In contrast, Kulbachinskii \emph{et al.}\ reported ferromagnetic (FM) ordering up to 12~K in (Fe,Bi)$_2$Se$_3$ single crystals.\cite{Kulbachinskii2002}

In Mn-doped TIs FM order has been observed in bulk crystals and thin films. In addition there are reports on the surface magnetism in Mn:Bi$_2$Se$_3$ with observations of both an enhanced moment and $T_\mathrm{C}$ due to surface segregation of the Mn ions,\cite{Teng2019} as well as a diminished surface moment and soft magnetic behavior with no hysteresis possibly caused by partial antiferromagnetic (AF) ordering near the surface.\cite{Collins-McIntyre2014} 
A band gap has been observed in Mn-doped TIs, but it is likely not of magnetic origin.\cite{Sanchez-Barriga2016}


Of all the single-ion dopants tried so far, Cr and V posses the most robust long-range FM order with out-of-plane anisotropy and a typical (doping concentration-dependent) $T_\mathrm{C}$ of 59~K (104~K) for Cr (V) concentrations of $x$=0.29 (0.25) in Sb$_{2-x}$(V/Cr)$_x$Te$_3$.\cite{Chang:2015aa}
In Table \ref{tab:dopants} we summarize the properties of the most common TI dopants.

In line with these observations, the QAHE effect has been observed in both Cr- and V-doped samples, but not in Mn or Fe doped ones (except for the intrinsic magnetic TIs based on MnBi$_2$Te$_4.$\cite{Otrokov2019, Deng2020}) However, the fact that even in these materials the QAHE is limited to very low temperatures, far below $T_\mathrm{C}$, calls the nature and robustness of the magnetic order into question. 

\begin{figure*}[ht!]
	\begin{center}
		\includegraphics[width=15cm]{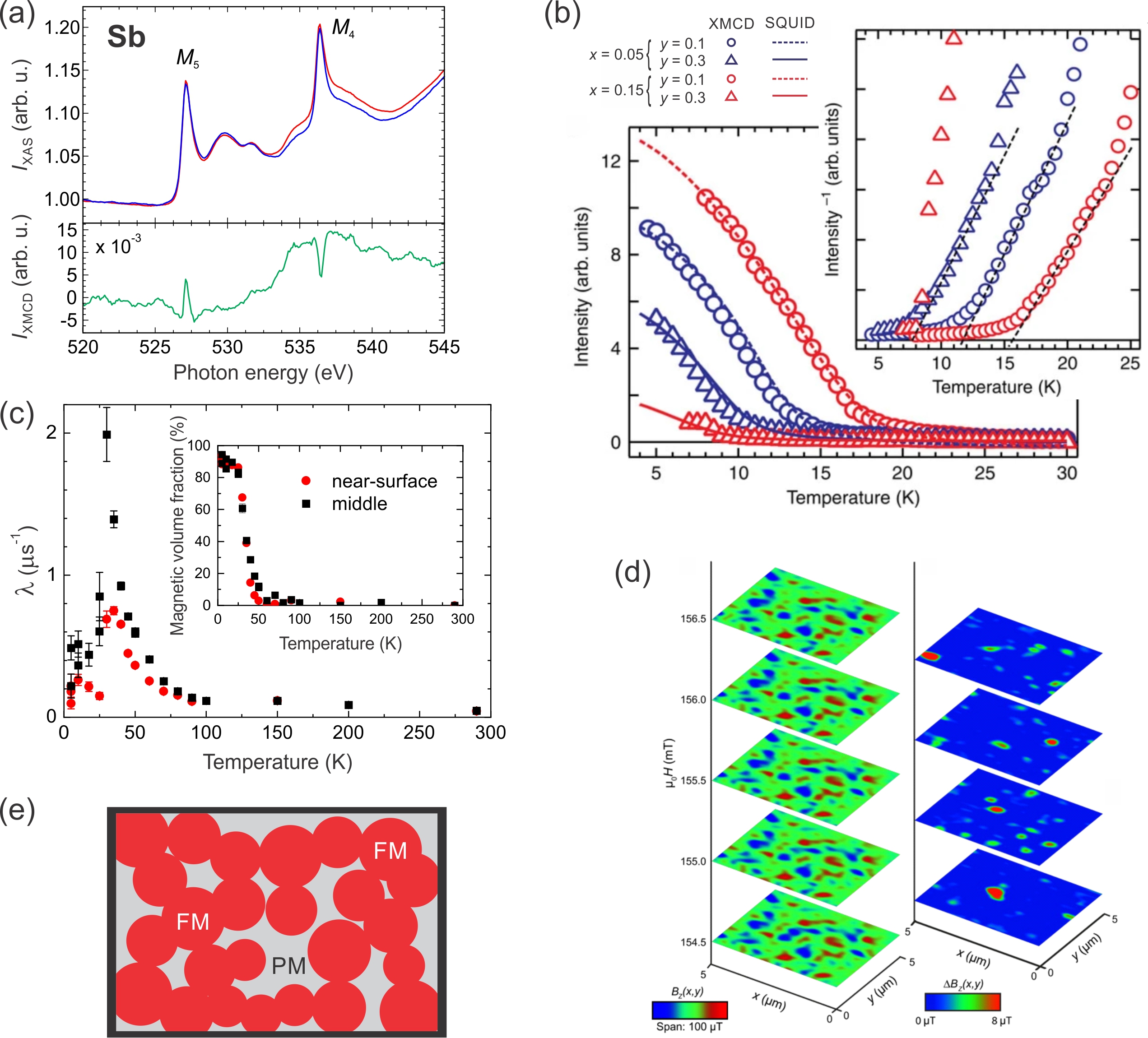}
        \caption{Magnetic properties of MTI films.
        {
		(a) X-ray absorption (XAS) and x-ray magnetic circular dichroism (XMCD) spectra at the Sb $M_{4,5}$ edges of an \emph{in-situ} cleaved Cr:Bi$_2$Te$_3$ film. The measurement was carried our in a field of 8~T and at a temperature of 3~K.
		The XMCD signal (green) is obtained by subtracting the XAS spectrum measured with right-circularly polarized x-rays (blue) from the one measured with left-circularly polarized x-rays (red).\cite{Duffy2017}
		(b) XMCD intensities (symbols) and SQUID magnetization data (lines) obtained for different Cr doping concentrations $x$ and different Sb to Bi ratios in counter-doped Cr$_x$(Sb$_{1-y}$Bi$_y$)$_{2-x}$Te$_3$ crystals.\cite{Ye_NComms_2015}
		From the plot of the inverse intensity (i.e., magnetization) in the panel on the right, the transition temperature is estimated, which shows a clear trend: $T_\mathrm{C}$ increases with $x$ whereas an increase in Bi:Sb ratio strongly reduces it.
		(c) Temperature-dependent transverse-field $\mu^+$SR measurement showing the magnetic transition in a Cr:Sb$_2$Te$_3$ film.\cite{crpuddles}
		The data shown in red represents the behavior of the near-surface region of the film, whereas the (higher muon energy) data shown in black includes the information from the middle of the film as well.
		The relaxation rate $\lambda$, which is a measure of the static and dynamic ($\mu$s) magnetic disorder in the system, is very broad across the magnetic transition.
		(d) SQUID-on-tip microscopy showing the magnetization reversal dynamics of a Cr:(Bi,Sb)$_2$Te$_3$ film.
		(Left) Magnetic images, i.e., the $B_z$ component of the film magnetization, obtained in applied magnetic fields increasing (0.5~mT field steps) at a temperature of 250~mK. 
		(Right) Change in the film's magnetic flux, illustrating the reversal of isolated domains.\cite{Lachman2015}
		(e) Illustration of the magnetic order in TM-doped TIs, showing magnetically ordered and paramagnetic regions coexisting at temperatures far below $T_\mathrm{C}$.
		Panel (b) reproduced from Ye \emph{et al.}, Nat.\ Commun.\ \textbf{6}, 8913 (2015);\cite{Ye_NComms_2015} $\copyright$ CC BY 4.0.
		Panel (d) adapted from Lachman \emph{et al.}, Sci.\ Adv.\ \textbf{1}, e1500740 (2015);\cite{Lachman2015} $\copyright$ CC BY 4.0.
		}
	}
		\vspace*{-0.5cm}
		\label{fig:doping}
	\end{center}
\end{figure*}

\begin{table*}[h]
    \centering
    \begin{tabular}{|c|c|c|c|c|c|c|c|c|l|l|}
    \hline
    \multirow{2}{*}{Dopant} & TI & Doping & Val- & Magnetic & Magnetic & Trans. & \multirow{2}{*}{MCA} &  Open & \multirow{2}{*}{Comments} & \multirow{2}{*}{Refs.}\\
    ~& host & conc.\ & ence & moment & order & temp. & ~ & loop & ~ & ~ \\ 
    ~ & ~ & [at.-\%] & ~ & [$\mu_\mathrm{B}$] & ~ & [K] & ~ & ~ & ~ & ~ \\ \hline\hline

Cr & Bi$_2$Se$_3$ & $\leq$5.2$^\ast$ & $<$3+ & 2$^\dagger$ & FM & 20$^\ddagger$ & - & yes & $^\ast$poor cryst.\ quality for $>$5\% Mn & [\onlinecite{Haazen2012}] \\
~ & ~ & ~ & ~ & ~ & ~ & ~ & ~ & ~ & $^\dagger$max.\ for 2\% Mn; $^\ddagger$max.\ for 5.2\% Mn & ~ \\\hline

Cr	& Bi$_2$Se$_3$ & 12 & 2+ & $\sim$2.1 &	FM	& - & OOP	& yes	& SUB \& INT	& [\onlinecite{Collins-McIntyre2014}] \\\hline

Cr & Bi$_2$Se$_3$ & 0.6$^\dagger$ & mixed & 2.49$^\ast$ & FM & 46$^\ast$ & OOP & ~ & $^\dagger$effective conc.\ for modulation doping & [\onlinecite{Liu2019}] \\
 ~ & ~ & ~ & 2+, 3+ & 1.3$^\ddagger$ & ~ & 31$^\ddagger$ & ~ & ~ & $^\ast$surface value; $^\ddagger$bulk value & ~ \\\hline

Cr	& Sb$_2$Te$_3$	& 7.5-21 &	2+ &	2.8 &	FM	& 28-125 & OOP	& yes &	SUB	& [\onlinecite{collins2016structural}] \\\hline

Cr	& (Bi,Sb)$_2$Te$_3$	& 5 & $<$3+	& 3.19	& FM	& 20	& OOP	& yes	& Zener-type $pd$-exchange interaction &	[\onlinecite{Tcakaev2020}] \\\hline

Cr	& (Bi,Sb)$_2$Te$_3$	& $\leq$15	& 3+ & 2.9 & FM & $\leq$59$^\ast$	& OOP	& yes & $^\ast$14-59~K for 5-15\% Cr & [\onlinecite{Chang:2015aa}] \\\hline

V	& Bi$_2$Se$_3$	& $\leq$12 & -	& - & FM	& $\leq$16$^\ast$ &	OOP &	yes & $^\ast$10~K for 1\%, 16~K for 6\% & [\onlinecite{Zhang2017}] \\\hline

V	& (Bi,Sb)$_2$Te$_3$	& $\leq$13 & mixed & 1.5 & FM & $\leq$104$^\ast$	& OOP & yes	& $^\ast$23-104~K for 4-13\% V	& [\onlinecite{Chang:2015aa}] \\
 ~ & ~ & ~ & 3+, 4+ & ~ & ~ & ~ &  & ~ & SUB & ~ \\
\hline

V	& Sb$_2$Te$_3$	& 5 & $<$3+	& 1.84 & FM	& 45 &	OOP &	yes & Zener-type $pd$-exchange interaction & [\onlinecite{Tcakaev2020}] \\\hline

Mn	& Bi$_2$Te$_3$	& $\leq$10	& -	& -	& FM$^\ast$	& $\leq$17 & OOP	& yes & SUB \& INT possible; FM$^\ast$: $\geq$2\% Mn &	[\onlinecite{Lee_PRB2014}] \\\hline

Mn	& Bi$_2$Se$_3$ & -	& mostly & 1.6$^\ast$	& FM	& 1.5 &	OOP	& no	& XAS:\cite{Figueroa:2015aa} SUB \& INT	& [\onlinecite{Liam2014_MnBS_AIP}] \\
~	& ~ & ~	& 2+	& 5.1$^\dagger$	& ~	& ~ &	~	& ~	& $^\ast$XMCD; $^\dagger$SQUID	& ~ \\\hline

Mn & Bi$_2$Te$_3$	& $\leq$13	& -	& -	& FM$^\ast$	& $\leq$15	& OOP	& yes	& vdW gap INT	& [\onlinecite{Ruzika2015}] \\ 
~ & ~ & ~ & ~ & ~ & ~ & ~ & ~ & ~ &  $^\ast$FM $\geq$3\% Mn; max $T_\mathrm{C}$ for  9\% & ~ \\\hline\hline

Gd	& Bi$_2$Te$_3$	& $\leq$30	& 3+ & 7	& AFM	& -2.5	& -	& no	& XAS:\cite{Figueroa-oxEXAFS} SUB & [\onlinecite{Harrison2014_Gd}], [\onlinecite{Li2010-MBE}]  \\ \hline
Dy	& Bi$_2$Te$_3$	& $\leq$36	& 3+ & 4.3-12.6	& AFM	& -1.2	& -	& no	& XAS:\cite{Figueroa-oxEXAFS} SUB	& [\onlinecite{Harrison2015-Dy}] \\ \hline
Ho	& Bi$_2$Te$_3$	& $\leq$21	& 3+ & 5.15	& AFM	& -0.837 & -	& no	& XAS:\cite{Figueroa-oxEXAFS} SUB & [\onlinecite{Harrison2015-Ho}] \\ \hline
Eu	& Bi$_2$Te$_3$	& $\leq$4 & 2+ & -	& -	& -	& -	& -	& SUB, EuTe for 9\% Eu	& [\onlinecite{Fornari2020}]
\\ \hline
Eu & Bi$_2$Se$_3$	& $\leq$21	& -	& -	& FM  & 8-64	& -	& yes & nonuniform Eu SUB; FM $\geq$10\% Eu & [\onlinecite{Aronzon2018}]
\\ \hline
    \end{tabular}
    \caption{Summary of the physical properties of transition metal and rare earth element doped TI thin films. The doping concentration is given in \% of the (Bi+Sb) sites for substitutional (SUB) doping, unless otherwise noted [e.g., for interstitial (INT) doping]. The magnetic moment is given per doping ion (in $\mu_\mathrm{B}$) and the transition temperature (trans.\ temp. $T_\mathrm{C}$ or $T_\mathrm{N}$ in K). The out-of-plane (OOP) magnetocrystalline anisotropy (MCA) is indicated when stated in the reference.}
    \label{tab:dopants}
\end{table*}


Early theoretical and experimental studies\cite{Yu2010,Chang2013,Li2015} on Cr- and V-doped TIs proposed that long-range magnetic ordering of the Van Vleck type could be established.\cite{VanVleck1953} 
This type of magnetic order is stabilized by intra-atomic mixing of the $d$-orbital ground state with higher energy excitations close in energy, which lead to a positive exchange integral.
Observations of long-range magnetic order in Cr- and V-doped thin films and crystals were first investigated by superconducting quantum interference device (SQUID) magnetometry, x-ray magnetic circular dichroism (XMCD) and polarized neutron reflectometry (PNR),\cite{Haazen2012, Collins-McIntyre2014, collins2016structural, Tcakaev2020} 
which reported a clear ferromagnetic transition and open hysteresis loops with out-of-plane anisotropy and, crucially, a doping concentration-dependent $T_\mathrm{C}$.
This is incompatible with the notion of magnetic order dominated by intra-atomic interactions and, instead, strongly hints at a magnetic ordering mechanism that is dependent on either carrier or spin concentrations. 

The exact features of the magnetic ordering appear to depend on the details of the modification of the electronic band structure due to the dopant. For instance, in V-doped samples, impurity bands caused by the V 3$d$ states are found near the Fermi energy, $E_\mathrm{F}$, and Dirac point of the system.\cite{Islam_PRB_2018,Vergniory2014}
These states appear to stabilize the magnetic ordering. Both V and Cr show significant \emph{pd} hybridization between the host 5$p$ states and the dopant 3$d$ states in $\mathrm{Sb}_2\mathrm{Te}_3$ based compounds as has been observed by a number of groups using XMCD.\cite{Duffy2017, Ye_NComms_2015, Islam_PRB_2018, Ye2019} Figure \ref{fig:doping}(a) shows the XMCD signal due to induced spin polarization in the Sb $p$-states for a Cr-doped Sb$_2$Te$_3$ thin film.
As the conduction and valence bands are mainly formed by the host lattice $p$-states, this leads to the question if it is not the carriers that mediate magnetic order in these systems. In the tetradymite dichalcogenides, near the center of the Brillouin zone, the valence band is formed by the anion $p$-states of Te or Se, whereas the conduction band is formed by the $p$-states of the Sb and Bi cations, though it has recently been shown that the bonds of magnetically doped compounds have a strong covalent character.\cite{Tcakaev2020} 

The important role carriers play in stabilizing the magnetic order was shown by the gate-voltage dependence of the magnetic relaxation time,\cite{Lachman2015} and strongly supported by the dependence of $T_\mathrm{C}$ on the concentration of Bi, acting as a counter dopant, in V- and Cr-doped films. Ye \emph{et al.} showed, by systematically varying dopant and Bi concentration in Cr:$\mathrm{(Sb,Bi)_{2}Te_3}$ and V:$\mathrm{(Sb,Bi)_{2}Te_3}$, not only that the magnetic ordering is stabilized with increasing doping concentration, but also that $T_\mathrm{C}$ is systematically suppressed when the carrier-hole concentration is decreased by substituting Sb with Bi.\cite{Ye_NComms_2015, Ye2019} 
Figure \ref{fig:doping}(b) shows Arrott plots from which the values of $T_\mathrm{C}$ are determined for systematically varied Cr and Bi concentrations in Cr$_x$(Sb$_{1-y}$Bi$_y$)$_{2-x}$Te$_3$ crystals. An increase in Cr concentration stabilizes the magnetic order whereas an increase in Bi strongly suppresses it. First principles calculations show that the extent of the magnetic coupling in the $c$-axis direction extents across the vdW gap. 
The picture that emerges is that of a dominant carrier-mediated interaction based on the hybridization of the $d$-states of the dopants with the carrier bands formed by the $p$-states of the host lattice, similar to the ordering in dilute magnetic semiconductors.\cite{Dietl2000, Storchak2008} 
In addition, it seems that the introduction of Bi weakens the \emph{pd}-hybridization between the 5$p$ states and the 3$d$ states.\cite{Ye_NComms_2015}
Apart from this carrier-mediated exchange, there is evidence of exchange interactions mediated by the host ions.\cite{Vergniory2014, Islam_PRB_2018} 
Notwithstanding, it is clear that the necessary tuning to reduce unwanted bulk carriers will always be detrimental to the long-range magnetic order in these compounds. 

It seems that the surface magnetism can differ from the bulk behavior in particular in Cr-based compounds. For instance, in Ref.\ \onlinecite{Fang2012} it was shown that the surface magnetization lies in-plane, rather than following the bulk out-of-plane anisotropy. 
However, Ye \emph{et al.}'s study using surface-sensitive XMCD in total-electron yield mode on Cr-doped $\mathrm{(Sb,Bi)_{2}Te_3}$ did not find any evidence for a variation of the Cr-magnetism at the surface, neither did Duffy \emph{et al.}\cite{Duffy2017}
In the work by Liu \emph{et al.},\cite{Liu2019} it was reported that $T_\mathrm{C}$ for modulation-doped Bi$_2$Te$_3$ increased if the Cr dopants were introduced at the sample surface.

Using muon spin rotation (\musr) techniques, the magnetic volume fraction can be directly tracked as a function of temperature in Cr- and V-doped thin films of $\mathrm{Sb}_2\mathrm{Te}_3$ and other TIs.\cite{Krieger_PRB2019,crpuddles}
These measurements show that ferromagnetic ordering develops rather gradually over a wide temperature range and that for lower doping concentrations, a significant fraction of the material remains paramagnetic. 
The magnetic transition can also be tracked through the the slow relaxation rate during the phase transition in $\mu$SR measurements [see Fig.\ \ref{fig:doping}(c)]. 
This relaxation rate is a measure of the static and $\mu$s-dynamic magnetic disorder in the system. This peak is very broad, and, for lower doping concentrations, disorder persists down to at least 4~K (the lowest temperature reachable in the system). 

Furthermore, a change of the internal magnetic field experienced by the muons in the samples provides evidence for a percolation transition, above which magnetically ordered clusters gradually appear in a paramagnetic sea, growing and percolating as the temperature is decreased.\cite{crpuddles}
For lower doping concentrations, the internal field remains shifted even at the lowest temperatures.\cite{Krieger_PRB2019, crpuddles} This behavior is likely due to residual paramagnetic regions, which remain between the ferromagnetic patches. 
Note that there can also be a large difference in percolation between surface and bulk due to different screening behavior.\cite{Fujii2011}
Lachman \emph{et al.}\cite{Lachman2015} investigated the magnetic domain pattern in a Cr-doped (Sb,Bi)$_2$Te$_3$ film using scanning SQUID-on-tip microscopy [see Fig.\ \ref{fig:doping}(d)].
In their measurements, the nearly non-interacting, slowly fluctuating, superparamagnetic domains were directly imaged. These measurements show directly that the magnetic order in TM-doped TIs is not robust over long lengthscales. Instead, a picture of superparamagnetic or superferromagnetic domains emerges which gradually form during a broad transition, and with significant disorder, and paramagnetic regions remaining intact until far below $T_\mathrm{C}$ [see illustration in Fig.\ \ref{fig:doping}(e)].
With an increase in doping concentration, the distances between the dopant ions shrink and additional carriers are introduced into the system. The net effect is the stabilization of the magnetic order, but at the cost of severely compromising the quantum properties of the material.

The phenomenon of nanoscale magnetic phase separation is known from dilute magnetic semiconductors, such as GaMnAs. There, depending on the extent of the localization of the (hole) carriers, either the magnetic transition follows the percolation dynamics of bound magnetic polarons in the case of strongly localized holes,\cite{Priour2006} or, in the case of hole states extended over distances longer than the average dopant distance, the Zener model becomes valid.\cite{Dietl2000} 
However, even in the Zener model description, paramagnetic regions may persist below $T_\mathrm{C}$ in hole-free regions.\cite{Storchak2008}

Rare earth dopants are extremely attractive given their large, localized magnetic moments and their similarity to the Bi and Sb ions in the host lattice. The ionic radius of Dy$^{3+}$ is 105~pm, much more similar to that of Bi$^{3+}$ (117~pm) than to Cr$^{2+}$ in the high (low) spin state of 94 (87)~pm. 
Doping with a high moment ion has the advantage that as the exchange gap is dependent on the size of the magnetization, the doping concentration can be reduced for the same gap size, thereby preserving crystal quality. Of the rare earth lanthanide series the high moment materials Gd, Dy, Ho were all successfully used to dope Bi$_2$Te$_3$.\cite{Harrison2014_Gd, Harrison2015-Dy, Harrison2015-Ho}
However, none of these materials display long-range magnetic order by themselves.\cite{Harrison2014_Gd,Harrison2015-Dy,Harrison2015-Ho,Harrison_SciRep2015,Figueroa_2017} 
Doping concentrations of up to 35\% were achieved in Gd-doped Bi$_2$Te$_3$ thin films; much above the bulk solubility limit.\cite{Li2013}
Details of the magnetic configurations are given in Table \ref{tab:dopants}. Of the investigated RE dopants, Dy shows the most interesting properties: it is the only ion that leads to a concentration-dependent moment, its susceptibility is large, and thin films can be magnetically saturated in fields as low as $\sim$2~T. PNR measurements have shown that the induced moment can be as large as $\sim$1/3 of the full Dy$^{3+}$ moment at fields as low as 0.65~T.\cite{Duffy2018_Dy} 
\musr~measurements showed that while no long-range order can be established, strong, short-range magnetic correlations establish themselves with decreasing temperature.\cite{Duffy2018_Dy} 
This makes Dy-doped TIs a prime candidate for further exploitation in heterostructures, where order can be introduced via proximity coupling.\cite{Duffy2018,Duffy2019}

\section{Magnetic heterostructures}\label{sec:heterostructures}
In the previous section, we have described how introducing magnetic
elements into TIs has been used to achieve
magnetically ordered materials. Incorporating TIs into magnetic
heterostructures has been explored both as an alternative to magnetic
doping and as an added materials design opportunity to tune the magnetic properties. 
Magnetic heterostructures offer the advantage of, e.g., opening an
exchange gap in the TI with less disorder than in doped systems since no extra scattering centers need to be introduced.
It is proposed that when a non-magnetic TI is combined with another magnetically ordered material,
proximity effects at the interface involving the surface Dirac
fermions will align spin moments of the TI band with the itinerant
carriers, and give rise to exchange splitting.
Similar exchange mechanisms are used in stacks with MTIs to alter their magnetic properties.
Engineering heterostructures by incorporating magnetic and
non-magnetic TIs has therefore opened up new avenues to access emergent
quantum phenomena.
In the following we review some of these aspects of
magnetic heterostructures incorporating TIs.

\begin{figure*}[ht!]
    \begin{center}
        \includegraphics[width=17cm]{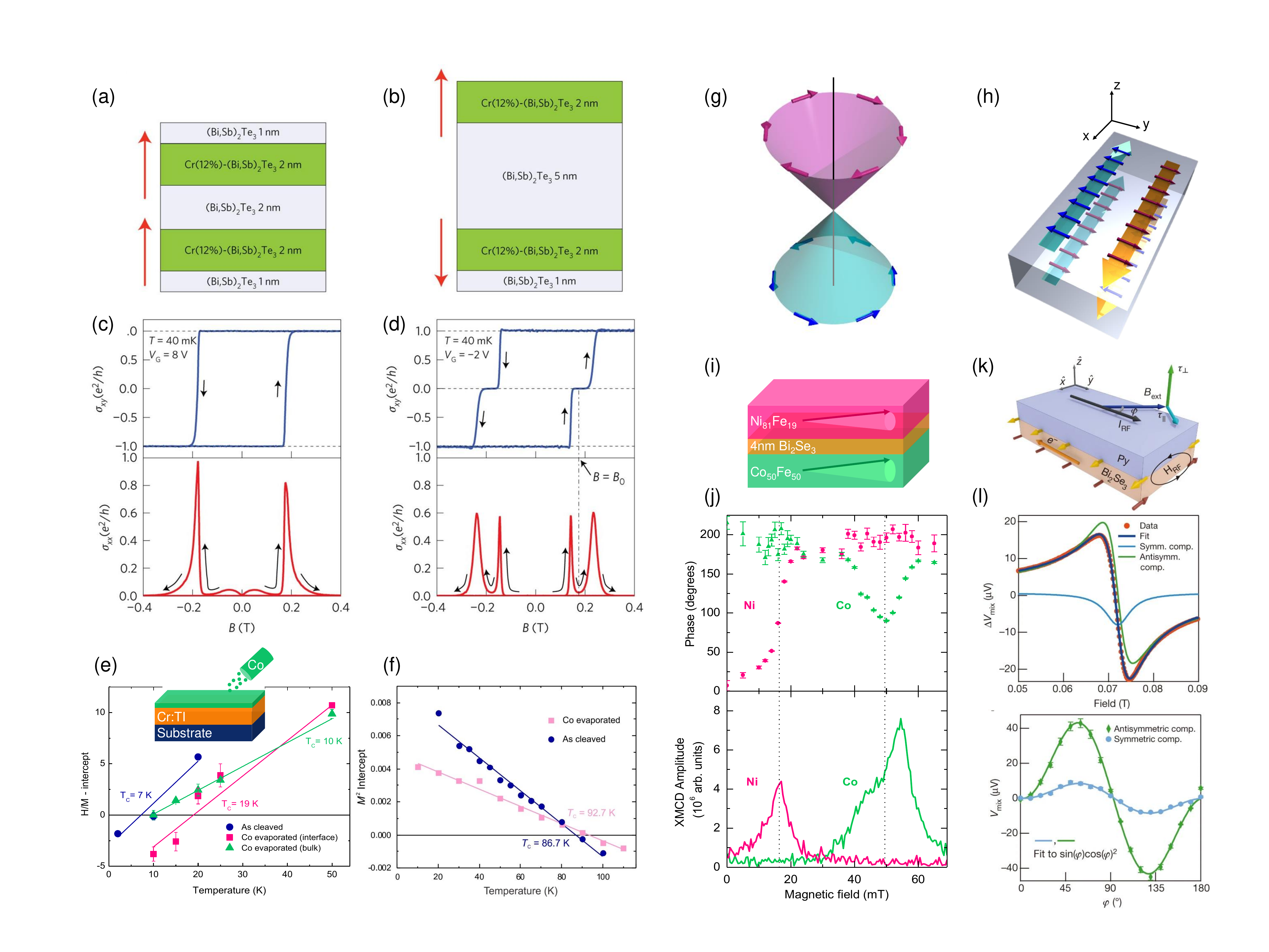}
        \vspace*{-0.5cm}
        \caption{Emergent phenomena in TI heterostructures.
        Schematic Cr-modulation-doped (a) quantum anomalous Hall and (b) axion insulators. 
        Panels (c) and (d) show the magnetic field ($B$) dependence of the Hall conductivity ($\sigma_\mathrm{xy}$) and longitudinal conductivity ($\sigma_\mathrm{xx}$) in the respective heterostructures films illustrated in (a) and (b).
        Panels (e) and (f) depict linear fits of parameters extracted from Arrott plots obtained from XMCD measurements in Cr-doped (e) Bi$_2$Se$_3$ and (f) Sb$_2$Te$_3$ thin films. The temperature at which the intercept goes through zero yields $T_\textrm{C}$. The inset illustrates the process of \textit{in-situ} evaporation of a thin Co layer on top of the Cr-doped TI film.
        (g) Schematic illustration of the spin-momentum locked helical spin texture of the TSSs in a TI. (h) Schematic of surface spin density on two opposite surfaces of a TI thin film for a charge current flowing along the -$x$ direction and for a charge current flowing along the +$x$ direction.
        (i) Schematic of a FM/Bi$_2$Se$_3$/FM heterostructure for spin pumping experiments with FMR and time-resolved FMR. 
        (j) Phase (top panel) and amplitude (bottom panel) of precession of magnetization for Py (pink circles) and Co$_{50}$Fe$_{50}$ (green triangles) layers for $t_{\textrm{\tiny{TI}}}=$4~nm. Dashed lines show the positions of the resonance amplitude peak.
        (k) Schematic diagram of SOT measurements in a Py/Bi$_2$Se$_3$ bilayer. The yellow and red arrows denote spin moment directions. (l) Measured field dependent (top panel) spin-torque FMR at room temperature for a Py (16~nm)/Bi$_2$Se$_3$ (8~nm) bilayer, together with fits of the symmetric and antisymmetric resonance components. 
        Bottom panel: Measured dependence on the magnetic field angle $\varphi$ for the symmetric and antisymmetric resonance components.
        Panels (a)-(d) reprinted with permission from Springer Nature, Mogi \emph{et al.}, Nature Mater.\ \textbf{16}, 516 (2017);\cite{Mogi2017} Copyright (2017).
        Panel (e) adapted and reprinted with permission from Baker \emph{et al.}, Phys.\ Rev.\ B \textbf{92}, 094420 (2015).\cite{baker2015magnetic}
        Copyright 2015 by the American Physical Society.
        Panel (f) adapted from Duffy \emph{et al.}, Phys.\ Rev.\ B \textbf{95}, 224422 (2017);\cite{Duffy2017} $\copyright$ CC BY 4.0.
        Panels (i) and (j) reprinted from J.\ Magn.\ Magn.\ Mater.\ \textbf{400}, 178 (2016),\cite{Figueroa_XFMR2016} with permission from Elsevier. 
        Panels (k)-(l) reprinted with permission from Springer Nature, Mellnik \emph{et al.}, Nature \textbf{511}, 449 (2014);\cite{Mellnik_Nature2014} Copyright (2014).
        }
        \vspace*{-0.5cm}
        \label{fig:heterostructures}
    \end{center}
\end{figure*}

Several strategies have been followed in order to exploit magnetic
proximity effects at the interface of such heterostructures with the goal of tuning and controlling the magnetic and quantum transport properties of MTIs. Modulation doping in heterostructures with Cr-doped (Bi,Sb)$_2$Te$_3$ films has increased the QAHE temperature from $\sim$30~mK up to 2~K.\cite{Mogi2015} 
The designed structures are illustrated in Figs.\ \ref{fig:heterostructures}(a) and \ref{fig:heterostructures}(b), and the characteristic magnetotransport data showing the QAHE are plotted in Fig.\ \ref{fig:heterostructures}(c). 
This approach has also allowed for the observation of the axion insulator state in MTIs\cite{Mogi2017} --- a novel magnetoelectric phenomenon characterized by a large longitudinal resistance and zero Hall plateau (where
both the Hall and longitudinal conductivity become zero), which is illustrated in Fig.\ \ref{fig:heterostructures}(d).

In order to raise the temperature at which quantum and magnetoelectric effects in MTIs can be observed, it seems natural to explore ways to increase their magnetic ordering temperature by, e.g., interfacial magnetic interactions. 
Growth of Cr-doped Bi$_2$Se$_3$ films onto the high $T_\mathrm{C}$ ($\sim$550~K) ferrimagnetic insulator Y$_3$Fe$_5$O$_{12}$ (YIG) was found to enhance the $T_\mathrm{C}$ of the MTI from $\sim$20~K up to $\sim$50~K.\cite{Liu2015}
The evaporation of a thin FM layer, such as Fe or Co, onto the pristine surface of \textit{in-vacuum} cleaved Mn-doped Bi$_2$Te$_3$ crystals or Cr-doped Bi$_2$Se$_3$, Cr-doped Sb$_2$Te$_3$ and Dy-doped Bi$_2$Te$_3$ thin films was also demonstrated to alter their $T_\mathrm{C}$ near the surface of the TI through proximity.\cite{Vobornik2011, baker2015magnetic, Duffy2017, Figueroa_2017}
Figures \ref{fig:heterostructures}(e) and \ref{fig:heterostructures}(f) depict the increase in $T_\mathrm{C}$ for Cr-doped Bi$_2$Se$_3$ and Sb$_2$Te$_3$ films with a Co overlayer, as obtained from XMCD measurements.
Growth of high-quality Dy:Bi$_2$Te$_3$/Cr:Sb$_2$Te$_3$ heterostructures resulted in successful introduction of long-range magnetic order in Dy:Bi$_2$Te$_3$ as a result of proximity coupling to the higher transition temperature Cr:Sb$_2$Te$_3$ layer.\cite{Duffy2018}

Alternative approaches to raise $T_\mathrm{C}$ of MTIs include their coupling to antiferromagnets.
An example are superlattices of the AF CrSb and Cr-doped (Bi,Sb)$_2$Te$_3$, where $T_\mathrm{C}$ was found to increase from $\sim$38~K up to $\sim$90~K.\cite{Hesjedal_hetero:2017, He_hetero_2017} 
The successful design of such structures is relying on the fact that these materials are chemically very compatible and well lattice-matched, resulting in abrupt interfaces which support the magnetic exchange across the interface.
This concept of using materials with related, or at least comparable crystal structure and similar atomic compositions has also been explored, both theoretically and experimentally, in heterostructures of MnBi$_2$Te$_4$/Bi$_2$Te$_3$, MnBi$_2$Se$_4$/Bi$_2$Se$_3$, MnSe/Bi$_2$Se$_3$, and similar compounds.\cite{Otrokov_JETP2017, Otrokov_2DMat2017, Hirahara_NanoLett2017, EremeevNanolett2018}
Especially heterostructures of, e.g., MnBi$_2$Te$_4$ and Bi$_2$Te$_3$, which are part of the (MnBi$_2$Te$_4$)(Bi$_2$Te$_3$)$_m$ homologous series,\cite{Klimovskikh2020} are particularly suited as there are no real interfaces (the MnBi$_2$Te$_4$ layer is effectively terminated by Bi$_2$Te$_3$ layers).
This makes the magnetic film a natural extension of the TI and allows for efficient TRS breaking.
In fact, some of these systems exhibit ferromagnetism up to room temperature and a clear Dirac cone gap opening of $\sim$100~meV.\cite{Hirahara_NanoLett2017}

The latter examples have an additional advantage in terms of the
electrical characteristic of the magnetic layer. 
They use magnetic insulators (MIs), instead of metallic films, which preserves the band structure of the TI, and thus the TSS, as they avoid hybridization with the bulk states of the AF or FM in contact. 
Using a MI is also convenient for the fabrication and incorporation of such structures into quantum devices.
Intensive investigation of proximity effects has been done in heterostructures comprising TIs and MIs, such as EuS/Bi$_{2}$Se$_3$,
EuS/(Bi,Sb)$_{2}$Te$_3$, (Bi,Sb)$_2$Te$_3$/Y$_3$Fe$_5$O$_{12}$,
(Bi,Sb)$_2$Te$_3$/Tm$_3$Fe$_5$O$_{12}$, Bi$_2$Te$_3$/Fe$_3$O$_{4}$,
(Bi,Sb)$_2$Te$_3$/Cr$_2$Ge$_2$Te$_6$, and
(Zn,Cr)Te/(Bi,Sb)$_2$Te$_3$/(Zn,Cr)Te.\cite{WeiPRL2013, KatmisNat2016, Lee2016-prox,JiangNanoLett2015, LiPRB2017,TangSciAdv2017,Krieger_PRB2019,Mogi2019, Watanabe2019, Pereira_PRM2020}
However, the QAHE has only been reported in
(Zn,Cr)Te/(Bi,Sb)$_2$Te$_3$/(Zn,Cr)Te,\cite{Watanabe2019}
which results from the appropriate design of the heterostructure to
guarantee smooth interfaces (as it satisfies the conditions
described above), as well as the maximization of the exchange gap at
the top and bottom surface states by using a MI/TI/MI sandwich
stack.

Proximity effects due to exchange interactions are not the only interesting phenomena that emerge at TI interfaces. 
The helical locking of the TSS, illustrated in Fig.\ \ref{fig:heterostructures}(g), and the strong spin-orbit coupling are very relevant for their application in topological spintronics. 
The existence of such a spin texture means that an electron current flowing on the surface of a TI can generate a non-equilibrium spin density [see Fig.\ \ref{fig:heterostructures}(h)] with both in-plane and out-of-plane components. 
The spin-polarized surface currents are expected to efficiently induce out-of-plane and in-plane torques in an adjacent magnetic layer. Experiments of spin-pumping using ferromagnetic resonance (FMR) techniques in FM/TI and FM/TI/FM stacks, as the one depicted in Fig.\ \ref{fig:heterostructures}(i), have confirmed the ability of a TI to absorb and transfer a pure spin current.\cite{Shiomi_PRL2014, Baker_XFMR2015, Jamali_2015, Figueroa_XFMR2016, Baker_XFMR2019}
Figure \ref{fig:heterostructures}(j) shows signatures of transfer of angular momentum between the FMs Co$_{50}$Fe$_{50}$ and Ni$_{81}$Fe$_{19}$ (Py, Permalloy), mediated by Bi$_2$Se$_3$, as observed in the amplitude and phase of precession of each FM layer recorded in time-resolved FMR measurements.\cite{Baker_XFMR2015, Figueroa_XFMR2016}

The absence of bulk conduction in ideal TIs further increases their spin-charge conversion efficiency. Exceptionally large spin-charge conversion values at TI/FM interfaces have been experimentally demonstrated by spin-orbit torque (SOT) measurements.\cite{Mellnik_Nature2014, Wang_PRL2015, Rojas_PRL2016, Wang_NatCom2017, Wu_PRL2019} 
Figure \ref{fig:heterostructures}(k) shows a diagram of a Py/Bi$_2$Se$_3$ bilayer, where the non-equilibrium magnetization produced by an in-plane charge current in SOT measurements is represented. The measured resonance line shape plotted in Fig.\ \ref{fig:heterostructures}(l) has two components: a symmetric, or antidamping-like SOT (in-plane), and an antisymmetric or field-like SOT (perpendicular). Analysis of these FMR curves yields spin-charge conversion parameters that are found to be much larger than those reported for heterostructures of FM with heavy metals.\cite{Mellnik_Nature2014}
SOT in such FM/TI heterostructures can be used to efficiently manipulate the magnetization of the FM --- a highly relevant feature for their applications in spintronic devices.
Effective magnetization switching by SOT at room temperature has been demonstrated in, e.g., Py/Bi$_2$Se$_3$,\cite{Wang_NatCom2017} CoTb/Bi$_2$Se$_3$, and CoTb/(Bi,Sb)$_2$Te$_3$ structures.\cite{Han_PRL2017}

\section{Discussion and Conclusions}

Despite the huge efforts striving to make practical use of the quantum and spin phenomena at the surfaces and interfaces of TIs, and despite the remarkable progress in the growth of thin films and complex heterostructures, there are a number of limitations intrinsic to these materials and many questions are still waiting to be addressed. 
Most of the experiments reviewed here rely on highly (charge) doped TIs in which the TSS coexist with bulk bands at the Fermi level, and for their analysis it is assumed that the FM/TI interface is perfectly abrupt. However, magnetic phase segregation and interdiffusion are common issues in doped TIs and at FM/TI interfaces, and only recently, the systematic analysis of FM/TI interfaces has started, including practical strategies for improving their quality.\cite{Wu_PRL2019, Bonell_NanoLett2020}
The fact that the QAHE, as well as other quantum, magnetoelectric and spin phenomena continue to be only observable at very low temperatures can, in fact, be associated with this disorder at the topological interface. 
Therefore, there is a clear path to improve the quality of these systems, and there is still real potential for significantly increasing the energy and temperature scales at which these intriguing quantum effects can be observed, opening the door for useful quantum devices in the future.\cite{Fei2020}  
So, coming back to the question we raised in the abstract, i.e., whether there is hope for magnetically doped 3D TIs in the (Sb,Bi)$_2$(Se,Te)$_3$ family, we hope to have been able to provide some evidence for answering the question with a cautious `yes'.\\

\section*{Acknowledgments}
Authors express their deep gratitude to the medical care personnel
during the Covid-19 pandemic. 
We acknowledge continuous support for the TI project at Oxford from the John Fell Fund (Oxford University Press), the synchrotron facilities Diamond Light Source, ALS, BESSYII, ALBA, and the ESRF and the neutron sources ISIS and ILL for beamtime leading to the reported insights. 
A.I.F.\ acknowledges funding from the European Union's Horizon 2020 research and innovation program under the Marie Sk\l{}odowska-Curie grant agreement No.\ 796925.

\section*{Data availability}
Data sharing is not applicable to this article as no new data were created or analyzed in this study.


\begin{thebibliography}{163}%
\makeatletter
\providecommand \@ifxundefined [1]{%
 \@ifx{#1\undefined}
}%
\providecommand \@ifnum [1]{%
 \ifnum #1\expandafter \@firstoftwo
 \else \expandafter \@secondoftwo
 \fi
}%
\providecommand \@ifx [1]{%
 \ifx #1\expandafter \@firstoftwo
 \else \expandafter \@secondoftwo
 \fi
}%
\providecommand \natexlab [1]{#1}%
\providecommand \enquote  [1]{``#1''}%
\providecommand \bibnamefont  [1]{#1}%
\providecommand \bibfnamefont [1]{#1}%
\providecommand \citenamefont [1]{#1}%
\providecommand \href@noop [0]{\@secondoftwo}%
\providecommand \href [0]{\begingroup \@sanitize@url \@href}%
\providecommand \@href[1]{\@@startlink{#1}\@@href}%
\providecommand \@@href[1]{\endgroup#1\@@endlink}%
\providecommand \@sanitize@url [0]{\catcode `\\12\catcode `\$12\catcode
  `\&12\catcode `\#12\catcode `\^12\catcode `\_12\catcode `\%12\relax}%
\providecommand \@@startlink[1]{}%
\providecommand \@@endlink[0]{}%
\providecommand \url  [0]{\begingroup\@sanitize@url \@url }%
\providecommand \@url [1]{\endgroup\@href {#1}{\urlprefix }}%
\providecommand \urlprefix  [0]{URL }%
\providecommand \Eprint [0]{\href }%
\providecommand \doibase [0]{http://dx.doi.org/}%
\providecommand \selectlanguage [0]{\@gobble}%
\providecommand \bibinfo  [0]{\@secondoftwo}%
\providecommand \bibfield  [0]{\@secondoftwo}%
\providecommand \translation [1]{[#1]}%
\providecommand \BibitemOpen [0]{}%
\providecommand \bibitemStop [0]{}%
\providecommand \bibitemNoStop [0]{.\EOS\space}%
\providecommand \EOS [0]{\spacefactor3000\relax}%
\providecommand \BibitemShut  [1]{\csname bibitem#1\endcsname}%
\let\auto@bib@innerbib\@empty
\bibitem [{\citenamefont {Fu}, \citenamefont {Kane},\ and\ \citenamefont
  {Mele}(2007)}]{Fu2007}%
  \BibitemOpen
  \bibfield  {author} {\bibinfo {author} {\bibfnamefont {L.}~\bibnamefont
  {Fu}}, \bibinfo {author} {\bibfnamefont {C.~L.}\ \bibnamefont {Kane}}, \ and\
  \bibinfo {author} {\bibfnamefont {E.~J.}\ \bibnamefont {Mele}},\ }\href@noop
  {} {\bibfield  {journal} {\bibinfo  {journal} {Phys. Rev. Lett.}\ }\textbf
  {\bibinfo {volume} {98}},\ \bibinfo {pages} {106803} (\bibinfo {year}
  {2007})}\BibitemShut {NoStop}%
\bibitem [{\citenamefont {Goldsmid}\ and\ \citenamefont
  {Douglas}(1954)}]{Goldsmid1954}%
  \BibitemOpen
  \bibfield  {author} {\bibinfo {author} {\bibfnamefont {H.~J.}\ \bibnamefont
  {Goldsmid}}\ and\ \bibinfo {author} {\bibfnamefont {R.~W.}\ \bibnamefont
  {Douglas}},\ }\href {\doibase 10.1088/0508-3443/5/11/303} {\bibfield
  {journal} {\bibinfo  {journal} {Br. J. Appl. Phys.}\ }\textbf {\bibinfo
  {volume} {5}},\ \bibinfo {pages} {386} (\bibinfo {year} {1954})}\BibitemShut
  {NoStop}%
\bibitem [{\citenamefont {Goldsmid}, \citenamefont {Sheard},\ and\
  \citenamefont {Wright}(1958)}]{Goldsmid_1958}%
  \BibitemOpen
  \bibfield  {author} {\bibinfo {author} {\bibfnamefont {H.~J.}\ \bibnamefont
  {Goldsmid}}, \bibinfo {author} {\bibfnamefont {A.~R.}\ \bibnamefont
  {Sheard}}, \ and\ \bibinfo {author} {\bibfnamefont {D.~A.}\ \bibnamefont
  {Wright}},\ }\href {\doibase 10.1088/0508-3443/9/9/306} {\bibfield  {journal}
  {\bibinfo  {journal} {Br. J. Appl. Phys.}\ }\textbf {\bibinfo {volume} {9}},\
  \bibinfo {pages} {365} (\bibinfo {year} {1958})}\BibitemShut {NoStop}%
\bibitem [{\citenamefont {Zhang}\ \emph
  {et~al.}(2009{\natexlab{a}})\citenamefont {Zhang}, \citenamefont {Liu},
  \citenamefont {Qi}, \citenamefont {Dai}, \citenamefont {Fang},\ and\
  \citenamefont {Zhang}}]{Zhang2009}%
  \BibitemOpen
  \bibfield  {author} {\bibinfo {author} {\bibfnamefont {H.}~\bibnamefont
  {Zhang}}, \bibinfo {author} {\bibfnamefont {C.-X.}\ \bibnamefont {Liu}},
  \bibinfo {author} {\bibfnamefont {X.-L.}\ \bibnamefont {Qi}}, \bibinfo
  {author} {\bibfnamefont {X.}~\bibnamefont {Dai}}, \bibinfo {author}
  {\bibfnamefont {Z.}~\bibnamefont {Fang}}, \ and\ \bibinfo {author}
  {\bibfnamefont {S.-C.}\ \bibnamefont {Zhang}},\ }\href@noop {} {\bibfield
  {journal} {\bibinfo  {journal} {Nat. Phys.}\ }\textbf {\bibinfo {volume}
  {5}},\ \bibinfo {pages} {438} (\bibinfo {year}
  {2009}{\natexlab{a}})}\BibitemShut {NoStop}%
\bibitem [{\citenamefont {Chen}\ \emph {et~al.}(2009)\citenamefont {Chen},
  \citenamefont {Analytis}, \citenamefont {Chu}, \citenamefont {Liu},
  \citenamefont {Mo}, \citenamefont {Qi}, \citenamefont {Zhang}, \citenamefont
  {Lu}, \citenamefont {Dai}, \citenamefont {Fang}, \citenamefont {Zhang},
  \citenamefont {Fisher}, \citenamefont {Hussain},\ and\ \citenamefont
  {Shen}}]{Chen2009}%
  \BibitemOpen
  \bibfield  {author} {\bibinfo {author} {\bibfnamefont {Y.~L.}\ \bibnamefont
  {Chen}}, \bibinfo {author} {\bibfnamefont {J.~G.}\ \bibnamefont {Analytis}},
  \bibinfo {author} {\bibfnamefont {J.-H.}\ \bibnamefont {Chu}}, \bibinfo
  {author} {\bibfnamefont {Z.~K.}\ \bibnamefont {Liu}}, \bibinfo {author}
  {\bibfnamefont {S.-K.}\ \bibnamefont {Mo}}, \bibinfo {author} {\bibfnamefont
  {X.-L.}\ \bibnamefont {Qi}}, \bibinfo {author} {\bibfnamefont {H.~J.}\
  \bibnamefont {Zhang}}, \bibinfo {author} {\bibfnamefont {D.~H.}\ \bibnamefont
  {Lu}}, \bibinfo {author} {\bibfnamefont {X.}~\bibnamefont {Dai}}, \bibinfo
  {author} {\bibfnamefont {Z.}~\bibnamefont {Fang}}, \bibinfo {author}
  {\bibfnamefont {S.~C.}\ \bibnamefont {Zhang}}, \bibinfo {author}
  {\bibfnamefont {I.~R.}\ \bibnamefont {Fisher}}, \bibinfo {author}
  {\bibfnamefont {Z.}~\bibnamefont {Hussain}}, \ and\ \bibinfo {author}
  {\bibfnamefont {Z.-X.}\ \bibnamefont {Shen}},\ }\href@noop {} {\bibfield
  {journal} {\bibinfo  {journal} {Science}\ }\textbf {\bibinfo {volume}
  {325}},\ \bibinfo {pages} {178} (\bibinfo {year} {2009})}\BibitemShut
  {NoStop}%
\bibitem [{\citenamefont {Hor}\ \emph {et~al.}(2009)\citenamefont {Hor},
  \citenamefont {Richardella}, \citenamefont {Roushan}, \citenamefont {Xia},
  \citenamefont {Checkelsky}, \citenamefont {Yazdani}, \citenamefont {Hasan},
  \citenamefont {Ong},\ and\ \citenamefont {Cava}}]{Hor2009}%
  \BibitemOpen
  \bibfield  {author} {\bibinfo {author} {\bibfnamefont {Y.~S.}\ \bibnamefont
  {Hor}}, \bibinfo {author} {\bibfnamefont {A.}~\bibnamefont {Richardella}},
  \bibinfo {author} {\bibfnamefont {P.}~\bibnamefont {Roushan}}, \bibinfo
  {author} {\bibfnamefont {Y.}~\bibnamefont {Xia}}, \bibinfo {author}
  {\bibfnamefont {J.~G.}\ \bibnamefont {Checkelsky}}, \bibinfo {author}
  {\bibfnamefont {A.}~\bibnamefont {Yazdani}}, \bibinfo {author} {\bibfnamefont
  {M.~Z.}\ \bibnamefont {Hasan}}, \bibinfo {author} {\bibfnamefont {N.~P.}\
  \bibnamefont {Ong}}, \ and\ \bibinfo {author} {\bibfnamefont {R.~J.}\
  \bibnamefont {Cava}},\ }\href@noop {} {\bibfield  {journal} {\bibinfo
  {journal} {{Phys. Rev. B}}\ }\textbf {\bibinfo {volume} {{79}}},\ \bibinfo
  {pages} {{195208}} (\bibinfo {year} {{2009}})}\BibitemShut {NoStop}%
\bibitem [{\citenamefont {Chang}\ \emph
  {et~al.}(2013{\natexlab{a}})\citenamefont {Chang}, \citenamefont {Zhang},
  \citenamefont {Feng}, \citenamefont {Shen}, \citenamefont {Zhang},
  \citenamefont {Guo}, \citenamefont {Li}, \citenamefont {Ou}, \citenamefont
  {Wei}, \citenamefont {Wang}, \citenamefont {Ji}, \citenamefont {Feng},
  \citenamefont {Ji}, \citenamefont {Chen}, \citenamefont {Jia}, \citenamefont
  {Dai}, \citenamefont {Fang}, \citenamefont {Zhang}, \citenamefont {He},
  \citenamefont {Wang}, \citenamefont {Lu}, \citenamefont {Ma},\ and\
  \citenamefont {Xue}}]{Chang2013}%
  \BibitemOpen
  \bibfield  {author} {\bibinfo {author} {\bibfnamefont {C.-Z.}\ \bibnamefont
  {Chang}}, \bibinfo {author} {\bibfnamefont {J.}~\bibnamefont {Zhang}},
  \bibinfo {author} {\bibfnamefont {X.}~\bibnamefont {Feng}}, \bibinfo {author}
  {\bibfnamefont {J.}~\bibnamefont {Shen}}, \bibinfo {author} {\bibfnamefont
  {Z.}~\bibnamefont {Zhang}}, \bibinfo {author} {\bibfnamefont
  {M.}~\bibnamefont {Guo}}, \bibinfo {author} {\bibfnamefont {K.}~\bibnamefont
  {Li}}, \bibinfo {author} {\bibfnamefont {Y.}~\bibnamefont {Ou}}, \bibinfo
  {author} {\bibfnamefont {P.}~\bibnamefont {Wei}}, \bibinfo {author}
  {\bibfnamefont {L.-L.}\ \bibnamefont {Wang}}, \bibinfo {author}
  {\bibfnamefont {Z.-Q.}\ \bibnamefont {Ji}}, \bibinfo {author} {\bibfnamefont
  {Y.}~\bibnamefont {Feng}}, \bibinfo {author} {\bibfnamefont {S.}~\bibnamefont
  {Ji}}, \bibinfo {author} {\bibfnamefont {X.}~\bibnamefont {Chen}}, \bibinfo
  {author} {\bibfnamefont {J.}~\bibnamefont {Jia}}, \bibinfo {author}
  {\bibfnamefont {X.}~\bibnamefont {Dai}}, \bibinfo {author} {\bibfnamefont
  {Z.}~\bibnamefont {Fang}}, \bibinfo {author} {\bibfnamefont {S.-C.}\
  \bibnamefont {Zhang}}, \bibinfo {author} {\bibfnamefont {K.}~\bibnamefont
  {He}}, \bibinfo {author} {\bibfnamefont {Y.}~\bibnamefont {Wang}}, \bibinfo
  {author} {\bibfnamefont {L.}~\bibnamefont {Lu}}, \bibinfo {author}
  {\bibfnamefont {X.-C.}\ \bibnamefont {Ma}}, \ and\ \bibinfo {author}
  {\bibfnamefont {Q.-K.}\ \bibnamefont {Xue}},\ }\href@noop {} {\bibfield
  {journal} {\bibinfo  {journal} {Science}\ }\textbf {\bibinfo {volume}
  {340}},\ \bibinfo {pages} {167} (\bibinfo {year}
  {2013}{\natexlab{a}})}\BibitemShut {NoStop}%
\bibitem [{\citenamefont {Qi}\ \emph {et~al.}(2009)\citenamefont {Qi},
  \citenamefont {Hughes}, \citenamefont {Raghu},\ and\ \citenamefont
  {Zhang}}]{QiPRL2009}%
  \BibitemOpen
  \bibfield  {author} {\bibinfo {author} {\bibfnamefont {X.-L.}\ \bibnamefont
  {Qi}}, \bibinfo {author} {\bibfnamefont {T.~L.}\ \bibnamefont {Hughes}},
  \bibinfo {author} {\bibfnamefont {S.}~\bibnamefont {Raghu}}, \ and\ \bibinfo
  {author} {\bibfnamefont {S.-C.}\ \bibnamefont {Zhang}},\ }\href {\doibase
  10.1103/PhysRevLett.102.187001} {\bibfield  {journal} {\bibinfo  {journal}
  {Phys. Rev. Lett.}\ }\textbf {\bibinfo {volume} {102}},\ \bibinfo {pages}
  {187001} (\bibinfo {year} {2009})}\BibitemShut {NoStop}%
\bibitem [{\citenamefont {Qi}\ and\ \citenamefont {Zhang}(2011)}]{Qi2011}%
  \BibitemOpen
  \bibfield  {author} {\bibinfo {author} {\bibfnamefont {X.-L.}\ \bibnamefont
  {Qi}}\ and\ \bibinfo {author} {\bibfnamefont {S.~C.}\ \bibnamefont {Zhang}},\
  }\href@noop {} {\bibfield  {journal} {\bibinfo  {journal} {{Rev. Mod.
  Phys.}}\ }\textbf {\bibinfo {volume} {{83}}},\ \bibinfo {pages} {1057}
  (\bibinfo {year} {2011})}\BibitemShut {NoStop}%
\bibitem [{\citenamefont {Pesin}\ and\ \citenamefont
  {MacDonald}(2012)}]{Pesin:2012kx}%
  \BibitemOpen
  \bibfield  {author} {\bibinfo {author} {\bibfnamefont {D.}~\bibnamefont
  {Pesin}}\ and\ \bibinfo {author} {\bibfnamefont {A.~H.}\ \bibnamefont
  {MacDonald}},\ }\href@noop {} {\bibfield  {journal} {\bibinfo  {journal}
  {Nat. Mater.}\ }\textbf {\bibinfo {volume} {11}},\ \bibinfo {pages} {409}
  (\bibinfo {year} {2012})}\BibitemShut {NoStop}%
\bibitem [{\citenamefont {Niu}\ \emph {et~al.}(2011)\citenamefont {Niu},
  \citenamefont {Dai}, \citenamefont {Guo}, \citenamefont {Wei}, \citenamefont
  {Ma},\ and\ \citenamefont {Huang}}]{Niu2011}%
  \BibitemOpen
  \bibfield  {author} {\bibinfo {author} {\bibfnamefont {C.}~\bibnamefont
  {Niu}}, \bibinfo {author} {\bibfnamefont {Y.}~\bibnamefont {Dai}}, \bibinfo
  {author} {\bibfnamefont {M.}~\bibnamefont {Guo}}, \bibinfo {author}
  {\bibfnamefont {W.}~\bibnamefont {Wei}}, \bibinfo {author} {\bibfnamefont
  {Y.}~\bibnamefont {Ma}}, \ and\ \bibinfo {author} {\bibfnamefont
  {B.}~\bibnamefont {Huang}},\ }\href@noop {} {\bibfield  {journal} {\bibinfo
  {journal} {{Appl. Phys. Lett.}}\ }\textbf {\bibinfo {volume} {98}},\ \bibinfo
  {pages} {252502} (\bibinfo {year} {2011})}\BibitemShut {NoStop}%
\bibitem [{\citenamefont {Elkholdi}\ \emph {et~al.}(1994)\citenamefont
  {Elkholdi}, \citenamefont {Averous}, \citenamefont {Charar}, \citenamefont
  {Fau}, \citenamefont {Brun}, \citenamefont {Ghoumaribouanani},\ and\
  \citenamefont {Deportes}}]{Elkholdi1994}%
  \BibitemOpen
  \bibfield  {author} {\bibinfo {author} {\bibfnamefont {M.}~\bibnamefont
  {Elkholdi}}, \bibinfo {author} {\bibfnamefont {M.}~\bibnamefont {Averous}},
  \bibinfo {author} {\bibfnamefont {S.}~\bibnamefont {Charar}}, \bibinfo
  {author} {\bibfnamefont {C.}~\bibnamefont {Fau}}, \bibinfo {author}
  {\bibfnamefont {G.}~\bibnamefont {Brun}}, \bibinfo {author} {\bibfnamefont
  {H.}~\bibnamefont {Ghoumaribouanani}}, \ and\ \bibinfo {author}
  {\bibfnamefont {J.}~\bibnamefont {Deportes}},\ }\href@noop {} {\bibfield
  {journal} {\bibinfo  {journal} {{Phys. Rev. B}}\ }\textbf {\bibinfo {volume}
  {{49}}},\ \bibinfo {pages} {{1711}} (\bibinfo {year} {{1994}})}\BibitemShut
  {NoStop}%
\bibitem [{\citenamefont {Song}\ \emph {et~al.}(2012)\citenamefont {Song},
  \citenamefont {Yang}, \citenamefont {Yao}, \citenamefont {Zhu}, \citenamefont
  {Miao}, \citenamefont {Xu}, \citenamefont {Wang}, \citenamefont {Li},
  \citenamefont {Yao}, \citenamefont {Ji}, \citenamefont {Qiao}, \citenamefont
  {Sun}, \citenamefont {Zhang}, \citenamefont {Gao}, \citenamefont {Liu},
  \citenamefont {Qian}, \citenamefont {Gao},\ and\ \citenamefont
  {Jia}}]{Song2012}%
  \BibitemOpen
  \bibfield  {author} {\bibinfo {author} {\bibfnamefont {Y.~R.}\ \bibnamefont
  {Song}}, \bibinfo {author} {\bibfnamefont {F.}~\bibnamefont {Yang}}, \bibinfo
  {author} {\bibfnamefont {M.-Y.}\ \bibnamefont {Yao}}, \bibinfo {author}
  {\bibfnamefont {F.}~\bibnamefont {Zhu}}, \bibinfo {author} {\bibfnamefont
  {L.}~\bibnamefont {Miao}}, \bibinfo {author} {\bibfnamefont {J.-P.}\
  \bibnamefont {Xu}}, \bibinfo {author} {\bibfnamefont {M.-X.}\ \bibnamefont
  {Wang}}, \bibinfo {author} {\bibfnamefont {H.}~\bibnamefont {Li}}, \bibinfo
  {author} {\bibfnamefont {X.}~\bibnamefont {Yao}}, \bibinfo {author}
  {\bibfnamefont {F.}~\bibnamefont {Ji}}, \bibinfo {author} {\bibfnamefont
  {S.}~\bibnamefont {Qiao}}, \bibinfo {author} {\bibfnamefont {Z.}~\bibnamefont
  {Sun}}, \bibinfo {author} {\bibfnamefont {G.~B.}\ \bibnamefont {Zhang}},
  \bibinfo {author} {\bibfnamefont {B.}~\bibnamefont {Gao}}, \bibinfo {author}
  {\bibfnamefont {C.}~\bibnamefont {Liu}}, \bibinfo {author} {\bibfnamefont
  {D.}~\bibnamefont {Qian}}, \bibinfo {author} {\bibfnamefont {C.~L.}\
  \bibnamefont {Gao}}, \ and\ \bibinfo {author} {\bibfnamefont {J.-F.}\
  \bibnamefont {Jia}},\ }\href@noop {} {\bibfield  {journal} {\bibinfo
  {journal} {{Appl. Phys. Lett.}}\ }\textbf {\bibinfo {volume} {{100}}},\
  \bibinfo {pages} {{242403}} (\bibinfo {year} {{2012}})}\BibitemShut {NoStop}%
\bibitem [{\citenamefont {Harrison}\ \emph
  {et~al.}(2014{\natexlab{a}})\citenamefont {Harrison}, \citenamefont
  {Collins-McIntyre}, \citenamefont {Li}, \citenamefont {Baker}, \citenamefont
  {Shelford}, \citenamefont {Huo}, \citenamefont {Pushp}, \citenamefont
  {Parkin}, \citenamefont {Harris}, \citenamefont {Arenholz}, \citenamefont
  {van~der Laan},\ and\ \citenamefont {Hesjedal}}]{Harrison2014_Gd}%
  \BibitemOpen
  \bibfield  {author} {\bibinfo {author} {\bibfnamefont {S.~E.}\ \bibnamefont
  {Harrison}}, \bibinfo {author} {\bibfnamefont {L.~J.}\ \bibnamefont
  {Collins-McIntyre}}, \bibinfo {author} {\bibfnamefont {S.}~\bibnamefont
  {Li}}, \bibinfo {author} {\bibfnamefont {A.~A.}\ \bibnamefont {Baker}},
  \bibinfo {author} {\bibfnamefont {L.~R.}\ \bibnamefont {Shelford}}, \bibinfo
  {author} {\bibfnamefont {Y.}~\bibnamefont {Huo}}, \bibinfo {author}
  {\bibfnamefont {A.}~\bibnamefont {Pushp}}, \bibinfo {author} {\bibfnamefont
  {S.~S.~P.}\ \bibnamefont {Parkin}}, \bibinfo {author} {\bibfnamefont {J.~S.}\
  \bibnamefont {Harris}}, \bibinfo {author} {\bibfnamefont {E.}~\bibnamefont
  {Arenholz}}, \bibinfo {author} {\bibfnamefont {G.}~\bibnamefont {van~der
  Laan}}, \ and\ \bibinfo {author} {\bibfnamefont {T.}~\bibnamefont
  {Hesjedal}},\ }\href@noop {} {\bibfield  {journal} {\bibinfo  {journal} {{J.
  Appl. Phys.}}\ }\textbf {\bibinfo {volume} {115}},\ \bibinfo {pages} {023904}
  (\bibinfo {year} {2014}{\natexlab{a}})}\BibitemShut {NoStop}%
\bibitem [{\citenamefont {Harrison}\ \emph
  {et~al.}(2015{\natexlab{a}})\citenamefont {Harrison}, \citenamefont
  {Collins-McIntyre}, \citenamefont {Zhang}, \citenamefont {Baker},
  \citenamefont {Figueroa}, \citenamefont {Kellock}, \citenamefont {Pushp},
  \citenamefont {Parkin}, \citenamefont {Harris}, \citenamefont {{van der
  Laan}},\ and\ \citenamefont {Hesjedal}}]{Harrison2015-Dy}%
  \BibitemOpen
  \bibfield  {author} {\bibinfo {author} {\bibfnamefont {S.~E.}\ \bibnamefont
  {Harrison}}, \bibinfo {author} {\bibfnamefont {L.~J.}\ \bibnamefont
  {Collins-McIntyre}}, \bibinfo {author} {\bibfnamefont {S.-L.}\ \bibnamefont
  {Zhang}}, \bibinfo {author} {\bibfnamefont {A.~A.}\ \bibnamefont {Baker}},
  \bibinfo {author} {\bibfnamefont {A.~I.}\ \bibnamefont {Figueroa}}, \bibinfo
  {author} {\bibfnamefont {A.~J.}\ \bibnamefont {Kellock}}, \bibinfo {author}
  {\bibfnamefont {A.}~\bibnamefont {Pushp}}, \bibinfo {author} {\bibfnamefont
  {S.~S.~P.}\ \bibnamefont {Parkin}}, \bibinfo {author} {\bibfnamefont {J.~S.}\
  \bibnamefont {Harris}}, \bibinfo {author} {\bibfnamefont {G.}~\bibnamefont
  {{van der Laan}}}, \ and\ \bibinfo {author} {\bibfnamefont {T.}~\bibnamefont
  {Hesjedal}},\ }\href {\doibase 10.1088/0953-8984/27/24/245602} {\bibfield
  {journal} {\bibinfo  {journal} {J. Phys.: Condens. Matter}\ }\textbf
  {\bibinfo {volume} {27}},\ \bibinfo {pages} {245602} (\bibinfo {year}
  {2015}{\natexlab{a}})}\BibitemShut {NoStop}%
\bibitem [{\citenamefont {Harrison}\ \emph
  {et~al.}(2015{\natexlab{b}})\citenamefont {Harrison}, \citenamefont
  {Collins-McIntyre}, \citenamefont {Zhang}, \citenamefont {Baker},
  \citenamefont {Figueroa}, \citenamefont {Kellock}, \citenamefont {Pushp},
  \citenamefont {Chen}, \citenamefont {Parkin}, \citenamefont {Harris},
  \citenamefont {{van der Laan}},\ and\ \citenamefont
  {Hesjedal}}]{Harrison2015-Ho}%
  \BibitemOpen
  \bibfield  {author} {\bibinfo {author} {\bibfnamefont {S.~E.}\ \bibnamefont
  {Harrison}}, \bibinfo {author} {\bibfnamefont {L.~J.}\ \bibnamefont
  {Collins-McIntyre}}, \bibinfo {author} {\bibfnamefont {S.-L.}\ \bibnamefont
  {Zhang}}, \bibinfo {author} {\bibfnamefont {A.~A.}\ \bibnamefont {Baker}},
  \bibinfo {author} {\bibfnamefont {A.~I.}\ \bibnamefont {Figueroa}}, \bibinfo
  {author} {\bibfnamefont {A.~J.}\ \bibnamefont {Kellock}}, \bibinfo {author}
  {\bibfnamefont {A.}~\bibnamefont {Pushp}}, \bibinfo {author} {\bibfnamefont
  {Y.}~\bibnamefont {Chen}}, \bibinfo {author} {\bibfnamefont {S.~S.~P.}\
  \bibnamefont {Parkin}}, \bibinfo {author} {\bibfnamefont {J.~S.}\
  \bibnamefont {Harris}}, \bibinfo {author} {\bibfnamefont {G.}~\bibnamefont
  {{van der Laan}}}, \ and\ \bibinfo {author} {\bibfnamefont {T.}~\bibnamefont
  {Hesjedal}},\ }\href@noop {} {\bibfield  {journal} {\bibinfo  {journal}
  {Appl. Phys. Lett.}\ }\textbf {\bibinfo {volume} {107}},\ \bibinfo {pages}
  {182406} (\bibinfo {year} {2015}{\natexlab{b}})}\BibitemShut {NoStop}%
\bibitem [{\citenamefont {Harrison}\ \emph
  {et~al.}(2015{\natexlab{c}})\citenamefont {Harrison}, \citenamefont
  {Collins-McIntyre}, \citenamefont {Sch{\"o}nherr}, \citenamefont {Vailionis},
  \citenamefont {Srot}, \citenamefont {van Aken}, \citenamefont {Kellock},
  \citenamefont {Pushp}, \citenamefont {Parkin}, \citenamefont {Harris},
  \citenamefont {Zhou}, \citenamefont {Chen},\ and\ \citenamefont
  {Hesjedal}}]{Harrison_SciRep2015}%
  \BibitemOpen
  \bibfield  {author} {\bibinfo {author} {\bibfnamefont {S.~E.}\ \bibnamefont
  {Harrison}}, \bibinfo {author} {\bibfnamefont {L.~J.}\ \bibnamefont
  {Collins-McIntyre}}, \bibinfo {author} {\bibfnamefont {P.}~\bibnamefont
  {Sch{\"o}nherr}}, \bibinfo {author} {\bibfnamefont {A.}~\bibnamefont
  {Vailionis}}, \bibinfo {author} {\bibfnamefont {V.}~\bibnamefont {Srot}},
  \bibinfo {author} {\bibfnamefont {P.~A.}\ \bibnamefont {van Aken}}, \bibinfo
  {author} {\bibfnamefont {A.~J.}\ \bibnamefont {Kellock}}, \bibinfo {author}
  {\bibfnamefont {A.}~\bibnamefont {Pushp}}, \bibinfo {author} {\bibfnamefont
  {S.~S.~P.}\ \bibnamefont {Parkin}}, \bibinfo {author} {\bibfnamefont {J.~S.}\
  \bibnamefont {Harris}}, \bibinfo {author} {\bibfnamefont {B.}~\bibnamefont
  {Zhou}}, \bibinfo {author} {\bibfnamefont {Y.~L.}\ \bibnamefont {Chen}}, \
  and\ \bibinfo {author} {\bibfnamefont {T.}~\bibnamefont {Hesjedal}},\ }\href
  {\doibase 10.1038/srep15767} {\bibfield  {journal} {\bibinfo  {journal} {Sci.
  Rep.}\ }\textbf {\bibinfo {volume} {5}},\ \bibinfo {pages} {15767} (\bibinfo
  {year} {2015}{\natexlab{c}})}\BibitemShut {NoStop}%
\bibitem [{\citenamefont {Collins-McIntyre}\ \emph
  {et~al.}(2014{\natexlab{a}})\citenamefont {Collins-McIntyre}, \citenamefont
  {Harrison}, \citenamefont {Sch{\"o}nherr}, \citenamefont {Steinke},
  \citenamefont {Kinane}, \citenamefont {Charlton}, \citenamefont
  {Alba-Veneroa}, \citenamefont {Pushp}, \citenamefont {Kellock}, \citenamefont
  {Parkin}, \citenamefont {Harris}, \citenamefont {Langridge}, \citenamefont
  {van~der Laan},\ and\ \citenamefont {Hesjedal}}]{Collins-McIntyre2014}%
  \BibitemOpen
  \bibfield  {author} {\bibinfo {author} {\bibfnamefont {L.~J.}\ \bibnamefont
  {Collins-McIntyre}}, \bibinfo {author} {\bibfnamefont {S.~E.}\ \bibnamefont
  {Harrison}}, \bibinfo {author} {\bibfnamefont {P.}~\bibnamefont
  {Sch{\"o}nherr}}, \bibinfo {author} {\bibfnamefont {N.-J.}\ \bibnamefont
  {Steinke}}, \bibinfo {author} {\bibfnamefont {C.~J.}\ \bibnamefont {Kinane}},
  \bibinfo {author} {\bibfnamefont {T.~R.}\ \bibnamefont {Charlton}}, \bibinfo
  {author} {\bibfnamefont {D.}~\bibnamefont {Alba-Veneroa}}, \bibinfo {author}
  {\bibfnamefont {A.}~\bibnamefont {Pushp}}, \bibinfo {author} {\bibfnamefont
  {A.~J.}\ \bibnamefont {Kellock}}, \bibinfo {author} {\bibfnamefont
  {S.~S.~P.}\ \bibnamefont {Parkin}}, \bibinfo {author} {\bibfnamefont {J.~S.}\
  \bibnamefont {Harris}}, \bibinfo {author} {\bibfnamefont {S.}~\bibnamefont
  {Langridge}}, \bibinfo {author} {\bibfnamefont {G.}~\bibnamefont {van~der
  Laan}}, \ and\ \bibinfo {author} {\bibfnamefont {T.}~\bibnamefont
  {Hesjedal}},\ }\href@noop {} {\bibfield  {journal} {\bibinfo  {journal} {EPL
  (Europhys. Lett.)}\ }\textbf {\bibinfo {volume} {107}},\ \bibinfo {pages}
  {57009} (\bibinfo {year} {2014}{\natexlab{a}})}\BibitemShut {NoStop}%
\bibitem [{\citenamefont {Collins-McIntyre}\ \emph
  {et~al.}(2014{\natexlab{b}})\citenamefont {Collins-McIntyre}, \citenamefont
  {Watson}, \citenamefont {Baker}, \citenamefont {Zhang}, \citenamefont
  {Coldea}, \citenamefont {Harrison}, \citenamefont {Pushp}, \citenamefont
  {Kellock}, \citenamefont {Parkin}, \citenamefont {van~der Laan},\ and\
  \citenamefont {Hesjedal}}]{Liam2014_MnBS_AIP}%
  \BibitemOpen
  \bibfield  {author} {\bibinfo {author} {\bibfnamefont {L.~J.}\ \bibnamefont
  {Collins-McIntyre}}, \bibinfo {author} {\bibfnamefont {M.~D.}\ \bibnamefont
  {Watson}}, \bibinfo {author} {\bibfnamefont {A.~A.}\ \bibnamefont {Baker}},
  \bibinfo {author} {\bibfnamefont {S.~L.}\ \bibnamefont {Zhang}}, \bibinfo
  {author} {\bibfnamefont {A.~I.}\ \bibnamefont {Coldea}}, \bibinfo {author}
  {\bibfnamefont {S.~E.}\ \bibnamefont {Harrison}}, \bibinfo {author}
  {\bibfnamefont {A.}~\bibnamefont {Pushp}}, \bibinfo {author} {\bibfnamefont
  {A.~J.}\ \bibnamefont {Kellock}}, \bibinfo {author} {\bibfnamefont
  {S.~S.~P.}\ \bibnamefont {Parkin}}, \bibinfo {author} {\bibfnamefont
  {G.}~\bibnamefont {van~der Laan}}, \ and\ \bibinfo {author} {\bibfnamefont
  {T.}~\bibnamefont {Hesjedal}},\ }\href {\doibase 10.1063/1.4904900}
  {\bibfield  {journal} {\bibinfo  {journal} {AIP Adv.}\ }\textbf {\bibinfo
  {volume} {4}},\ \bibinfo {pages} {127136} (\bibinfo {year}
  {2014}{\natexlab{b}})}\BibitemShut {NoStop}%
\bibitem [{\citenamefont {Duffy}\ \emph {et~al.}(2017)\citenamefont {Duffy},
  \citenamefont {Figueroa}, \citenamefont {G\l{}adczuk}, \citenamefont
  {Steinke}, \citenamefont {Kummer}, \citenamefont {van~der Laan},\ and\
  \citenamefont {Hesjedal}}]{Duffy2017}%
  \BibitemOpen
  \bibfield  {author} {\bibinfo {author} {\bibfnamefont {L.~B.}\ \bibnamefont
  {Duffy}}, \bibinfo {author} {\bibfnamefont {A.~I.}\ \bibnamefont {Figueroa}},
  \bibinfo {author} {\bibfnamefont {L.}~\bibnamefont {G\l{}adczuk}}, \bibinfo
  {author} {\bibfnamefont {N.-J.}\ \bibnamefont {Steinke}}, \bibinfo {author}
  {\bibfnamefont {K.}~\bibnamefont {Kummer}}, \bibinfo {author} {\bibfnamefont
  {G.}~\bibnamefont {van~der Laan}}, \ and\ \bibinfo {author} {\bibfnamefont
  {T.}~\bibnamefont {Hesjedal}},\ }\href {\doibase 10.1103/PhysRevB.95.224422}
  {\bibfield  {journal} {\bibinfo  {journal} {Phys. Rev. B}\ }\textbf {\bibinfo
  {volume} {95}},\ \bibinfo {pages} {224422} (\bibinfo {year}
  {2017})}\BibitemShut {NoStop}%
\bibitem [{\citenamefont {Deng}\ \emph {et~al.}(2020)\citenamefont {Deng},
  \citenamefont {Yu}, \citenamefont {Shi}, \citenamefont {Guo}, \citenamefont
  {Xu}, \citenamefont {Wang}, \citenamefont {Chen},\ and\ \citenamefont
  {Zhang}}]{Deng2020}%
  \BibitemOpen
  \bibfield  {author} {\bibinfo {author} {\bibfnamefont {Y.}~\bibnamefont
  {Deng}}, \bibinfo {author} {\bibfnamefont {Y.}~\bibnamefont {Yu}}, \bibinfo
  {author} {\bibfnamefont {M.~Z.}\ \bibnamefont {Shi}}, \bibinfo {author}
  {\bibfnamefont {Z.}~\bibnamefont {Guo}}, \bibinfo {author} {\bibfnamefont
  {Z.}~\bibnamefont {Xu}}, \bibinfo {author} {\bibfnamefont {J.}~\bibnamefont
  {Wang}}, \bibinfo {author} {\bibfnamefont {X.~H.}\ \bibnamefont {Chen}}, \
  and\ \bibinfo {author} {\bibfnamefont {Y.}~\bibnamefont {Zhang}},\ }\href
  {\doibase 10.1126/science.aax8156} {\bibfield  {journal} {\bibinfo  {journal}
  {Science}\ }\textbf {\bibinfo {volume} {367}},\ \bibinfo {pages} {895}
  (\bibinfo {year} {2020})}\BibitemShut {NoStop}%
\bibitem [{\citenamefont {Chang}\ \emph
  {et~al.}(2013{\natexlab{b}})\citenamefont {Chang}, \citenamefont {Zhang},
  \citenamefont {Liu}, \citenamefont {Zhang}, \citenamefont {Feng},
  \citenamefont {Li}, \citenamefont {Wang}, \citenamefont {Chen}, \citenamefont
  {Dai}, \citenamefont {Fang}, \citenamefont {Qi}, \citenamefont {Zhang},
  \citenamefont {Wang}, \citenamefont {He}, \citenamefont {Ma},\ and\
  \citenamefont {Xue}}]{Chang2011}%
  \BibitemOpen
  \bibfield  {author} {\bibinfo {author} {\bibfnamefont {C.-Z.}\ \bibnamefont
  {Chang}}, \bibinfo {author} {\bibfnamefont {J.}~\bibnamefont {Zhang}},
  \bibinfo {author} {\bibfnamefont {M.}~\bibnamefont {Liu}}, \bibinfo {author}
  {\bibfnamefont {Z.}~\bibnamefont {Zhang}}, \bibinfo {author} {\bibfnamefont
  {X.}~\bibnamefont {Feng}}, \bibinfo {author} {\bibfnamefont {K.}~\bibnamefont
  {Li}}, \bibinfo {author} {\bibfnamefont {L.-L.}\ \bibnamefont {Wang}},
  \bibinfo {author} {\bibfnamefont {X.}~\bibnamefont {Chen}}, \bibinfo {author}
  {\bibfnamefont {X.}~\bibnamefont {Dai}}, \bibinfo {author} {\bibfnamefont
  {Z.}~\bibnamefont {Fang}}, \bibinfo {author} {\bibfnamefont {X.-L.}\
  \bibnamefont {Qi}}, \bibinfo {author} {\bibfnamefont {S.-C.}\ \bibnamefont
  {Zhang}}, \bibinfo {author} {\bibfnamefont {Y.}~\bibnamefont {Wang}},
  \bibinfo {author} {\bibfnamefont {K.}~\bibnamefont {He}}, \bibinfo {author}
  {\bibfnamefont {X.-C.}\ \bibnamefont {Ma}}, \ and\ \bibinfo {author}
  {\bibfnamefont {Q.-K.}\ \bibnamefont {Xue}},\ }\href {\doibase
  10.1002/adma.201203493} {\bibfield  {journal} {\bibinfo  {journal} {Adv.
  Mater.}\ }\textbf {\bibinfo {volume} {25}},\ \bibinfo {pages} {1065}
  (\bibinfo {year} {2013}{\natexlab{b}})}\BibitemShut {NoStop}%
\bibitem [{\citenamefont {Chang}\ \emph {et~al.}(2015)\citenamefont {Chang},
  \citenamefont {Zhao}, \citenamefont {Kim}, \citenamefont {Zhang},
  \citenamefont {Assaf}, \citenamefont {Heiman}, \citenamefont {Zhang},
  \citenamefont {Liu}, \citenamefont {Chan},\ and\ \citenamefont
  {Moodera}}]{Chang:2015aa}%
  \BibitemOpen
  \bibfield  {author} {\bibinfo {author} {\bibfnamefont {C.-Z.}\ \bibnamefont
  {Chang}}, \bibinfo {author} {\bibfnamefont {W.}~\bibnamefont {Zhao}},
  \bibinfo {author} {\bibfnamefont {D.~Y.}\ \bibnamefont {Kim}}, \bibinfo
  {author} {\bibfnamefont {H.}~\bibnamefont {Zhang}}, \bibinfo {author}
  {\bibfnamefont {B.~A.}\ \bibnamefont {Assaf}}, \bibinfo {author}
  {\bibfnamefont {D.}~\bibnamefont {Heiman}}, \bibinfo {author} {\bibfnamefont
  {S.-C.}\ \bibnamefont {Zhang}}, \bibinfo {author} {\bibfnamefont
  {C.}~\bibnamefont {Liu}}, \bibinfo {author} {\bibfnamefont {M.~H.~W.}\
  \bibnamefont {Chan}}, \ and\ \bibinfo {author} {\bibfnamefont {J.~S.}\
  \bibnamefont {Moodera}},\ }\href@noop {} {\bibfield  {journal} {\bibinfo
  {journal} {Nat. Mater.}\ }\textbf {\bibinfo {volume} {14}},\ \bibinfo {pages}
  {473} (\bibinfo {year} {2015})}\BibitemShut {NoStop}%
\bibitem [{\citenamefont {Tokura}, \citenamefont {Yasuda},\ and\ \citenamefont
  {Tsukazaki}(2019)}]{Tokura2019}%
  \BibitemOpen
  \bibfield  {author} {\bibinfo {author} {\bibfnamefont {Y.}~\bibnamefont
  {Tokura}}, \bibinfo {author} {\bibfnamefont {K.}~\bibnamefont {Yasuda}}, \
  and\ \bibinfo {author} {\bibfnamefont {A.}~\bibnamefont {Tsukazaki}},\ }\href
  {\doibase 10.1038/s42254-018-0011-5} {\bibfield  {journal} {\bibinfo
  {journal} {Nat. Rev. Phys.}\ }\textbf {\bibinfo {volume} {1}},\ \bibinfo
  {pages} {126} (\bibinfo {year} {2019})}\BibitemShut {NoStop}%
\bibitem [{\citenamefont {He}, \citenamefont {Sun},\ and\ \citenamefont
  {He}(2019)}]{He_Rev2019}%
  \BibitemOpen
  \bibfield  {author} {\bibinfo {author} {\bibfnamefont {M.}~\bibnamefont
  {He}}, \bibinfo {author} {\bibfnamefont {H.}~\bibnamefont {Sun}}, \ and\
  \bibinfo {author} {\bibfnamefont {Q.~L.}\ \bibnamefont {He}},\ }\href
  {\doibase 10.1007/s11467-019-0893-4} {\bibfield  {journal} {\bibinfo
  {journal} {Front. Phys.}\ }\textbf {\bibinfo {volume} {14}},\ \bibinfo
  {pages} {43401} (\bibinfo {year} {2019})}\BibitemShut {NoStop}%
\bibitem [{\citenamefont {Fei}\ \emph {et~al.}(2020)\citenamefont {Fei},
  \citenamefont {Zhang}, \citenamefont {Zhang}, \citenamefont {Shah},
  \citenamefont {Song}, \citenamefont {Wang},\ and\ \citenamefont
  {Wang}}]{Fei2020}%
  \BibitemOpen
  \bibfield  {author} {\bibinfo {author} {\bibfnamefont {F.}~\bibnamefont
  {Fei}}, \bibinfo {author} {\bibfnamefont {S.}~\bibnamefont {Zhang}}, \bibinfo
  {author} {\bibfnamefont {M.}~\bibnamefont {Zhang}}, \bibinfo {author}
  {\bibfnamefont {S.~A.}\ \bibnamefont {Shah}}, \bibinfo {author}
  {\bibfnamefont {F.}~\bibnamefont {Song}}, \bibinfo {author} {\bibfnamefont
  {X.}~\bibnamefont {Wang}}, \ and\ \bibinfo {author} {\bibfnamefont
  {B.}~\bibnamefont {Wang}},\ }\href {\doibase 10.1002/adma.201904593}
  {\bibfield  {journal} {\bibinfo  {journal} {Adv. Mater.}\ }\textbf {\bibinfo
  {volume} {32}},\ \bibinfo {pages} {1904593} (\bibinfo {year}
  {2020})}\BibitemShut {NoStop}%
\bibitem [{\citenamefont {K\"ohler}(1973)}]{Koehler1973}%
  \BibitemOpen
  \bibfield  {author} {\bibinfo {author} {\bibfnamefont {H.}~\bibnamefont
  {K\"ohler}},\ }\href {\doibase 10.1002/pssb.2220580109} {\bibfield  {journal}
  {\bibinfo  {journal} {Phys. Stat. Sol. (b)}\ }\textbf {\bibinfo {volume}
  {58}},\ \bibinfo {pages} {91} (\bibinfo {year} {1973})}\BibitemShut {NoStop}%
\bibitem [{\citenamefont {Watson}\ \emph {et~al.}(2013)\citenamefont {Watson},
  \citenamefont {Collins-McIntyre}, \citenamefont {Shelford}, \citenamefont
  {Coldea}, \citenamefont {Prabhakaran}, \citenamefont {Speller}, \citenamefont
  {Mousavi}, \citenamefont {Grovenor}, \citenamefont {Salman}, \citenamefont
  {Giblin}, \citenamefont {van~der Laan},\ and\ \citenamefont
  {Hesjedal}}]{Watson2013}%
  \BibitemOpen
  \bibfield  {author} {\bibinfo {author} {\bibfnamefont {M.~D.}\ \bibnamefont
  {Watson}}, \bibinfo {author} {\bibfnamefont {L.~J.}\ \bibnamefont
  {Collins-McIntyre}}, \bibinfo {author} {\bibfnamefont {L.~R.}\ \bibnamefont
  {Shelford}}, \bibinfo {author} {\bibfnamefont {A.~I.}\ \bibnamefont
  {Coldea}}, \bibinfo {author} {\bibfnamefont {D.}~\bibnamefont {Prabhakaran}},
  \bibinfo {author} {\bibfnamefont {S.~C.}\ \bibnamefont {Speller}}, \bibinfo
  {author} {\bibfnamefont {T.}~\bibnamefont {Mousavi}}, \bibinfo {author}
  {\bibfnamefont {C.~R.~M.}\ \bibnamefont {Grovenor}}, \bibinfo {author}
  {\bibfnamefont {Z.}~\bibnamefont {Salman}}, \bibinfo {author} {\bibfnamefont
  {S.~R.}\ \bibnamefont {Giblin}}, \bibinfo {author} {\bibfnamefont
  {G.}~\bibnamefont {van~der Laan}}, \ and\ \bibinfo {author} {\bibfnamefont
  {T.}~\bibnamefont {Hesjedal}},\ }\href {\doibase
  10.1088/1367-2630/15/10/103016} {\bibfield  {journal} {\bibinfo  {journal}
  {New J. Phys.}\ }\textbf {\bibinfo {volume} {15}},\ \bibinfo {pages} {103016}
  (\bibinfo {year} {2013})}\BibitemShut {NoStop}%
\bibitem [{\citenamefont {Li}\ \emph {et~al.}(2006)\citenamefont {Li},
  \citenamefont {Toprak}, \citenamefont {Soliman}, \citenamefont {Zhou},
  \citenamefont {Muhammed}, \citenamefont {Platzek},\ and\ \citenamefont
  {Mueller}}]{Echem_2006}%
  \BibitemOpen
  \bibfield  {author} {\bibinfo {author} {\bibfnamefont {S.}~\bibnamefont
  {Li}}, \bibinfo {author} {\bibfnamefont {M.~S.}\ \bibnamefont {Toprak}},
  \bibinfo {author} {\bibfnamefont {H.~M.~A.}\ \bibnamefont {Soliman}},
  \bibinfo {author} {\bibfnamefont {J.}~\bibnamefont {Zhou}}, \bibinfo {author}
  {\bibfnamefont {M.}~\bibnamefont {Muhammed}}, \bibinfo {author}
  {\bibfnamefont {D.}~\bibnamefont {Platzek}}, \ and\ \bibinfo {author}
  {\bibfnamefont {E.}~\bibnamefont {Mueller}},\ }\href {\doibase
  10.1021/cm060171o} {\bibfield  {journal} {\bibinfo  {journal} {Chem. Mater.}\
  }\textbf {\bibinfo {volume} {18}},\ \bibinfo {pages} {3627} (\bibinfo {year}
  {2006})}\BibitemShut {NoStop}%
\bibitem [{\citenamefont {Alegria}\ \emph {et~al.}(2012)\citenamefont
  {Alegria}, \citenamefont {Schroer}, \citenamefont {Chatterjee}, \citenamefont
  {Poirier}, \citenamefont {Pretko}, \citenamefont {Patel},\ and\ \citenamefont
  {Petta}}]{Alegria_NS_MOCVD_2012}%
  \BibitemOpen
  \bibfield  {author} {\bibinfo {author} {\bibfnamefont {L.~D.}\ \bibnamefont
  {Alegria}}, \bibinfo {author} {\bibfnamefont {M.~D.}\ \bibnamefont
  {Schroer}}, \bibinfo {author} {\bibfnamefont {A.}~\bibnamefont {Chatterjee}},
  \bibinfo {author} {\bibfnamefont {G.~R.}\ \bibnamefont {Poirier}}, \bibinfo
  {author} {\bibfnamefont {M.}~\bibnamefont {Pretko}}, \bibinfo {author}
  {\bibfnamefont {S.~K.}\ \bibnamefont {Patel}}, \ and\ \bibinfo {author}
  {\bibfnamefont {J.~R.}\ \bibnamefont {Petta}},\ }\href {\doibase
  10.1021/nl302108r} {\bibfield  {journal} {\bibinfo  {journal} {Nano Lett.}\
  }\textbf {\bibinfo {volume} {12}},\ \bibinfo {pages} {4711} (\bibinfo {year}
  {2012})}\BibitemShut {NoStop}%
\bibitem [{\citenamefont {Harrison}\ \emph
  {et~al.}(2014{\natexlab{b}})\citenamefont {Harrison}, \citenamefont
  {Schoenherr}, \citenamefont {Huo}, \citenamefont {Harris},\ and\
  \citenamefont {Hesjedal}}]{Harrison_nanowire_2014}%
  \BibitemOpen
  \bibfield  {author} {\bibinfo {author} {\bibfnamefont {S.~E.}\ \bibnamefont
  {Harrison}}, \bibinfo {author} {\bibfnamefont {P.}~\bibnamefont
  {Schoenherr}}, \bibinfo {author} {\bibfnamefont {Y.}~\bibnamefont {Huo}},
  \bibinfo {author} {\bibfnamefont {J.~S.}\ \bibnamefont {Harris}}, \ and\
  \bibinfo {author} {\bibfnamefont {T.}~\bibnamefont {Hesjedal}},\ }\href
  {\doibase 10.1063/1.4898816} {\bibfield  {journal} {\bibinfo  {journal}
  {Appl. Phys. Lett.}\ }\textbf {\bibinfo {volume} {105}},\ \bibinfo {pages}
  {153114} (\bibinfo {year} {2014}{\natexlab{b}})}\BibitemShut {NoStop}%
\bibitem [{\citenamefont {Schoenherr}\ \emph
  {et~al.}(2014{\natexlab{a}})\citenamefont {Schoenherr}, \citenamefont
  {Prabhakaran}, \citenamefont {Jones}, \citenamefont {Dimitratos},
  \citenamefont {Bowker},\ and\ \citenamefont {Hesjedal}}]{Schoenherr_2014}%
  \BibitemOpen
  \bibfield  {author} {\bibinfo {author} {\bibfnamefont {P.}~\bibnamefont
  {Schoenherr}}, \bibinfo {author} {\bibfnamefont {D.}~\bibnamefont
  {Prabhakaran}}, \bibinfo {author} {\bibfnamefont {W.}~\bibnamefont {Jones}},
  \bibinfo {author} {\bibfnamefont {N.}~\bibnamefont {Dimitratos}}, \bibinfo
  {author} {\bibfnamefont {M.}~\bibnamefont {Bowker}}, \ and\ \bibinfo {author}
  {\bibfnamefont {T.}~\bibnamefont {Hesjedal}},\ }\href {\doibase
  10.1063/1.4885217} {\bibfield  {journal} {\bibinfo  {journal} {Appl. Phys.
  Lett.}\ }\textbf {\bibinfo {volume} {104}},\ \bibinfo {pages} {253103}
  (\bibinfo {year} {2014}{\natexlab{a}})}\BibitemShut {NoStop}%
\bibitem [{\citenamefont {Schoenherr}\ \emph
  {et~al.}(2014{\natexlab{b}})\citenamefont {Schoenherr}, \citenamefont
  {Collins-Mclntyre}, \citenamefont {Zhang}, \citenamefont {Kusch},
  \citenamefont {Reich}, \citenamefont {Giles}, \citenamefont {Daisenberger},
  \citenamefont {Prabhakaran},\ and\ \citenamefont
  {Hesjedal}}]{Schoenherr_2014b}%
  \BibitemOpen
  \bibfield  {author} {\bibinfo {author} {\bibfnamefont {P.}~\bibnamefont
  {Schoenherr}}, \bibinfo {author} {\bibfnamefont {L.~J.}\ \bibnamefont
  {Collins-Mclntyre}}, \bibinfo {author} {\bibfnamefont {S.}~\bibnamefont
  {Zhang}}, \bibinfo {author} {\bibfnamefont {P.}~\bibnamefont {Kusch}},
  \bibinfo {author} {\bibfnamefont {S.}~\bibnamefont {Reich}}, \bibinfo
  {author} {\bibfnamefont {T.}~\bibnamefont {Giles}}, \bibinfo {author}
  {\bibfnamefont {D.}~\bibnamefont {Daisenberger}}, \bibinfo {author}
  {\bibfnamefont {D.}~\bibnamefont {Prabhakaran}}, \ and\ \bibinfo {author}
  {\bibfnamefont {T.}~\bibnamefont {Hesjedal}},\ }\href {\doibase
  10.1186/1556-276X-9-127} {\bibfield  {journal} {\bibinfo  {journal}
  {Nanoscale Res. Lett.}\ }\textbf {\bibinfo {volume} {9}},\ \bibinfo {pages}
  {{127}} (\bibinfo {year} {2014}{\natexlab{b}})}\BibitemShut {NoStop}%
\bibitem [{\citenamefont {Cecchini}\ \emph {et~al.}(2019)\citenamefont
  {Cecchini}, \citenamefont {Gajjela}, \citenamefont {Martella}, \citenamefont
  {Wiemer}, \citenamefont {Lamperti}, \citenamefont {Nasi}, \citenamefont
  {Lazzarini}, \citenamefont {Nobili},\ and\ \citenamefont
  {Longo}}]{Cecchini2019}%
  \BibitemOpen
  \bibfield  {author} {\bibinfo {author} {\bibfnamefont {R.}~\bibnamefont
  {Cecchini}}, \bibinfo {author} {\bibfnamefont {R.~S.~R.}\ \bibnamefont
  {Gajjela}}, \bibinfo {author} {\bibfnamefont {C.}~\bibnamefont {Martella}},
  \bibinfo {author} {\bibfnamefont {C.}~\bibnamefont {Wiemer}}, \bibinfo
  {author} {\bibfnamefont {A.}~\bibnamefont {Lamperti}}, \bibinfo {author}
  {\bibfnamefont {L.}~\bibnamefont {Nasi}}, \bibinfo {author} {\bibfnamefont
  {L.}~\bibnamefont {Lazzarini}}, \bibinfo {author} {\bibfnamefont {L.~G.}\
  \bibnamefont {Nobili}}, \ and\ \bibinfo {author} {\bibfnamefont
  {M.}~\bibnamefont {Longo}},\ }\href {\doibase 10.1002/smll.201901743}
  {\bibfield  {journal} {\bibinfo  {journal} {Small}\ }\textbf {\bibinfo
  {volume} {15}},\ \bibinfo {pages} {1901743} (\bibinfo {year}
  {2019})}\BibitemShut {NoStop}%
\bibitem [{\citenamefont {Ferhat}\ \emph {et~al.}(1996)\citenamefont {Ferhat},
  \citenamefont {Liautard}, \citenamefont {Brun}, \citenamefont {Tedenac},
  \citenamefont {Nouaoura},\ and\ \citenamefont {Lassabatere}}]{Ferhat1996}%
  \BibitemOpen
  \bibfield  {author} {\bibinfo {author} {\bibfnamefont {M.}~\bibnamefont
  {Ferhat}}, \bibinfo {author} {\bibfnamefont {B.}~\bibnamefont {Liautard}},
  \bibinfo {author} {\bibfnamefont {G.}~\bibnamefont {Brun}}, \bibinfo {author}
  {\bibfnamefont {J.}~\bibnamefont {Tedenac}}, \bibinfo {author} {\bibfnamefont
  {M.}~\bibnamefont {Nouaoura}}, \ and\ \bibinfo {author} {\bibfnamefont
  {L.}~\bibnamefont {Lassabatere}},\ }\href {\doibase
  10.1016/0022-0248(96)00247-3} {\bibfield  {journal} {\bibinfo  {journal} {J.
  Cryst. Growth}\ }\textbf {\bibinfo {volume} {167}},\ \bibinfo {pages} {122}
  (\bibinfo {year} {1996})}\BibitemShut {NoStop}%
\bibitem [{\citenamefont {Ferhata}, \citenamefont {Tedenaca},\ and\
  \citenamefont {Nagaob}(2000)}]{Ferhat2000}%
  \BibitemOpen
  \bibfield  {author} {\bibinfo {author} {\bibfnamefont {M.}~\bibnamefont
  {Ferhata}}, \bibinfo {author} {\bibfnamefont {J.~C.}\ \bibnamefont
  {Tedenaca}}, \ and\ \bibinfo {author} {\bibfnamefont {J.}~\bibnamefont
  {Nagaob}},\ }\href@noop {} {\bibfield  {journal} {\bibinfo  {journal} {J.
  Cryst. Growth}\ }\textbf {\bibinfo {volume} {218}},\ \bibinfo {pages} {250}
  (\bibinfo {year} {2000})}\BibitemShut {NoStop}%
\bibitem [{\citenamefont {Cao}\ \emph {et~al.}(2012)\citenamefont {Cao},
  \citenamefont {Venkatasubramanian}, \citenamefont {Liu}, \citenamefont
  {Pierce}, \citenamefont {Yang}, \citenamefont {Hasan}, \citenamefont {Wu},\
  and\ \citenamefont {Chen}}]{Cao_MOCVD_films_2012}%
  \BibitemOpen
  \bibfield  {author} {\bibinfo {author} {\bibfnamefont {H.}~\bibnamefont
  {Cao}}, \bibinfo {author} {\bibfnamefont {R.}~\bibnamefont
  {Venkatasubramanian}}, \bibinfo {author} {\bibfnamefont {C.}~\bibnamefont
  {Liu}}, \bibinfo {author} {\bibfnamefont {J.}~\bibnamefont {Pierce}},
  \bibinfo {author} {\bibfnamefont {H.}~\bibnamefont {Yang}}, \bibinfo {author}
  {\bibfnamefont {M.~Z.}\ \bibnamefont {Hasan}}, \bibinfo {author}
  {\bibfnamefont {Y.}~\bibnamefont {Wu}}, \ and\ \bibinfo {author}
  {\bibfnamefont {Y.~P.}\ \bibnamefont {Chen}},\ }\href {\doibase
  {10.1063/1.4760226}} {\bibfield  {journal} {\bibinfo  {journal} {Appl. Phys.
  Lett.}\ }\textbf {\bibinfo {volume} {101}},\ \bibinfo {pages} {162104}
  (\bibinfo {year} {2012})}\BibitemShut {NoStop}%
\bibitem [{\citenamefont {Wang}, \citenamefont {Gao},\ and\ \citenamefont
  {Li}(2016)}]{Wang2016}%
  \BibitemOpen
  \bibfield  {author} {\bibinfo {author} {\bibfnamefont {W.~J.}\ \bibnamefont
  {Wang}}, \bibinfo {author} {\bibfnamefont {K.~H.}\ \bibnamefont {Gao}}, \
  and\ \bibinfo {author} {\bibfnamefont {Z.~Q.}\ \bibnamefont {Li}},\ }\href
  {\doibase 10.1038/srep25291} {\bibfield  {journal} {\bibinfo  {journal} {Sci.
  Rep.}\ }\textbf {\bibinfo {volume} {6}},\ \bibinfo {pages} {25291} (\bibinfo
  {year} {2016})}\BibitemShut {NoStop}%
\bibitem [{\citenamefont {Rusek}\ \emph {et~al.}(2017)\citenamefont {Rusek},
  \citenamefont {Komossa}, \citenamefont {Bendt},\ and\ \citenamefont
  {Schulz}}]{Rusek_ALD_2017}%
  \BibitemOpen
  \bibfield  {author} {\bibinfo {author} {\bibfnamefont {M.}~\bibnamefont
  {Rusek}}, \bibinfo {author} {\bibfnamefont {T.}~\bibnamefont {Komossa}},
  \bibinfo {author} {\bibfnamefont {G.}~\bibnamefont {Bendt}}, \ and\ \bibinfo
  {author} {\bibfnamefont {S.}~\bibnamefont {Schulz}},\ }\href {\doibase
  10.1016/j.jcrysgro.2017.04.019} {\bibfield  {journal} {\bibinfo  {journal}
  {J. Cryst. Growth}\ }\textbf {\bibinfo {volume} {470}},\ \bibinfo {pages}
  {128} (\bibinfo {year} {2017})}\BibitemShut {NoStop}%
\bibitem [{\citenamefont {Liao}\ \emph {et~al.}(2019)\citenamefont {Liao},
  \citenamefont {Brahlek}, \citenamefont {Ok}, \citenamefont {Nuckols},
  \citenamefont {Sharma}, \citenamefont {Lu}, \citenamefont {Zhang},\ and\
  \citenamefont {Lee}}]{Liao2019}%
  \BibitemOpen
  \bibfield  {author} {\bibinfo {author} {\bibfnamefont {Z.}~\bibnamefont
  {Liao}}, \bibinfo {author} {\bibfnamefont {M.}~\bibnamefont {Brahlek}},
  \bibinfo {author} {\bibfnamefont {J.~M.}\ \bibnamefont {Ok}}, \bibinfo
  {author} {\bibfnamefont {L.}~\bibnamefont {Nuckols}}, \bibinfo {author}
  {\bibfnamefont {Y.}~\bibnamefont {Sharma}}, \bibinfo {author} {\bibfnamefont
  {Q.}~\bibnamefont {Lu}}, \bibinfo {author} {\bibfnamefont {Y.}~\bibnamefont
  {Zhang}}, \ and\ \bibinfo {author} {\bibfnamefont {H.~N.}\ \bibnamefont
  {Lee}},\ }\href {\doibase 10.1063/1.5088190} {\bibfield  {journal} {\bibinfo
  {journal} {APL Mater.}\ }\textbf {\bibinfo {volume} {7}},\ \bibinfo {pages}
  {041101} (\bibinfo {year} {2019})}\BibitemShut {NoStop}%
\bibitem [{\citenamefont {Charles}, \citenamefont {Groubert},\ and\
  \citenamefont {Boyer}(1988)}]{Charles1988}%
  \BibitemOpen
  \bibfield  {author} {\bibinfo {author} {\bibfnamefont {E.}~\bibnamefont
  {Charles}}, \bibinfo {author} {\bibfnamefont {E.}~\bibnamefont {Groubert}}, \
  and\ \bibinfo {author} {\bibfnamefont {A.}~\bibnamefont {Boyer}},\ }\href
  {\doibase 10.1007/BF01730298} {\bibfield  {journal} {\bibinfo  {journal} {J.
  Mater. Sci. Lett.}\ }\textbf {\bibinfo {volume} {7}},\ \bibinfo {pages} {575}
  (\bibinfo {year} {1988})}\BibitemShut {NoStop}%
\bibitem [{\citenamefont {Iwata}\ \emph {et~al.}(1999)\citenamefont {Iwata},
  \citenamefont {Kobayashi}, \citenamefont {Kikuchi}, \citenamefont {Hatta},\
  and\ \citenamefont {Mukasa}}]{Iwata1999}%
  \BibitemOpen
  \bibfield  {author} {\bibinfo {author} {\bibfnamefont {Y.}~\bibnamefont
  {Iwata}}, \bibinfo {author} {\bibfnamefont {H.}~\bibnamefont {Kobayashi}},
  \bibinfo {author} {\bibfnamefont {S.}~\bibnamefont {Kikuchi}}, \bibinfo
  {author} {\bibfnamefont {E.}~\bibnamefont {Hatta}}, \ and\ \bibinfo {author}
  {\bibfnamefont {K.}~\bibnamefont {Mukasa}},\ }\href@noop {} {\bibfield
  {journal} {\bibinfo  {journal} {J. Cryst. Growth}\ }\textbf {\bibinfo
  {volume} {203}},\ \bibinfo {pages} {125} (\bibinfo {year}
  {1999})}\BibitemShut {NoStop}%
\bibitem [{\citenamefont {Wang}, \citenamefont {Endicott},\ and\ \citenamefont
  {Uher}(2011)}]{Wang_Review_2011}%
  \BibitemOpen
  \bibfield  {author} {\bibinfo {author} {\bibfnamefont {G.}~\bibnamefont
  {Wang}}, \bibinfo {author} {\bibfnamefont {L.}~\bibnamefont {Endicott}}, \
  and\ \bibinfo {author} {\bibfnamefont {C.}~\bibnamefont {Uher}},\ }\href
  {\doibase 10.1166/sam.2011.1182} {\bibfield  {journal} {\bibinfo  {journal}
  {Sci. Adv. Mater.}\ }\textbf {\bibinfo {volume} {3}},\ \bibinfo {pages} {539}
  (\bibinfo {year} {2011})}\BibitemShut {NoStop}%
\bibitem [{\citenamefont {He}, \citenamefont {Kou},\ and\ \citenamefont
  {Wang}(2013)}]{He2013}%
  \BibitemOpen
  \bibfield  {author} {\bibinfo {author} {\bibfnamefont {L.}~\bibnamefont
  {He}}, \bibinfo {author} {\bibfnamefont {X.}~\bibnamefont {Kou}}, \ and\
  \bibinfo {author} {\bibfnamefont {K.~L.}\ \bibnamefont {Wang}},\ }\href@noop
  {} {\bibfield  {journal} {\bibinfo  {journal} {{Phys. Status Solidi - Rapid
  Res. Lett.}}\ }\textbf {\bibinfo {volume} {{7}}},\ \bibinfo {pages} {{50}}
  (\bibinfo {year} {{2013}})}\BibitemShut {NoStop}%
\bibitem [{\citenamefont {Ginley}, \citenamefont {Wang},\ and\ \citenamefont
  {Law}(2016)}]{Ginley2016}%
  \BibitemOpen
  \bibfield  {author} {\bibinfo {author} {\bibfnamefont {T.~P.}\ \bibnamefont
  {Ginley}}, \bibinfo {author} {\bibfnamefont {Y.}~\bibnamefont {Wang}}, \ and\
  \bibinfo {author} {\bibfnamefont {S.}~\bibnamefont {Law}},\ }\href {\doibase
  10.3390/cryst6110154} {\bibfield  {journal} {\bibinfo  {journal} {Crystals}\
  }\textbf {\bibinfo {volume} {6}},\ \bibinfo {pages} {154} (\bibinfo {year}
  {2016})}\BibitemShut {NoStop}%
\bibitem [{\citenamefont {Zhang}\ \emph
  {et~al.}(2009{\natexlab{b}})\citenamefont {Zhang}, \citenamefont {Qin},
  \citenamefont {Teng}, \citenamefont {Guo}, \citenamefont {Guo}, \citenamefont
  {Dai}, \citenamefont {Fang},\ and\ \citenamefont {Wu}}]{ZhangAPL2009}%
  \BibitemOpen
  \bibfield  {author} {\bibinfo {author} {\bibfnamefont {G.}~\bibnamefont
  {Zhang}}, \bibinfo {author} {\bibfnamefont {H.}~\bibnamefont {Qin}}, \bibinfo
  {author} {\bibfnamefont {J.}~\bibnamefont {Teng}}, \bibinfo {author}
  {\bibfnamefont {J.}~\bibnamefont {Guo}}, \bibinfo {author} {\bibfnamefont
  {Q.}~\bibnamefont {Guo}}, \bibinfo {author} {\bibfnamefont {X.}~\bibnamefont
  {Dai}}, \bibinfo {author} {\bibfnamefont {Z.}~\bibnamefont {Fang}}, \ and\
  \bibinfo {author} {\bibfnamefont {K.}~\bibnamefont {Wu}},\ }\href {\doibase
  10.1063/1.3200237} {\bibfield  {journal} {\bibinfo  {journal} {Appl. Phys.
  Lett.}\ }\textbf {\bibinfo {volume} {95}},\ \bibinfo {pages} {053114}
  (\bibinfo {year} {2009}{\natexlab{b}})}\BibitemShut {NoStop}%
\bibitem [{\citenamefont {Li}\ \emph {et~al.}(2010{\natexlab{a}})\citenamefont
  {Li}, \citenamefont {Wang}, \citenamefont {Zhu}, \citenamefont {Liu},
  \citenamefont {Ye}, \citenamefont {Chen}, \citenamefont {Wang}, \citenamefont
  {He}, \citenamefont {Wang}, \citenamefont {Ma}, \citenamefont {Zhang},
  \citenamefont {Dai}, \citenamefont {Fang}, \citenamefont {Xie}, \citenamefont
  {Liu}, \citenamefont {Qi}, \citenamefont {Jia}, \citenamefont {Zhang},\ and\
  \citenamefont {Xue}}]{Li2010}%
  \BibitemOpen
  \bibfield  {author} {\bibinfo {author} {\bibfnamefont {Y.-Y.}\ \bibnamefont
  {Li}}, \bibinfo {author} {\bibfnamefont {G.}~\bibnamefont {Wang}}, \bibinfo
  {author} {\bibfnamefont {X.-G.}\ \bibnamefont {Zhu}}, \bibinfo {author}
  {\bibfnamefont {M.-H.}\ \bibnamefont {Liu}}, \bibinfo {author} {\bibfnamefont
  {C.}~\bibnamefont {Ye}}, \bibinfo {author} {\bibfnamefont {X.}~\bibnamefont
  {Chen}}, \bibinfo {author} {\bibfnamefont {Y.-Y.}\ \bibnamefont {Wang}},
  \bibinfo {author} {\bibfnamefont {K.}~\bibnamefont {He}}, \bibinfo {author}
  {\bibfnamefont {L.-L.}\ \bibnamefont {Wang}}, \bibinfo {author}
  {\bibfnamefont {X.-C.}\ \bibnamefont {Ma}}, \bibinfo {author} {\bibfnamefont
  {H.-J.}\ \bibnamefont {Zhang}}, \bibinfo {author} {\bibfnamefont
  {X.}~\bibnamefont {Dai}}, \bibinfo {author} {\bibfnamefont {Z.}~\bibnamefont
  {Fang}}, \bibinfo {author} {\bibfnamefont {X.-C.}\ \bibnamefont {Xie}},
  \bibinfo {author} {\bibfnamefont {Y.}~\bibnamefont {Liu}}, \bibinfo {author}
  {\bibfnamefont {X.-L.}\ \bibnamefont {Qi}}, \bibinfo {author} {\bibfnamefont
  {J.-F.}\ \bibnamefont {Jia}}, \bibinfo {author} {\bibfnamefont {S.-C.}\
  \bibnamefont {Zhang}}, \ and\ \bibinfo {author} {\bibfnamefont {Q.-K.}\
  \bibnamefont {Xue}},\ }\href@noop {} {\bibfield  {journal} {\bibinfo
  {journal} {{Adv. Mater.}}\ }\textbf {\bibinfo {volume} {{22}}},\ \bibinfo
  {pages} {{4002}} (\bibinfo {year} {{2010}}{\natexlab{a}})}\BibitemShut
  {NoStop}%
\bibitem [{\citenamefont {He}\ \emph {et~al.}(2011)\citenamefont {He},
  \citenamefont {Xiu}, \citenamefont {Wang}, \citenamefont {Fedorov},
  \citenamefont {Huang}, \citenamefont {Kou}, \citenamefont {Lang},
  \citenamefont {Beyermann}, \citenamefont {Zou},\ and\ \citenamefont
  {Wang}}]{He2011}%
  \BibitemOpen
  \bibfield  {author} {\bibinfo {author} {\bibfnamefont {L.}~\bibnamefont
  {He}}, \bibinfo {author} {\bibfnamefont {F.}~\bibnamefont {Xiu}}, \bibinfo
  {author} {\bibfnamefont {Y.}~\bibnamefont {Wang}}, \bibinfo {author}
  {\bibfnamefont {A.~V.}\ \bibnamefont {Fedorov}}, \bibinfo {author}
  {\bibfnamefont {G.}~\bibnamefont {Huang}}, \bibinfo {author} {\bibfnamefont
  {X.}~\bibnamefont {Kou}}, \bibinfo {author} {\bibfnamefont {M.}~\bibnamefont
  {Lang}}, \bibinfo {author} {\bibfnamefont {W.~P.}\ \bibnamefont {Beyermann}},
  \bibinfo {author} {\bibfnamefont {J.}~\bibnamefont {Zou}}, \ and\ \bibinfo
  {author} {\bibfnamefont {K.~L.}\ \bibnamefont {Wang}},\ }\href@noop {}
  {\bibfield  {journal} {\bibinfo  {journal} {J. Appl. Phys.}\ }\textbf
  {\bibinfo {volume} {109}},\ \bibinfo {pages} {103702} (\bibinfo {year}
  {2011})}\BibitemShut {NoStop}%
\bibitem [{\citenamefont {Krumrain}\ \emph {et~al.}(2011)\citenamefont
  {Krumrain}, \citenamefont {Mussler}, \citenamefont {Borisova}, \citenamefont
  {Stoica}, \citenamefont {Plucinski}, \citenamefont {Schneider},\ and\
  \citenamefont {Gr{\"u}tzmacher}}]{Krumrain2011}%
  \BibitemOpen
  \bibfield  {author} {\bibinfo {author} {\bibfnamefont {J.}~\bibnamefont
  {Krumrain}}, \bibinfo {author} {\bibfnamefont {G.}~\bibnamefont {Mussler}},
  \bibinfo {author} {\bibfnamefont {S.}~\bibnamefont {Borisova}}, \bibinfo
  {author} {\bibfnamefont {T.}~\bibnamefont {Stoica}}, \bibinfo {author}
  {\bibfnamefont {L.}~\bibnamefont {Plucinski}}, \bibinfo {author}
  {\bibfnamefont {C.}~\bibnamefont {Schneider}}, \ and\ \bibinfo {author}
  {\bibfnamefont {D.}~\bibnamefont {Gr{\"u}tzmacher}},\ }\href {\doibase
  10.1016/j.jcrysgro.2011.03.008} {\bibfield  {journal} {\bibinfo  {journal}
  {J. Cryst. Growth}\ }\textbf {\bibinfo {volume} {324}},\ \bibinfo {pages}
  {115} (\bibinfo {year} {2011})}\BibitemShut {NoStop}%
\bibitem [{\citenamefont {Liu}\ \emph {et~al.}(2011{\natexlab{a}})\citenamefont
  {Liu}, \citenamefont {Smith}, \citenamefont {Fan}, \citenamefont {Zhang},
  \citenamefont {Cao}, \citenamefont {Chen}, \citenamefont {Leiner},
  \citenamefont {Kirby}, \citenamefont {Dobrowolska},\ and\ \citenamefont
  {Furdyna}}]{Liu2013}%
  \BibitemOpen
  \bibfield  {author} {\bibinfo {author} {\bibfnamefont {X.}~\bibnamefont
  {Liu}}, \bibinfo {author} {\bibfnamefont {D.~J.}\ \bibnamefont {Smith}},
  \bibinfo {author} {\bibfnamefont {J.}~\bibnamefont {Fan}}, \bibinfo {author}
  {\bibfnamefont {Y.-H.}\ \bibnamefont {Zhang}}, \bibinfo {author}
  {\bibfnamefont {H.}~\bibnamefont {Cao}}, \bibinfo {author} {\bibfnamefont
  {Y.~P.}\ \bibnamefont {Chen}}, \bibinfo {author} {\bibfnamefont
  {J.}~\bibnamefont {Leiner}}, \bibinfo {author} {\bibfnamefont {B.~J.}\
  \bibnamefont {Kirby}}, \bibinfo {author} {\bibfnamefont {M.}~\bibnamefont
  {Dobrowolska}}, \ and\ \bibinfo {author} {\bibfnamefont {J.~K.}\ \bibnamefont
  {Furdyna}},\ }\href@noop {} {\bibfield  {journal} {\bibinfo  {journal} {Appl.
  Phys. Lett.}\ }\textbf {\bibinfo {volume} {99}},\ \bibinfo {pages} {171903}
  (\bibinfo {year} {2011}{\natexlab{a}})}\BibitemShut {NoStop}%
\bibitem [{\citenamefont {Liu}\ \emph {et~al.}(2012)\citenamefont {Liu},
  \citenamefont {Smith}, \citenamefont {Cao}, \citenamefont {Chen},
  \citenamefont {Fan}, \citenamefont {Zhang}, \citenamefont {Pimpinella},
  \citenamefont {Dobrowolska},\ and\ \citenamefont {Furdyna}}]{Liu2012JVST}%
  \BibitemOpen
  \bibfield  {author} {\bibinfo {author} {\bibfnamefont {X.}~\bibnamefont
  {Liu}}, \bibinfo {author} {\bibfnamefont {D.~J.}\ \bibnamefont {Smith}},
  \bibinfo {author} {\bibfnamefont {H.}~\bibnamefont {Cao}}, \bibinfo {author}
  {\bibfnamefont {Y.~P.}\ \bibnamefont {Chen}}, \bibinfo {author}
  {\bibfnamefont {J.}~\bibnamefont {Fan}}, \bibinfo {author} {\bibfnamefont
  {Y.-H.}\ \bibnamefont {Zhang}}, \bibinfo {author} {\bibfnamefont {R.~E.}\
  \bibnamefont {Pimpinella}}, \bibinfo {author} {\bibfnamefont
  {M.}~\bibnamefont {Dobrowolska}}, \ and\ \bibinfo {author} {\bibfnamefont
  {J.~K.}\ \bibnamefont {Furdyna}},\ }\href@noop {} {\bibfield  {journal}
  {\bibinfo  {journal} {J. Vac. Sci. Technol. B}\ }\textbf {\bibinfo {volume}
  {30}},\ \bibinfo {pages} {02B103} (\bibinfo {year} {2012})}\BibitemShut
  {NoStop}%
\bibitem [{\citenamefont {Taskin}\ \emph
  {et~al.}(2012{\natexlab{a}})\citenamefont {Taskin}, \citenamefont {Sasaki},
  \citenamefont {Segawa},\ and\ \citenamefont {Ando}}]{TaskinPRL2012}%
  \BibitemOpen
  \bibfield  {author} {\bibinfo {author} {\bibfnamefont {A.~A.}\ \bibnamefont
  {Taskin}}, \bibinfo {author} {\bibfnamefont {S.}~\bibnamefont {Sasaki}},
  \bibinfo {author} {\bibfnamefont {K.}~\bibnamefont {Segawa}}, \ and\ \bibinfo
  {author} {\bibfnamefont {Y.}~\bibnamefont {Ando}},\ }\href {\doibase
  10.1103/PhysRevLett.109.066803} {\bibfield  {journal} {\bibinfo  {journal}
  {Phys. Rev. Lett.}\ }\textbf {\bibinfo {volume} {109}},\ \bibinfo {pages}
  {066803} (\bibinfo {year} {2012}{\natexlab{a}})}\BibitemShut {NoStop}%
\bibitem [{\citenamefont {Lee}\ \emph {et~al.}(2012)\citenamefont {Lee},
  \citenamefont {Schmitt}, \citenamefont {Moore}, \citenamefont {Vishik},
  \citenamefont {Ma},\ and\ \citenamefont {Shen}}]{Lee2012APL}%
  \BibitemOpen
  \bibfield  {author} {\bibinfo {author} {\bibfnamefont {J.~J.}\ \bibnamefont
  {Lee}}, \bibinfo {author} {\bibfnamefont {F.~T.}\ \bibnamefont {Schmitt}},
  \bibinfo {author} {\bibfnamefont {R.~G.}\ \bibnamefont {Moore}}, \bibinfo
  {author} {\bibfnamefont {I.~M.}\ \bibnamefont {Vishik}}, \bibinfo {author}
  {\bibfnamefont {Y.}~\bibnamefont {Ma}}, \ and\ \bibinfo {author}
  {\bibfnamefont {Z.~X.}\ \bibnamefont {Shen}},\ }\href@noop {} {\bibfield
  {journal} {\bibinfo  {journal} {{Appl. Phys. Lett.}}\ }\textbf {\bibinfo
  {volume} {{101}}},\ \bibinfo {pages} {{013118}} (\bibinfo {year}
  {{2012}})}\BibitemShut {NoStop}%
\bibitem [{\citenamefont {Bansal}\ \emph {et~al.}(2012)\citenamefont {Bansal},
  \citenamefont {Kim}, \citenamefont {Brahlek}, \citenamefont {Edrey},\ and\
  \citenamefont {Oh}}]{BansalPRL2012}%
  \BibitemOpen
  \bibfield  {author} {\bibinfo {author} {\bibfnamefont {N.}~\bibnamefont
  {Bansal}}, \bibinfo {author} {\bibfnamefont {Y.~S.}\ \bibnamefont {Kim}},
  \bibinfo {author} {\bibfnamefont {M.}~\bibnamefont {Brahlek}}, \bibinfo
  {author} {\bibfnamefont {E.}~\bibnamefont {Edrey}}, \ and\ \bibinfo {author}
  {\bibfnamefont {S.}~\bibnamefont {Oh}},\ }\href {\doibase
  10.1103/PhysRevLett.109.116804} {\bibfield  {journal} {\bibinfo  {journal}
  {Phys. Rev. Lett.}\ }\textbf {\bibinfo {volume} {109}},\ \bibinfo {pages}
  {116804} (\bibinfo {year} {2012})}\BibitemShut {NoStop}%
\bibitem [{\citenamefont {Harrison}\ \emph {et~al.}(2013)\citenamefont
  {Harrison}, \citenamefont {Li}, \citenamefont {Huo}, \citenamefont {Zhou},
  \citenamefont {Chen},\ and\ \citenamefont {Harris}}]{Harrison2013}%
  \BibitemOpen
  \bibfield  {author} {\bibinfo {author} {\bibfnamefont {S.~E.}\ \bibnamefont
  {Harrison}}, \bibinfo {author} {\bibfnamefont {S.}~\bibnamefont {Li}},
  \bibinfo {author} {\bibfnamefont {Y.}~\bibnamefont {Huo}}, \bibinfo {author}
  {\bibfnamefont {B.}~\bibnamefont {Zhou}}, \bibinfo {author} {\bibfnamefont
  {Y.~L.}\ \bibnamefont {Chen}}, \ and\ \bibinfo {author} {\bibfnamefont
  {J.~S.}\ \bibnamefont {Harris}},\ }\href@noop {} {\bibfield  {journal}
  {\bibinfo  {journal} {Appl. Phys. Lett.}\ }\textbf {\bibinfo {volume}
  {102}},\ \bibinfo {pages} {171906} (\bibinfo {year} {2013})}\BibitemShut
  {NoStop}%
\bibitem [{\citenamefont {Zhao}\ \emph {et~al.}(2013)\citenamefont {Zhao},
  \citenamefont {Chang}, \citenamefont {Jiang}, \citenamefont {DaSilva},
  \citenamefont {Sun}, \citenamefont {Wang}, \citenamefont {Xing},
  \citenamefont {Wang}, \citenamefont {He}, \citenamefont {Ma}, \citenamefont
  {Xue},\ and\ \citenamefont {Wang}}]{Zhao2013}%
  \BibitemOpen
  \bibfield  {author} {\bibinfo {author} {\bibfnamefont {Y.}~\bibnamefont
  {Zhao}}, \bibinfo {author} {\bibfnamefont {C.-Z.}\ \bibnamefont {Chang}},
  \bibinfo {author} {\bibfnamefont {Y.}~\bibnamefont {Jiang}}, \bibinfo
  {author} {\bibfnamefont {A.}~\bibnamefont {DaSilva}}, \bibinfo {author}
  {\bibfnamefont {Y.}~\bibnamefont {Sun}}, \bibinfo {author} {\bibfnamefont
  {H.}~\bibnamefont {Wang}}, \bibinfo {author} {\bibfnamefont {Y.}~\bibnamefont
  {Xing}}, \bibinfo {author} {\bibfnamefont {Y.}~\bibnamefont {Wang}}, \bibinfo
  {author} {\bibfnamefont {K.}~\bibnamefont {He}}, \bibinfo {author}
  {\bibfnamefont {X.}~\bibnamefont {Ma}}, \bibinfo {author} {\bibfnamefont
  {Q.-K.~X.}\ \bibnamefont {Xue}}, \ and\ \bibinfo {author} {\bibfnamefont
  {J.}~\bibnamefont {Wang}},\ }\href@noop {} {\bibfield  {journal} {\bibinfo
  {journal} {{Sci. Rep.}}\ }\textbf {\bibinfo {volume} {3}},\ \bibinfo {pages}
  {3060} (\bibinfo {year} {2013})}\BibitemShut {NoStop}%
\bibitem [{\citenamefont {Richardella}\ \emph {et~al.}(2010)\citenamefont
  {Richardella}, \citenamefont {Zhang}, \citenamefont {Lee}, \citenamefont
  {Koser}, \citenamefont {Rench}, \citenamefont {Yeats}, \citenamefont
  {Buckley}, \citenamefont {Awschalom},\ and\ \citenamefont
  {Samarth}}]{RichardellaAPL2010}%
  \BibitemOpen
  \bibfield  {author} {\bibinfo {author} {\bibfnamefont {A.}~\bibnamefont
  {Richardella}}, \bibinfo {author} {\bibfnamefont {D.~M.}\ \bibnamefont
  {Zhang}}, \bibinfo {author} {\bibfnamefont {J.~S.}\ \bibnamefont {Lee}},
  \bibinfo {author} {\bibfnamefont {A.}~\bibnamefont {Koser}}, \bibinfo
  {author} {\bibfnamefont {D.~W.}\ \bibnamefont {Rench}}, \bibinfo {author}
  {\bibfnamefont {A.~L.}\ \bibnamefont {Yeats}}, \bibinfo {author}
  {\bibfnamefont {B.~B.}\ \bibnamefont {Buckley}}, \bibinfo {author}
  {\bibfnamefont {D.~D.}\ \bibnamefont {Awschalom}}, \ and\ \bibinfo {author}
  {\bibfnamefont {N.}~\bibnamefont {Samarth}},\ }\href {\doibase
  10.1063/1.3532845} {\bibfield  {journal} {\bibinfo  {journal} {Appl. Phys.
  Lett.}\ }\textbf {\bibinfo {volume} {97}},\ \bibinfo {pages} {262104}
  (\bibinfo {year} {2010})}\BibitemShut {NoStop}%
\bibitem [{\citenamefont {Chen}\ \emph {et~al.}(2014)\citenamefont {Chen},
  \citenamefont {Garcia}, \citenamefont {De~Jesus}, \citenamefont {Zhao},
  \citenamefont {Deng}, \citenamefont {Secor}, \citenamefont {Begliarbekov},
  \citenamefont {Krusin-Elbaum},\ and\ \citenamefont {Tamargo}}]{Chen2014}%
  \BibitemOpen
  \bibfield  {author} {\bibinfo {author} {\bibfnamefont {Z.}~\bibnamefont
  {Chen}}, \bibinfo {author} {\bibfnamefont {T.~A.}\ \bibnamefont {Garcia}},
  \bibinfo {author} {\bibfnamefont {J.}~\bibnamefont {De~Jesus}}, \bibinfo
  {author} {\bibfnamefont {L.}~\bibnamefont {Zhao}}, \bibinfo {author}
  {\bibfnamefont {H.}~\bibnamefont {Deng}}, \bibinfo {author} {\bibfnamefont
  {J.}~\bibnamefont {Secor}}, \bibinfo {author} {\bibfnamefont
  {M.}~\bibnamefont {Begliarbekov}}, \bibinfo {author} {\bibfnamefont
  {L.}~\bibnamefont {Krusin-Elbaum}}, \ and\ \bibinfo {author} {\bibfnamefont
  {M.~C.}\ \bibnamefont {Tamargo}},\ }\href@noop {} {\bibfield  {journal}
  {\bibinfo  {journal} {{J. Electron. Mater.}}\ }\textbf {\bibinfo {volume}
  {{43}}},\ \bibinfo {pages} {{909}} (\bibinfo {year} {{2014}})}\BibitemShut
  {NoStop}%
\bibitem [{\citenamefont {Eddrief}\ \emph {et~al.}(2014)\citenamefont
  {Eddrief}, \citenamefont {Atkinson}, \citenamefont {Etgens},\ and\
  \citenamefont {Jusserand}}]{Eddrief2014}%
  \BibitemOpen
  \bibfield  {author} {\bibinfo {author} {\bibfnamefont {M.}~\bibnamefont
  {Eddrief}}, \bibinfo {author} {\bibfnamefont {P.}~\bibnamefont {Atkinson}},
  \bibinfo {author} {\bibfnamefont {V.}~\bibnamefont {Etgens}}, \ and\ \bibinfo
  {author} {\bibfnamefont {B.}~\bibnamefont {Jusserand}},\ }\href@noop {}
  {\bibfield  {journal} {\bibinfo  {journal} {{Nanotechnology}}\ }\textbf
  {\bibinfo {volume} {{25}}},\ \bibinfo {pages} {245701} (\bibinfo {year}
  {{2014}})}\BibitemShut {NoStop}%
\bibitem [{\citenamefont {Liu}\ \emph {et~al.}(2011{\natexlab{b}})\citenamefont
  {Liu}, \citenamefont {Smith}, \citenamefont {Fan}, \citenamefont {Zhang},
  \citenamefont {Cao}, \citenamefont {Chen}, \citenamefont {Kirby},
  \citenamefont {Sun}, \citenamefont {Ruggiero}, \citenamefont {Leiner},
  \citenamefont {Pimpinella}, \citenamefont {Hagmann}, \citenamefont
  {Tivakornsasithorn}, \citenamefont {Dobrowolska},\ and\ \citenamefont
  {Furdyna}}]{Liu2011}%
  \BibitemOpen
  \bibfield  {author} {\bibinfo {author} {\bibfnamefont {X.}~\bibnamefont
  {Liu}}, \bibinfo {author} {\bibfnamefont {D.~J.}\ \bibnamefont {Smith}},
  \bibinfo {author} {\bibfnamefont {J.}~\bibnamefont {Fan}}, \bibinfo {author}
  {\bibfnamefont {Y.}~\bibnamefont {Zhang}}, \bibinfo {author} {\bibfnamefont
  {H.}~\bibnamefont {Cao}}, \bibinfo {author} {\bibfnamefont {Y.~P.}\
  \bibnamefont {Chen}}, \bibinfo {author} {\bibfnamefont {B.~J.}\ \bibnamefont
  {Kirby}}, \bibinfo {author} {\bibfnamefont {N.}~\bibnamefont {Sun}}, \bibinfo
  {author} {\bibfnamefont {S.~T.}\ \bibnamefont {Ruggiero}}, \bibinfo {author}
  {\bibfnamefont {J.}~\bibnamefont {Leiner}}, \bibinfo {author} {\bibfnamefont
  {R.~E.}\ \bibnamefont {Pimpinella}}, \bibinfo {author} {\bibfnamefont
  {J.}~\bibnamefont {Hagmann}}, \bibinfo {author} {\bibfnamefont
  {K.}~\bibnamefont {Tivakornsasithorn}}, \bibinfo {author} {\bibfnamefont
  {M.}~\bibnamefont {Dobrowolska}}, \ and\ \bibinfo {author} {\bibfnamefont
  {J.~K.}\ \bibnamefont {Furdyna}},\ }\href {\doibase 10.1063/1.3671709}
  {\bibfield  {journal} {\bibinfo  {journal} {AIP Conf. Proc.}\ }\textbf
  {\bibinfo {volume} {1416}},\ \bibinfo {pages} {105} (\bibinfo {year}
  {2011}{\natexlab{b}})}\BibitemShut {NoStop}%
\bibitem [{\citenamefont {Guillet}\ \emph {et~al.}(2018)\citenamefont
  {Guillet}, \citenamefont {Marty}, \citenamefont {Beigne}, \citenamefont
  {Vergnaud}, \citenamefont {Dau}, \citenamefont {Noel}, \citenamefont
  {Frigerio}, \citenamefont {Isella},\ and\ \citenamefont
  {Jamet}}]{Guillet2018}%
  \BibitemOpen
  \bibfield  {author} {\bibinfo {author} {\bibfnamefont {T.}~\bibnamefont
  {Guillet}}, \bibinfo {author} {\bibfnamefont {A.}~\bibnamefont {Marty}},
  \bibinfo {author} {\bibfnamefont {C.}~\bibnamefont {Beigne}}, \bibinfo
  {author} {\bibfnamefont {C.}~\bibnamefont {Vergnaud}}, \bibinfo {author}
  {\bibfnamefont {M.~T.}\ \bibnamefont {Dau}}, \bibinfo {author} {\bibfnamefont
  {P.}~\bibnamefont {Noel}}, \bibinfo {author} {\bibfnamefont {J.}~\bibnamefont
  {Frigerio}}, \bibinfo {author} {\bibfnamefont {G.}~\bibnamefont {Isella}}, \
  and\ \bibinfo {author} {\bibfnamefont {M.}~\bibnamefont {Jamet}},\ }\href
  {\doibase 10.1063/1.5048547} {\bibfield  {journal} {\bibinfo  {journal} {AIP
  Adv.}\ }\textbf {\bibinfo {volume} {8}},\ \bibinfo {pages} {115125} (\bibinfo
  {year} {2018})}\BibitemShut {NoStop}%
\bibitem [{\citenamefont {Kim}\ \emph {et~al.}(2018)\citenamefont {Kim},
  \citenamefont {Lee}, \citenamefont {Woo},\ and\ \citenamefont
  {Lee}}]{Kim2018}%
  \BibitemOpen
  \bibfield  {author} {\bibinfo {author} {\bibfnamefont {S.}~\bibnamefont
  {Kim}}, \bibinfo {author} {\bibfnamefont {S.}~\bibnamefont {Lee}}, \bibinfo
  {author} {\bibfnamefont {J.}~\bibnamefont {Woo}}, \ and\ \bibinfo {author}
  {\bibfnamefont {G.}~\bibnamefont {Lee}},\ }\href {\doibase
  10.1016/j.apsusc.2017.03.029} {\bibfield  {journal} {\bibinfo  {journal}
  {Appl. Surf. Sci.}\ }\textbf {\bibinfo {volume} {432}},\ \bibinfo {pages}
  {152} (\bibinfo {year} {2018})}\BibitemShut {NoStop}%
\bibitem [{\citenamefont {Kou}\ \emph {et~al.}(2011)\citenamefont {Kou},
  \citenamefont {He}, \citenamefont {Xiu}, \citenamefont {Lang}, \citenamefont
  {Liao}, \citenamefont {Wang}, \citenamefont {Fedorov}, \citenamefont {Yu},
  \citenamefont {Tang}, \citenamefont {Huang}, \citenamefont {Jiang},
  \citenamefont {Zhu}, \citenamefont {Zou},\ and\ \citenamefont
  {Wang}}]{Kou2011}%
  \BibitemOpen
  \bibfield  {author} {\bibinfo {author} {\bibfnamefont {X.~F.}\ \bibnamefont
  {Kou}}, \bibinfo {author} {\bibfnamefont {L.}~\bibnamefont {He}}, \bibinfo
  {author} {\bibfnamefont {F.~X.}\ \bibnamefont {Xiu}}, \bibinfo {author}
  {\bibfnamefont {M.~R.}\ \bibnamefont {Lang}}, \bibinfo {author}
  {\bibfnamefont {Z.~M.}\ \bibnamefont {Liao}}, \bibinfo {author}
  {\bibfnamefont {Y.}~\bibnamefont {Wang}}, \bibinfo {author} {\bibfnamefont
  {A.~V.}\ \bibnamefont {Fedorov}}, \bibinfo {author} {\bibfnamefont {X.~X.}\
  \bibnamefont {Yu}}, \bibinfo {author} {\bibfnamefont {J.~S.}\ \bibnamefont
  {Tang}}, \bibinfo {author} {\bibfnamefont {G.}~\bibnamefont {Huang}},
  \bibinfo {author} {\bibfnamefont {X.~W.}\ \bibnamefont {Jiang}}, \bibinfo
  {author} {\bibfnamefont {J.~F.}\ \bibnamefont {Zhu}}, \bibinfo {author}
  {\bibfnamefont {J.}~\bibnamefont {Zou}}, \ and\ \bibinfo {author}
  {\bibfnamefont {K.~L.}\ \bibnamefont {Wang}},\ }\href {\doibase
  10.1063/1.3599540} {\bibfield  {journal} {\bibinfo  {journal} {Appl. Phys.
  Lett.}\ }\textbf {\bibinfo {volume} {98}},\ \bibinfo {pages} {242102}
  (\bibinfo {year} {2011})}\BibitemShut {NoStop}%
\bibitem [{\citenamefont {Zhang}\ \emph {et~al.}(2011)\citenamefont {Zhang},
  \citenamefont {Qin}, \citenamefont {Chen}, \citenamefont {He}, \citenamefont
  {Lu}, \citenamefont {Li},\ and\ \citenamefont {Wu}}]{Zhang_STO_2011}%
  \BibitemOpen
  \bibfield  {author} {\bibinfo {author} {\bibfnamefont {G.}~\bibnamefont
  {Zhang}}, \bibinfo {author} {\bibfnamefont {H.}~\bibnamefont {Qin}}, \bibinfo
  {author} {\bibfnamefont {J.}~\bibnamefont {Chen}}, \bibinfo {author}
  {\bibfnamefont {X.}~\bibnamefont {He}}, \bibinfo {author} {\bibfnamefont
  {L.}~\bibnamefont {Lu}}, \bibinfo {author} {\bibfnamefont {Y.}~\bibnamefont
  {Li}}, \ and\ \bibinfo {author} {\bibfnamefont {K.}~\bibnamefont {Wu}},\
  }\href {\doibase 10.1002/adfm.201002667} {\bibfield  {journal} {\bibinfo
  {journal} {Adv. Funct. Mater.}\ }\textbf {\bibinfo {volume} {21}},\ \bibinfo
  {pages} {2351} (\bibinfo {year} {2011})}\BibitemShut {NoStop}%
\bibitem [{\citenamefont {Zhang}\ \emph {et~al.}(2010)\citenamefont {Zhang},
  \citenamefont {He}, \citenamefont {Chang}, \citenamefont {Song},
  \citenamefont {Wang}, \citenamefont {Chen}, \citenamefont {Jia},
  \citenamefont {Fang}, \citenamefont {Dai}, \citenamefont {Shan},
  \citenamefont {Shen}, \citenamefont {Niu}, \citenamefont {Qi}, \citenamefont
  {Zhang}, \citenamefont {Ma},\ and\ \citenamefont {Xue}}]{Zhang2010NatPhys}%
  \BibitemOpen
  \bibfield  {author} {\bibinfo {author} {\bibfnamefont {Y.}~\bibnamefont
  {Zhang}}, \bibinfo {author} {\bibfnamefont {K.}~\bibnamefont {He}}, \bibinfo
  {author} {\bibfnamefont {C.-Z.}\ \bibnamefont {Chang}}, \bibinfo {author}
  {\bibfnamefont {C.-L.}\ \bibnamefont {Song}}, \bibinfo {author}
  {\bibfnamefont {L.-L.}\ \bibnamefont {Wang}}, \bibinfo {author}
  {\bibfnamefont {X.}~\bibnamefont {Chen}}, \bibinfo {author} {\bibfnamefont
  {J.-F.}\ \bibnamefont {Jia}}, \bibinfo {author} {\bibfnamefont
  {Z.}~\bibnamefont {Fang}}, \bibinfo {author} {\bibfnamefont {X.}~\bibnamefont
  {Dai}}, \bibinfo {author} {\bibfnamefont {W.-Y.}\ \bibnamefont {Shan}},
  \bibinfo {author} {\bibfnamefont {S.-Q.}\ \bibnamefont {Shen}}, \bibinfo
  {author} {\bibfnamefont {Q.}~\bibnamefont {Niu}}, \bibinfo {author}
  {\bibfnamefont {X.-L.}\ \bibnamefont {Qi}}, \bibinfo {author} {\bibfnamefont
  {S.-C.}\ \bibnamefont {Zhang}}, \bibinfo {author} {\bibfnamefont {X.-C.}\
  \bibnamefont {Ma}}, \ and\ \bibinfo {author} {\bibfnamefont {Q.-K.}\
  \bibnamefont {Xue}},\ }\href {\doibase 10.1038/nphys1689} {\bibfield
  {journal} {\bibinfo  {journal} {Nat. Phys.}\ }\textbf {\bibinfo {volume}
  {6}},\ \bibinfo {pages} {584} (\bibinfo {year} {2010})}\BibitemShut {NoStop}%
\bibitem [{\citenamefont {Schreyeck}\ \emph {et~al.}(2013)\citenamefont
  {Schreyeck}, \citenamefont {Tarakina}, \citenamefont {Karczewski},
  \citenamefont {Schumacher}, \citenamefont {Borzenko}, \citenamefont {Brune},
  \citenamefont {Buhmann}, \citenamefont {Gould}, \citenamefont {Brunner},\
  and\ \citenamefont {Molenkamp}}]{Schreyeck2013}%
  \BibitemOpen
  \bibfield  {author} {\bibinfo {author} {\bibfnamefont {S.}~\bibnamefont
  {Schreyeck}}, \bibinfo {author} {\bibfnamefont {N.~V.}\ \bibnamefont
  {Tarakina}}, \bibinfo {author} {\bibfnamefont {G.}~\bibnamefont
  {Karczewski}}, \bibinfo {author} {\bibfnamefont {C.}~\bibnamefont
  {Schumacher}}, \bibinfo {author} {\bibfnamefont {T.}~\bibnamefont
  {Borzenko}}, \bibinfo {author} {\bibfnamefont {C.}~\bibnamefont {Brune}},
  \bibinfo {author} {\bibfnamefont {H.}~\bibnamefont {Buhmann}}, \bibinfo
  {author} {\bibfnamefont {C.}~\bibnamefont {Gould}}, \bibinfo {author}
  {\bibfnamefont {K.}~\bibnamefont {Brunner}}, \ and\ \bibinfo {author}
  {\bibfnamefont {L.~W.}\ \bibnamefont {Molenkamp}},\ }\href {\doibase
  10.1063/1.4789775} {\bibfield  {journal} {\bibinfo  {journal} {{Appl. Phys.
  Lett.}}\ }\textbf {\bibinfo {volume} {102}},\ \bibinfo {pages} {041914}
  (\bibinfo {year} {2013})}\BibitemShut {NoStop}%
\bibitem [{\citenamefont {Guo}\ \emph {et~al.}(2013)\citenamefont {Guo},
  \citenamefont {Xu}, \citenamefont {Liu}, \citenamefont {Zhao}, \citenamefont
  {Dai}, \citenamefont {He}, \citenamefont {Wang}, \citenamefont {Liu},
  \citenamefont {Ho},\ and\ \citenamefont {Xie}}]{Guo2013}%
  \BibitemOpen
  \bibfield  {author} {\bibinfo {author} {\bibfnamefont {X.}~\bibnamefont
  {Guo}}, \bibinfo {author} {\bibfnamefont {Z.~J.}\ \bibnamefont {Xu}},
  \bibinfo {author} {\bibfnamefont {H.~C.}\ \bibnamefont {Liu}}, \bibinfo
  {author} {\bibfnamefont {B.}~\bibnamefont {Zhao}}, \bibinfo {author}
  {\bibfnamefont {X.~Q.}\ \bibnamefont {Dai}}, \bibinfo {author} {\bibfnamefont
  {H.~T.}\ \bibnamefont {He}}, \bibinfo {author} {\bibfnamefont {J.~N.}\
  \bibnamefont {Wang}}, \bibinfo {author} {\bibfnamefont {H.~J.}\ \bibnamefont
  {Liu}}, \bibinfo {author} {\bibfnamefont {W.~K.}\ \bibnamefont {Ho}}, \ and\
  \bibinfo {author} {\bibfnamefont {M.~H.}\ \bibnamefont {Xie}},\ }\href@noop
  {} {\bibfield  {journal} {\bibinfo  {journal} {Appl. Phys. Lett.}\ }\textbf
  {\bibinfo {volume} {102}},\ \bibinfo {pages} {151604} (\bibinfo {year}
  {2013})}\BibitemShut {NoStop}%
\bibitem [{\citenamefont {Bonell}\ \emph {et~al.}(2017)\citenamefont {Bonell},
  \citenamefont {Cuxart}, \citenamefont {Song}, \citenamefont {Robles},
  \citenamefont {Ordejon}, \citenamefont {Roche}, \citenamefont {Mugarza},\
  and\ \citenamefont {Valenzuela}}]{Bonell_2017}%
  \BibitemOpen
  \bibfield  {author} {\bibinfo {author} {\bibfnamefont {F.}~\bibnamefont
  {Bonell}}, \bibinfo {author} {\bibfnamefont {M.~G.}\ \bibnamefont {Cuxart}},
  \bibinfo {author} {\bibfnamefont {K.}~\bibnamefont {Song}}, \bibinfo {author}
  {\bibfnamefont {R.}~\bibnamefont {Robles}}, \bibinfo {author} {\bibfnamefont
  {P.}~\bibnamefont {Ordejon}}, \bibinfo {author} {\bibfnamefont
  {S.}~\bibnamefont {Roche}}, \bibinfo {author} {\bibfnamefont
  {A.}~\bibnamefont {Mugarza}}, \ and\ \bibinfo {author} {\bibfnamefont
  {S.~O.}\ \bibnamefont {Valenzuela}},\ }\href {\doibase
  10.1021/acs.cgd.7b00525} {\bibfield  {journal} {\bibinfo  {journal} {Cryst.
  Growth Des.}\ }\textbf {\bibinfo {volume} {17}},\ \bibinfo {pages} {4655}
  (\bibinfo {year} {2017})}\BibitemShut {NoStop}%
\bibitem [{\citenamefont {Jerng}\ \emph {et~al.}(2013)\citenamefont {Jerng},
  \citenamefont {Joo}, \citenamefont {Kim}, \citenamefont {Yoon}, \citenamefont
  {Lee}, \citenamefont {Kim}, \citenamefont {Kim}, \citenamefont {Yoon},
  \citenamefont {Chun},\ and\ \citenamefont {Kim}}]{Jerng2013}%
  \BibitemOpen
  \bibfield  {author} {\bibinfo {author} {\bibfnamefont {S.}~\bibnamefont
  {Jerng}}, \bibinfo {author} {\bibfnamefont {K.}~\bibnamefont {Joo}}, \bibinfo
  {author} {\bibfnamefont {Y.}~\bibnamefont {Kim}}, \bibinfo {author}
  {\bibfnamefont {S.}~\bibnamefont {Yoon}}, \bibinfo {author} {\bibfnamefont
  {J.}~\bibnamefont {Lee}}, \bibinfo {author} {\bibfnamefont {M.}~\bibnamefont
  {Kim}}, \bibinfo {author} {\bibfnamefont {J.}~\bibnamefont {Kim}}, \bibinfo
  {author} {\bibfnamefont {E.}~\bibnamefont {Yoon}}, \bibinfo {author}
  {\bibfnamefont {S.}~\bibnamefont {Chun}}, \ and\ \bibinfo {author}
  {\bibfnamefont {Y.}~\bibnamefont {Kim}},\ }\href@noop {} {\bibfield
  {journal} {\bibinfo  {journal} {Nanoscale}\ }\textbf {\bibinfo {volume}
  {5}},\ \bibinfo {pages} {10618} (\bibinfo {year} {2013})}\BibitemShut
  {NoStop}%
\bibitem [{\citenamefont {Bansal}\ \emph {et~al.}(2014)\citenamefont {Bansal},
  \citenamefont {Cho}, \citenamefont {Brahlek}, \citenamefont {Koirala},
  \citenamefont {Horibe}, \citenamefont {Chen}, \citenamefont {Wu},
  \citenamefont {Park},\ and\ \citenamefont {Oh}}]{Bansal2014}%
  \BibitemOpen
  \bibfield  {author} {\bibinfo {author} {\bibfnamefont {N.}~\bibnamefont
  {Bansal}}, \bibinfo {author} {\bibfnamefont {M.~R.}\ \bibnamefont {Cho}},
  \bibinfo {author} {\bibfnamefont {M.}~\bibnamefont {Brahlek}}, \bibinfo
  {author} {\bibfnamefont {N.}~\bibnamefont {Koirala}}, \bibinfo {author}
  {\bibfnamefont {Y.}~\bibnamefont {Horibe}}, \bibinfo {author} {\bibfnamefont
  {J.}~\bibnamefont {Chen}}, \bibinfo {author} {\bibfnamefont {W.}~\bibnamefont
  {Wu}}, \bibinfo {author} {\bibfnamefont {Y.~D.}\ \bibnamefont {Park}}, \ and\
  \bibinfo {author} {\bibfnamefont {S.}~\bibnamefont {Oh}},\ }\href@noop {}
  {\bibfield  {journal} {\bibinfo  {journal} {Nano Lett.}\ }\textbf {\bibinfo
  {volume} {14}},\ \bibinfo {pages} {1343} (\bibinfo {year}
  {2014})}\BibitemShut {NoStop}%
\bibitem [{\citenamefont {Jeon}\ \emph {et~al.}(2014)\citenamefont {Jeon},
  \citenamefont {Song}, \citenamefont {Kim}, \citenamefont {Jang},
  \citenamefont {Park}, \citenamefont {Yoon},\ and\ \citenamefont
  {Kahng}}]{Jeon2014}%
  \BibitemOpen
  \bibfield  {author} {\bibinfo {author} {\bibfnamefont {J.}~\bibnamefont
  {Jeon}}, \bibinfo {author} {\bibfnamefont {M.}~\bibnamefont {Song}}, \bibinfo
  {author} {\bibfnamefont {H.}~\bibnamefont {Kim}}, \bibinfo {author}
  {\bibfnamefont {W.}~\bibnamefont {Jang}}, \bibinfo {author} {\bibfnamefont
  {J.}~\bibnamefont {Park}}, \bibinfo {author} {\bibfnamefont {S.}~\bibnamefont
  {Yoon}}, \ and\ \bibinfo {author} {\bibfnamefont {S.}~\bibnamefont {Kahng}},\
  }\href@noop {} {\bibfield  {journal} {\bibinfo  {journal} {Appl. Surf. Sci.}\
  }\textbf {\bibinfo {volume} {316}},\ \bibinfo {pages} {42} (\bibinfo {year}
  {2014})}\BibitemShut {NoStop}%
\bibitem [{\citenamefont {Collins-McIntyre}\ \emph {et~al.}(2015)\citenamefont
  {Collins-McIntyre}, \citenamefont {Wang}, \citenamefont {Zhou}, \citenamefont
  {Speller}, \citenamefont {Chen},\ and\ \citenamefont
  {Hesjedal}}]{Collins_silica}%
  \BibitemOpen
  \bibfield  {author} {\bibinfo {author} {\bibfnamefont {L.~J.}\ \bibnamefont
  {Collins-McIntyre}}, \bibinfo {author} {\bibfnamefont {W.}~\bibnamefont
  {Wang}}, \bibinfo {author} {\bibfnamefont {B.}~\bibnamefont {Zhou}}, \bibinfo
  {author} {\bibfnamefont {S.~C.}\ \bibnamefont {Speller}}, \bibinfo {author}
  {\bibfnamefont {Y.~L.}\ \bibnamefont {Chen}}, \ and\ \bibinfo {author}
  {\bibfnamefont {T.}~\bibnamefont {Hesjedal}},\ }\href {\doibase
  10.1002/pssb.201552003} {\bibfield  {journal} {\bibinfo  {journal} {Phys.
  Status Solidi B}\ }\textbf {\bibinfo {volume} {252}},\ \bibinfo {pages}
  {1334} (\bibinfo {year} {2015})}\BibitemShut {NoStop}%
\bibitem [{Note1()}]{Note1}%
  \BibitemOpen
  \bibinfo {note} {For achieving in-plane ordered films, the substrates should
  have hexagonal symmetry.}\BibitemShut {Stop}%
\bibitem [{\citenamefont {Koma}, \citenamefont {Sunouchi},\ and\ \citenamefont
  {Miyajima}(1985)}]{Koma1985}%
  \BibitemOpen
  \bibfield  {author} {\bibinfo {author} {\bibfnamefont {A.}~\bibnamefont
  {Koma}}, \bibinfo {author} {\bibfnamefont {K.}~\bibnamefont {Sunouchi}}, \
  and\ \bibinfo {author} {\bibfnamefont {T.}~\bibnamefont {Miyajima}},\
  }\href@noop {} {\bibfield  {journal} {\bibinfo  {journal} {J. Vac. Sci.
  Technol. B}\ }\textbf {\bibinfo {volume} {3}},\ \bibinfo {pages} {724}
  (\bibinfo {year} {1985})}\BibitemShut {NoStop}%
\bibitem [{\citenamefont {Koma}(1992)}]{Koma1992}%
  \BibitemOpen
  \bibfield  {author} {\bibinfo {author} {\bibfnamefont {A.}~\bibnamefont
  {Koma}},\ }\href@noop {} {\bibfield  {journal} {\bibinfo  {journal} {Thin
  Sol. Films}\ }\textbf {\bibinfo {volume} {216}},\ \bibinfo {pages} {72}
  (\bibinfo {year} {1992})}\BibitemShut {NoStop}%
\bibitem [{\citenamefont {Koma}(1999)}]{Koma1999}%
  \BibitemOpen
  \bibfield  {author} {\bibinfo {author} {\bibfnamefont {A.}~\bibnamefont
  {Koma}},\ }\href {\doibase 10.1016/S0022-0248(98)01329-3} {\bibfield
  {journal} {\bibinfo  {journal} {J. Cryst. Growth}\ }\textbf {\bibinfo
  {volume} {201--202}},\ \bibinfo {pages} {236 } (\bibinfo {year}
  {1999})}\BibitemShut {NoStop}%
\bibitem [{\citenamefont {Walsh}\ and\ \citenamefont
  {Hinkle}(2017)}]{Walsh_vdWepi_2017}%
  \BibitemOpen
  \bibfield  {author} {\bibinfo {author} {\bibfnamefont {L.~A.}\ \bibnamefont
  {Walsh}}\ and\ \bibinfo {author} {\bibfnamefont {C.~L.}\ \bibnamefont
  {Hinkle}},\ }\href {\doibase 10.1016/j.apmt.2017.09.010} {\bibfield
  {journal} {\bibinfo  {journal} {Appl. Mater. Today}\ }\textbf {\bibinfo
  {volume} {9}},\ \bibinfo {pages} {504} (\bibinfo {year} {2017})}\BibitemShut
  {NoStop}%
\bibitem [{\citenamefont {Chen}\ \emph {et~al.}(2011)\citenamefont {Chen},
  \citenamefont {Ma}, \citenamefont {He}, \citenamefont {Jia},\ and\
  \citenamefont {Xue}}]{Chen2011}%
  \BibitemOpen
  \bibfield  {author} {\bibinfo {author} {\bibfnamefont {X.}~\bibnamefont
  {Chen}}, \bibinfo {author} {\bibfnamefont {X.-C.}\ \bibnamefont {Ma}},
  \bibinfo {author} {\bibfnamefont {K.}~\bibnamefont {He}}, \bibinfo {author}
  {\bibfnamefont {J.-F.}\ \bibnamefont {Jia}}, \ and\ \bibinfo {author}
  {\bibfnamefont {Q.-K.}\ \bibnamefont {Xue}},\ }\href@noop {} {\bibfield
  {journal} {\bibinfo  {journal} {Adv. Mater.}\ }\textbf {\bibinfo {volume}
  {23}},\ \bibinfo {pages} {1162} (\bibinfo {year} {2011})}\BibitemShut
  {NoStop}%
\bibitem [{\citenamefont {Taskin}\ \emph
  {et~al.}(2012{\natexlab{b}})\citenamefont {Taskin}, \citenamefont {Sasaki},
  \citenamefont {Segawa},\ and\ \citenamefont {Ando}}]{Taskin2012}%
  \BibitemOpen
  \bibfield  {author} {\bibinfo {author} {\bibfnamefont {A.~A.}\ \bibnamefont
  {Taskin}}, \bibinfo {author} {\bibfnamefont {S.}~\bibnamefont {Sasaki}},
  \bibinfo {author} {\bibfnamefont {K.}~\bibnamefont {Segawa}}, \ and\ \bibinfo
  {author} {\bibfnamefont {Y.}~\bibnamefont {Ando}},\ }\href {\doibase
  10.1002/adma.201201827} {\bibfield  {journal} {\bibinfo  {journal} {Adv.
  Mater.}\ }\textbf {\bibinfo {volume} {24}},\ \bibinfo {pages} {5581}
  (\bibinfo {year} {2012}{\natexlab{b}})}\BibitemShut {NoStop}%
\bibitem [{\citenamefont {Haazen}\ \emph {et~al.}(2012)\citenamefont {Haazen},
  \citenamefont {Lalo{\"e}}, \citenamefont {Nummy}, \citenamefont {Swagten},
  \citenamefont {Jarillo-Herrero}, \citenamefont {Heiman},\ and\ \citenamefont
  {Moodera}}]{Haazen2012}%
  \BibitemOpen
  \bibfield  {author} {\bibinfo {author} {\bibfnamefont {P.~P.~J.}\
  \bibnamefont {Haazen}}, \bibinfo {author} {\bibfnamefont {J.-B.}\
  \bibnamefont {Lalo{\"e}}}, \bibinfo {author} {\bibfnamefont {T.~J.}\
  \bibnamefont {Nummy}}, \bibinfo {author} {\bibfnamefont {H.~J.~M.}\
  \bibnamefont {Swagten}}, \bibinfo {author} {\bibfnamefont {P.}~\bibnamefont
  {Jarillo-Herrero}}, \bibinfo {author} {\bibfnamefont {D.}~\bibnamefont
  {Heiman}}, \ and\ \bibinfo {author} {\bibfnamefont {J.~S.}\ \bibnamefont
  {Moodera}},\ }\href {\doibase 10.1063/1.3688043} {\bibfield  {journal}
  {\bibinfo  {journal} {Appl. Phys. Lett.}\ }\textbf {\bibinfo {volume}
  {100}},\ \bibinfo {pages} {082404} (\bibinfo {year} {2012})}\BibitemShut
  {NoStop}%
\bibitem [{\citenamefont {Li}\ \emph {et~al.}(2013)\citenamefont {Li},
  \citenamefont {Harrison}, \citenamefont {Huo}, \citenamefont {Pushp},
  \citenamefont {Yuan}, \citenamefont {Zhou}, \citenamefont {Kellock},
  \citenamefont {Parkin}, \citenamefont {Chen}, \citenamefont {Hesjedal},\ and\
  \citenamefont {Harris}}]{Li2013}%
  \BibitemOpen
  \bibfield  {author} {\bibinfo {author} {\bibfnamefont {S.}~\bibnamefont
  {Li}}, \bibinfo {author} {\bibfnamefont {S.}~\bibnamefont {Harrison}},
  \bibinfo {author} {\bibfnamefont {Y.}~\bibnamefont {Huo}}, \bibinfo {author}
  {\bibfnamefont {A.}~\bibnamefont {Pushp}}, \bibinfo {author} {\bibfnamefont
  {H.~T.}\ \bibnamefont {Yuan}}, \bibinfo {author} {\bibfnamefont
  {B.}~\bibnamefont {Zhou}}, \bibinfo {author} {\bibfnamefont {A.~J.}\
  \bibnamefont {Kellock}}, \bibinfo {author} {\bibfnamefont {S.~S.~P.}\
  \bibnamefont {Parkin}}, \bibinfo {author} {\bibfnamefont {Y.-L.}\
  \bibnamefont {Chen}}, \bibinfo {author} {\bibfnamefont {T.}~\bibnamefont
  {Hesjedal}}, \ and\ \bibinfo {author} {\bibfnamefont {J.~S.}\ \bibnamefont
  {Harris}},\ }\href@noop {} {\bibfield  {journal} {\bibinfo  {journal} {{Appl.
  Phys. Lett.}}\ }\textbf {\bibinfo {volume} {102}},\ \bibinfo {pages} {242412}
  (\bibinfo {year} {2013})}\BibitemShut {NoStop}%
\bibitem [{\citenamefont {Kim}\ \emph {et~al.}(2016)\citenamefont {Kim},
  \citenamefont {Lee}, \citenamefont {Kim}, \citenamefont {Choi}, \citenamefont
  {Chang}, \citenamefont {Won}, \citenamefont {Kwon}, \citenamefont {Kim},
  \citenamefont {Hyun}, \citenamefont {Kim}, \citenamefont {Koo}, \citenamefont
  {Choi}, \citenamefont {Kim}, \citenamefont {Kim},\ and\ \citenamefont
  {Baek}}]{Kim2016}%
  \BibitemOpen
  \bibfield  {author} {\bibinfo {author} {\bibfnamefont {K.-C.}\ \bibnamefont
  {Kim}}, \bibinfo {author} {\bibfnamefont {J.}~\bibnamefont {Lee}}, \bibinfo
  {author} {\bibfnamefont {B.~K.}\ \bibnamefont {Kim}}, \bibinfo {author}
  {\bibfnamefont {W.~Y.}\ \bibnamefont {Choi}}, \bibinfo {author}
  {\bibfnamefont {H.~J.}\ \bibnamefont {Chang}}, \bibinfo {author}
  {\bibfnamefont {S.~O.}\ \bibnamefont {Won}}, \bibinfo {author} {\bibfnamefont
  {B.}~\bibnamefont {Kwon}}, \bibinfo {author} {\bibfnamefont {S.~K.}\
  \bibnamefont {Kim}}, \bibinfo {author} {\bibfnamefont {D.-B.}\ \bibnamefont
  {Hyun}}, \bibinfo {author} {\bibfnamefont {H.~J.}\ \bibnamefont {Kim}},
  \bibinfo {author} {\bibfnamefont {H.~C.}\ \bibnamefont {Koo}}, \bibinfo
  {author} {\bibfnamefont {J.-H.}\ \bibnamefont {Choi}}, \bibinfo {author}
  {\bibfnamefont {D.-I.}\ \bibnamefont {Kim}}, \bibinfo {author} {\bibfnamefont
  {J.-S.}\ \bibnamefont {Kim}}, \ and\ \bibinfo {author} {\bibfnamefont
  {S.-H.}\ \bibnamefont {Baek}},\ }\href {\doibase 10.1038/ncomms12449}
  {\bibfield  {journal} {\bibinfo  {journal} {Nat. Commun.}\ }\textbf {\bibinfo
  {volume} {7}},\ \bibinfo {pages} {12449} (\bibinfo {year}
  {2016})}\BibitemShut {NoStop}%
\bibitem [{\citenamefont {Karma}\ and\ \citenamefont
  {Plapp}(1998)}]{Karma1998}%
  \BibitemOpen
  \bibfield  {author} {\bibinfo {author} {\bibfnamefont {A.}~\bibnamefont
  {Karma}}\ and\ \bibinfo {author} {\bibfnamefont {M.}~\bibnamefont {Plapp}},\
  }\href@noop {} {\bibfield  {journal} {\bibinfo  {journal} {Phys. Rev. Lett.}\
  }\textbf {\bibinfo {volume} {81}},\ \bibinfo {pages} {4444} (\bibinfo {year}
  {1998})}\BibitemShut {NoStop}%
\bibitem [{\citenamefont {Liu}, \citenamefont {Weinert},\ and\ \citenamefont
  {Li}(2012)}]{Liu2012}%
  \BibitemOpen
  \bibfield  {author} {\bibinfo {author} {\bibfnamefont {Y.}~\bibnamefont
  {Liu}}, \bibinfo {author} {\bibfnamefont {M.}~\bibnamefont {Weinert}}, \ and\
  \bibinfo {author} {\bibfnamefont {L.}~\bibnamefont {Li}},\ }\href@noop {}
  {\bibfield  {journal} {\bibinfo  {journal} {Phys. Rev. Lett.}\ }\textbf
  {\bibinfo {volume} {108}},\ \bibinfo {pages} {115501} (\bibinfo {year}
  {2012})}\BibitemShut {NoStop}%
\bibitem [{\citenamefont {Dingle}\ \emph {et~al.}(1978)\citenamefont {Dingle},
  \citenamefont {St\"ormer}, \citenamefont {Gossard},\ and\ \citenamefont
  {Wiegmann}}]{Dingle1978}%
  \BibitemOpen
  \bibfield  {author} {\bibinfo {author} {\bibfnamefont {R.}~\bibnamefont
  {Dingle}}, \bibinfo {author} {\bibfnamefont {H.~L.}\ \bibnamefont
  {St\"ormer}}, \bibinfo {author} {\bibfnamefont {A.~C.}\ \bibnamefont
  {Gossard}}, \ and\ \bibinfo {author} {\bibfnamefont {W.}~\bibnamefont
  {Wiegmann}},\ }\href {\doibase 10.1063/1.90457} {\bibfield  {journal}
  {\bibinfo  {journal} {Appl. Phys. Lett.}\ }\textbf {\bibinfo {volume} {33}},\
  \bibinfo {pages} {665} (\bibinfo {year} {1978})}\BibitemShut {NoStop}%
\bibitem [{\citenamefont {St\"ormer}\ \emph {et~al.}(1981)\citenamefont
  {St\"ormer}, \citenamefont {Pinczuk}, \citenamefont {Gossard},\ and\
  \citenamefont {Wiegmann}}]{Stoermer1981}%
  \BibitemOpen
  \bibfield  {author} {\bibinfo {author} {\bibfnamefont {H.~L.}\ \bibnamefont
  {St\"ormer}}, \bibinfo {author} {\bibfnamefont {A.}~\bibnamefont {Pinczuk}},
  \bibinfo {author} {\bibfnamefont {A.~C.}\ \bibnamefont {Gossard}}, \ and\
  \bibinfo {author} {\bibfnamefont {W.}~\bibnamefont {Wiegmann}},\ }\href
  {\doibase 10.1063/1.92481} {\bibfield  {journal} {\bibinfo  {journal} {Appl.
  Phys. Lett.}\ }\textbf {\bibinfo {volume} {38}},\ \bibinfo {pages} {691}
  (\bibinfo {year} {1981})}\BibitemShut {NoStop}%
\bibitem [{\citenamefont {Carlson}(1960)}]{Carlson1960}%
  \BibitemOpen
  \bibfield  {author} {\bibinfo {author} {\bibfnamefont {R.~O.}\ \bibnamefont
  {Carlson}},\ }\href {\doibase 10.1016/0022-3697(60)90127-X} {\bibfield
  {journal} {\bibinfo  {journal} {J. Phys. Chem. Solids}\ }\textbf {\bibinfo
  {volume} {13}},\ \bibinfo {pages} {65} (\bibinfo {year} {1960})}\BibitemShut
  {NoStop}%
\bibitem [{\citenamefont {Korzhuev}\ and\ \citenamefont
  {Svechnikov}(1991)}]{Korzhuev1991}%
  \BibitemOpen
  \bibfield  {author} {\bibinfo {author} {\bibfnamefont {M.~A.}\ \bibnamefont
  {Korzhuev}}\ and\ \bibinfo {author} {\bibfnamefont {T.~E.}\ \bibnamefont
  {Svechnikov}},\ }\href@noop {} {\bibfield  {journal} {\bibinfo  {journal}
  {Sov. Phys. Semicond.}\ }\textbf {\bibinfo {volume} {25}},\ \bibinfo {pages}
  {1288} (\bibinfo {year} {1991})}\BibitemShut {NoStop}%
\bibitem [{\citenamefont {Koski}\ \emph {et~al.}(2012)\citenamefont {Koski},
  \citenamefont {Wessells}, \citenamefont {Reed}, \citenamefont {Cha},
  \citenamefont {Kong},\ and\ \citenamefont {Cui}}]{Koski2012_IntercalBS}%
  \BibitemOpen
  \bibfield  {author} {\bibinfo {author} {\bibfnamefont {K.~J.}\ \bibnamefont
  {Koski}}, \bibinfo {author} {\bibfnamefont {C.~D.}\ \bibnamefont {Wessells}},
  \bibinfo {author} {\bibfnamefont {B.~W.}\ \bibnamefont {Reed}}, \bibinfo
  {author} {\bibfnamefont {J.~J.}\ \bibnamefont {Cha}}, \bibinfo {author}
  {\bibfnamefont {D.}~\bibnamefont {Kong}}, \ and\ \bibinfo {author}
  {\bibfnamefont {Y.}~\bibnamefont {Cui}},\ }\href@noop {} {\bibfield
  {journal} {\bibinfo  {journal} {J. Am. Chem. Soc.}\ }\textbf {\bibinfo
  {volume} {134}},\ \bibinfo {pages} {13773} (\bibinfo {year}
  {2012})}\BibitemShut {NoStop}%
\bibitem [{\citenamefont {Music}\ \emph {et~al.}(2018)\citenamefont {Music},
  \citenamefont {Chen}, \citenamefont {Holzapfel}, \citenamefont {Bilyalova},
  \citenamefont {Helvaci}, \citenamefont {Heymann}, \citenamefont {Aghda},
  \citenamefont {Maron}, \citenamefont {Ravensburg}, \citenamefont {S\"alker},
  \citenamefont {Schnelle},\ and\ \citenamefont {Woeste}}]{Music2018}%
  \BibitemOpen
  \bibfield  {author} {\bibinfo {author} {\bibfnamefont {D.}~\bibnamefont
  {Music}}, \bibinfo {author} {\bibfnamefont {X.}~\bibnamefont {Chen}},
  \bibinfo {author} {\bibfnamefont {D.~M.}\ \bibnamefont {Holzapfel}}, \bibinfo
  {author} {\bibfnamefont {H.~M.}\ \bibnamefont {Bilyalova}}, \bibinfo {author}
  {\bibfnamefont {M.}~\bibnamefont {Helvaci}}, \bibinfo {author} {\bibfnamefont
  {A.~O.~D.}\ \bibnamefont {Heymann}}, \bibinfo {author} {\bibfnamefont
  {S.~K.}\ \bibnamefont {Aghda}}, \bibinfo {author} {\bibfnamefont
  {T.}~\bibnamefont {Maron}}, \bibinfo {author} {\bibfnamefont {A.~L.}\
  \bibnamefont {Ravensburg}}, \bibinfo {author} {\bibfnamefont {J.~A.}\
  \bibnamefont {S\"alker}}, \bibinfo {author} {\bibfnamefont {L.}~\bibnamefont
  {Schnelle}}, \ and\ \bibinfo {author} {\bibfnamefont {L.~A.}\ \bibnamefont
  {Woeste}},\ }\href {\doibase 10.1063/1.5050558} {\bibfield  {journal}
  {\bibinfo  {journal} {J. Appl. Phys.}\ }\textbf {\bibinfo {volume} {124}},\
  \bibinfo {pages} {185106} (\bibinfo {year} {2018})}\BibitemShut {NoStop}%
\bibitem [{\citenamefont {Shaughnessy}\ \emph {et~al.}(2014)\citenamefont
  {Shaughnessy}, \citenamefont {Bartelt}, \citenamefont {Zimmerman},\ and\
  \citenamefont {Sugar}}]{Shaughnessy2014}%
  \BibitemOpen
  \bibfield  {author} {\bibinfo {author} {\bibfnamefont {M.~C.}\ \bibnamefont
  {Shaughnessy}}, \bibinfo {author} {\bibfnamefont {N.~C.}\ \bibnamefont
  {Bartelt}}, \bibinfo {author} {\bibfnamefont {J.~A.}\ \bibnamefont
  {Zimmerman}}, \ and\ \bibinfo {author} {\bibfnamefont {J.~D.}\ \bibnamefont
  {Sugar}},\ }\href {\doibase 10.1063/1.4865735} {\bibfield  {journal}
  {\bibinfo  {journal} {J. Appl. Phys.}\ }\textbf {\bibinfo {volume} {115}},\
  \bibinfo {pages} {063705} (\bibinfo {year} {2014})}\BibitemShut {NoStop}%
\bibitem [{\citenamefont {Lan}\ \emph {et~al.}(2008)\citenamefont {Lan},
  \citenamefont {Wang}, \citenamefont {Chen},\ and\ \citenamefont
  {Ren}}]{Lan2008}%
  \BibitemOpen
  \bibfield  {author} {\bibinfo {author} {\bibfnamefont {Y.~C.}\ \bibnamefont
  {Lan}}, \bibinfo {author} {\bibfnamefont {D.~Z.}\ \bibnamefont {Wang}},
  \bibinfo {author} {\bibfnamefont {G.}~\bibnamefont {Chen}}, \ and\ \bibinfo
  {author} {\bibfnamefont {Z.~F.}\ \bibnamefont {Ren}},\ }\href {\doibase
  10.1063/1.2896310} {\bibfield  {journal} {\bibinfo  {journal} {Appl. Phys.
  Lett.}\ }\textbf {\bibinfo {volume} {92}},\ \bibinfo {pages} {101910}
  (\bibinfo {year} {2008})}\BibitemShut {NoStop}%
\bibitem [{\citenamefont {Keys}\ and\ \citenamefont {Dutton}(1963)}]{Keys1963}%
  \BibitemOpen
  \bibfield  {author} {\bibinfo {author} {\bibfnamefont {J.~D.}\ \bibnamefont
  {Keys}}\ and\ \bibinfo {author} {\bibfnamefont {H.~M.}\ \bibnamefont
  {Dutton}},\ }\href {\doibase 10.1063/1.1702694} {\bibfield  {journal}
  {\bibinfo  {journal} {J. Appl. Phys.}\ }\textbf {\bibinfo {volume} {34}},\
  \bibinfo {pages} {1830} (\bibinfo {year} {1963})}\BibitemShut {NoStop}%
\bibitem [{\citenamefont {Zhou}\ \emph {et~al.}(2006)\citenamefont {Zhou},
  \citenamefont {Zabeik}, \citenamefont {Lostak},\ and\ \citenamefont
  {Uher}}]{Zhou2006_FeSb2Te3}%
  \BibitemOpen
  \bibfield  {author} {\bibinfo {author} {\bibfnamefont {Z.}~\bibnamefont
  {Zhou}}, \bibinfo {author} {\bibfnamefont {M.}~\bibnamefont {Zabeik}},
  \bibinfo {author} {\bibfnamefont {P.}~\bibnamefont {Lostak}}, \ and\ \bibinfo
  {author} {\bibfnamefont {C.}~\bibnamefont {Uher}},\ }\href {\doibase
  10.1063/1.2171787} {\bibfield  {journal} {\bibinfo  {journal} {J. Appl.
  Phys.}\ }\textbf {\bibinfo {volume} {99}},\ \bibinfo {pages} {043901}
  (\bibinfo {year} {2006})}\BibitemShut {NoStop}%
\bibitem [{\citenamefont {Sugama}\ \emph {et~al.}(2001)\citenamefont {Sugama},
  \citenamefont {Hayashi}, \citenamefont {Nakagawa}, \citenamefont {Miura},\
  and\ \citenamefont {Kulbachnskii}}]{Sugama2001}%
  \BibitemOpen
  \bibfield  {author} {\bibinfo {author} {\bibfnamefont {Y.}~\bibnamefont
  {Sugama}}, \bibinfo {author} {\bibfnamefont {T.}~\bibnamefont {Hayashi}},
  \bibinfo {author} {\bibfnamefont {H.}~\bibnamefont {Nakagawa}}, \bibinfo
  {author} {\bibfnamefont {N.}~\bibnamefont {Miura}}, \ and\ \bibinfo {author}
  {\bibfnamefont {V.~A.}\ \bibnamefont {Kulbachnskii}},\ }\href@noop {}
  {\bibfield  {journal} {\bibinfo  {journal} {Low Temperat. Phys.}\ }\textbf
  {\bibinfo {volume} {298}},\ \bibinfo {pages} {531} (\bibinfo {year}
  {2001})}\BibitemShut {NoStop}%
\bibitem [{\citenamefont {Rodriguez}(2019)}]{Rodriguez2019}%
  \BibitemOpen
  \bibfield  {author} {\bibinfo {author} {\bibfnamefont {J.~H.}\ \bibnamefont
  {Rodriguez}},\ }\href {\doibase 10.1103/PhysRevB.100.165113} {\bibfield
  {journal} {\bibinfo  {journal} {Phys. Rev. B}\ }\textbf {\bibinfo {volume}
  {100}},\ \bibinfo {pages} {165113} (\bibinfo {year} {2019})}\BibitemShut
  {NoStop}%
\bibitem [{\citenamefont {Kulbachinskii}\ \emph {et~al.}(2002)\citenamefont
  {Kulbachinskii}, \citenamefont {Kaminskii}, \citenamefont {Kindo},
  \citenamefont {Narumi}, \citenamefont {Suga}, \citenamefont {Lostak},\ and\
  \citenamefont {Svanda}}]{Kulbachinskii2002}%
  \BibitemOpen
  \bibfield  {author} {\bibinfo {author} {\bibfnamefont {V.}~\bibnamefont
  {Kulbachinskii}}, \bibinfo {author} {\bibfnamefont {A.~Y.}\ \bibnamefont
  {Kaminskii}}, \bibinfo {author} {\bibfnamefont {K.}~\bibnamefont {Kindo}},
  \bibinfo {author} {\bibfnamefont {Y.}~\bibnamefont {Narumi}}, \bibinfo
  {author} {\bibfnamefont {K.}~\bibnamefont {Suga}}, \bibinfo {author}
  {\bibfnamefont {P.}~\bibnamefont {Lostak}}, \ and\ \bibinfo {author}
  {\bibfnamefont {P.}~\bibnamefont {Svanda}},\ }\href@noop {} {\bibfield
  {journal} {\bibinfo  {journal} {Physica B}\ }\textbf {\bibinfo {volume}
  {311}},\ \bibinfo {pages} {292} (\bibinfo {year} {2002})}\BibitemShut
  {NoStop}%
\bibitem [{\citenamefont {Teng}, \citenamefont {Liu},\ and\ \citenamefont
  {Li}(2019)}]{Teng2019}%
  \BibitemOpen
  \bibfield  {author} {\bibinfo {author} {\bibfnamefont {J.}~\bibnamefont
  {Teng}}, \bibinfo {author} {\bibfnamefont {N.}~\bibnamefont {Liu}}, \ and\
  \bibinfo {author} {\bibfnamefont {Y.}~\bibnamefont {Li}},\ }\href {\doibase
  10.1088/1674-4926/40/8/081507} {\bibfield  {journal} {\bibinfo  {journal} {J.
  Semicond.}\ }\textbf {\bibinfo {volume} {40}},\ \bibinfo {pages} {081507}
  (\bibinfo {year} {2019})}\BibitemShut {NoStop}%
\bibitem [{\citenamefont {S{\'{a}}nchez-Barriga}\ \emph
  {et~al.}(2016)\citenamefont {S{\'{a}}nchez-Barriga}, \citenamefont
  {Varykhalov}, \citenamefont {Springholz}, \citenamefont {Steiner},
  \citenamefont {Kirchschlager}, \citenamefont {Bauer}, \citenamefont {Caha},
  \citenamefont {Schierle}, \citenamefont {Weschke}, \citenamefont
  {{\"{U}}nal}, \citenamefont {Valencia}, \citenamefont {Dunst}, \citenamefont
  {Braun}, \citenamefont {Ebert}, \citenamefont {Min{\'{a}}r}, \citenamefont
  {Golias}, \citenamefont {Yashina}, \citenamefont {Ney}, \citenamefont
  {Hol{\'{y}}},\ and\ \citenamefont {Rader}}]{Sanchez-Barriga2016}%
  \BibitemOpen
  \bibfield  {author} {\bibinfo {author} {\bibfnamefont {J.}~\bibnamefont
  {S{\'{a}}nchez-Barriga}}, \bibinfo {author} {\bibfnamefont {A.}~\bibnamefont
  {Varykhalov}}, \bibinfo {author} {\bibfnamefont {G.}~\bibnamefont
  {Springholz}}, \bibinfo {author} {\bibfnamefont {H.}~\bibnamefont {Steiner}},
  \bibinfo {author} {\bibfnamefont {R.}~\bibnamefont {Kirchschlager}}, \bibinfo
  {author} {\bibfnamefont {G.}~\bibnamefont {Bauer}}, \bibinfo {author}
  {\bibfnamefont {O.}~\bibnamefont {Caha}}, \bibinfo {author} {\bibfnamefont
  {E.}~\bibnamefont {Schierle}}, \bibinfo {author} {\bibfnamefont
  {E.}~\bibnamefont {Weschke}}, \bibinfo {author} {\bibfnamefont {A.~A.}\
  \bibnamefont {{\"{U}}nal}}, \bibinfo {author} {\bibfnamefont
  {S.}~\bibnamefont {Valencia}}, \bibinfo {author} {\bibfnamefont
  {M.}~\bibnamefont {Dunst}}, \bibinfo {author} {\bibfnamefont
  {J.}~\bibnamefont {Braun}}, \bibinfo {author} {\bibfnamefont
  {H.}~\bibnamefont {Ebert}}, \bibinfo {author} {\bibfnamefont
  {J.}~\bibnamefont {Min{\'{a}}r}}, \bibinfo {author} {\bibfnamefont
  {E.}~\bibnamefont {Golias}}, \bibinfo {author} {\bibfnamefont {L.~V.}\
  \bibnamefont {Yashina}}, \bibinfo {author} {\bibfnamefont {A.}~\bibnamefont
  {Ney}}, \bibinfo {author} {\bibfnamefont {V.}~\bibnamefont {Hol{\'{y}}}}, \
  and\ \bibinfo {author} {\bibfnamefont {O.}~\bibnamefont {Rader}},\ }\href
  {\doibase 10.1038/ncomms10559} {\bibfield  {journal} {\bibinfo  {journal}
  {Nat. Commun.}\ }\textbf {\bibinfo {volume} {7}},\ \bibinfo {pages} {1}
  (\bibinfo {year} {2016})}\BibitemShut {NoStop}%
\bibitem [{\citenamefont {Otrokov}\ \emph {et~al.}(2019)\citenamefont
  {Otrokov}, \citenamefont {Klimovskikh}, \citenamefont {Bentmann},
  \citenamefont {Estyunin}, \citenamefont {Zeugner}, \citenamefont {Aliev},
  \citenamefont {Ga{\ss}}, \citenamefont {Wolter}, \citenamefont {Koroleva},
  \citenamefont {Shikin}, \citenamefont {Blanco-Rey}, \citenamefont {Hoffmann},
  \citenamefont {Rusinov}, \citenamefont {Eremeev}, \citenamefont {Koroteev},
  \citenamefont {Kuznetsov}, \citenamefont {Freyse}, \citenamefont
  {S{\'{a}}nchez-Barriga}, \citenamefont {Amiraslanov}, \citenamefont
  {Babanly}, \citenamefont {Mamedov}, \citenamefont {Abdullayev}, \citenamefont
  {Zverev}, \citenamefont {Alfonsov}, \citenamefont {Kataev}, \citenamefont
  {B\"uchner}, \citenamefont {Schwier}, \citenamefont {Kumar}, \citenamefont
  {Kimura}, \citenamefont {Petaccia}, \citenamefont {Di~Santo}, \citenamefont
  {Vidal}, \citenamefont {Schatz}, \citenamefont {Ki{\ss}ner}, \citenamefont
  {{\"{U}}nzelmann}, \citenamefont {Min}, \citenamefont {Moser}, \citenamefont
  {Peixoto}, \citenamefont {Reinert}, \citenamefont {Ernst}, \citenamefont
  {Echenique}, \citenamefont {Isaeva},\ and\ \citenamefont
  {Chulkov}}]{Otrokov2019}%
  \BibitemOpen
  \bibfield  {author} {\bibinfo {author} {\bibfnamefont {M.~M.}\ \bibnamefont
  {Otrokov}}, \bibinfo {author} {\bibfnamefont {I.~I.}\ \bibnamefont
  {Klimovskikh}}, \bibinfo {author} {\bibfnamefont {H.}~\bibnamefont
  {Bentmann}}, \bibinfo {author} {\bibfnamefont {D.}~\bibnamefont {Estyunin}},
  \bibinfo {author} {\bibfnamefont {A.}~\bibnamefont {Zeugner}}, \bibinfo
  {author} {\bibfnamefont {Z.~S.}\ \bibnamefont {Aliev}}, \bibinfo {author}
  {\bibfnamefont {S.}~\bibnamefont {Ga{\ss}}}, \bibinfo {author} {\bibfnamefont
  {A.~U.~B.}\ \bibnamefont {Wolter}}, \bibinfo {author} {\bibfnamefont {A.~V.}\
  \bibnamefont {Koroleva}}, \bibinfo {author} {\bibfnamefont {A.~M.}\
  \bibnamefont {Shikin}}, \bibinfo {author} {\bibfnamefont {M.}~\bibnamefont
  {Blanco-Rey}}, \bibinfo {author} {\bibfnamefont {M.}~\bibnamefont
  {Hoffmann}}, \bibinfo {author} {\bibfnamefont {A.~Y.}\ \bibnamefont
  {Rusinov}, \bibfnamefont {I.~P.~Vyazovskaya}}, \bibinfo {author}
  {\bibfnamefont {S.~V.}\ \bibnamefont {Eremeev}}, \bibinfo {author}
  {\bibfnamefont {Y.~M.}\ \bibnamefont {Koroteev}}, \bibinfo {author}
  {\bibfnamefont {V.~M.}\ \bibnamefont {Kuznetsov}}, \bibinfo {author}
  {\bibfnamefont {F.}~\bibnamefont {Freyse}}, \bibinfo {author} {\bibfnamefont
  {J.}~\bibnamefont {S{\'{a}}nchez-Barriga}}, \bibinfo {author} {\bibfnamefont
  {I.~R.}\ \bibnamefont {Amiraslanov}}, \bibinfo {author} {\bibfnamefont
  {M.~B.}\ \bibnamefont {Babanly}}, \bibinfo {author} {\bibfnamefont {N.~T.}\
  \bibnamefont {Mamedov}}, \bibinfo {author} {\bibfnamefont {N.~A.}\
  \bibnamefont {Abdullayev}}, \bibinfo {author} {\bibfnamefont {V.~N.}\
  \bibnamefont {Zverev}}, \bibinfo {author} {\bibfnamefont {A.}~\bibnamefont
  {Alfonsov}}, \bibinfo {author} {\bibfnamefont {V.}~\bibnamefont {Kataev}},
  \bibinfo {author} {\bibfnamefont {B.}~\bibnamefont {B\"uchner}}, \bibinfo
  {author} {\bibfnamefont {E.~F.}\ \bibnamefont {Schwier}}, \bibinfo {author}
  {\bibfnamefont {S.}~\bibnamefont {Kumar}}, \bibinfo {author} {\bibfnamefont
  {A.}~\bibnamefont {Kimura}}, \bibinfo {author} {\bibfnamefont
  {L.}~\bibnamefont {Petaccia}}, \bibinfo {author} {\bibfnamefont
  {G.}~\bibnamefont {Di~Santo}}, \bibinfo {author} {\bibfnamefont {R.~C.}\
  \bibnamefont {Vidal}}, \bibinfo {author} {\bibfnamefont {S.}~\bibnamefont
  {Schatz}}, \bibinfo {author} {\bibfnamefont {K.}~\bibnamefont {Ki{\ss}ner}},
  \bibinfo {author} {\bibfnamefont {M.}~\bibnamefont {{\"{U}}nzelmann}},
  \bibinfo {author} {\bibfnamefont {C.~H.}\ \bibnamefont {Min}}, \bibinfo
  {author} {\bibfnamefont {S.}~\bibnamefont {Moser}}, \bibinfo {author}
  {\bibfnamefont {T.~R.~F.}\ \bibnamefont {Peixoto}}, \bibinfo {author}
  {\bibfnamefont {F.}~\bibnamefont {Reinert}}, \bibinfo {author} {\bibfnamefont
  {A.}~\bibnamefont {Ernst}}, \bibinfo {author} {\bibfnamefont {P.~M.}\
  \bibnamefont {Echenique}}, \bibinfo {author} {\bibfnamefont {A.}~\bibnamefont
  {Isaeva}}, \ and\ \bibinfo {author} {\bibfnamefont {E.~V.}\ \bibnamefont
  {Chulkov}},\ }\href {\doibase 10.1038/s41586-019-1840-9} {\bibfield
  {journal} {\bibinfo  {journal} {Nature}\ }\textbf {\bibinfo {volume} {576}},\
  \bibinfo {pages} {416} (\bibinfo {year} {2019})}\BibitemShut {NoStop}%
\bibitem [{\citenamefont {Ye}\ \emph {et~al.}(2015)\citenamefont {Ye},
  \citenamefont {Li}, \citenamefont {Zhu}, \citenamefont {Takeda},
  \citenamefont {Saitoh}, \citenamefont {Wang}, \citenamefont {Pan},
  \citenamefont {Nurmamat}, \citenamefont {Sumida}, \citenamefont {Ji},
  \citenamefont {Liu}, \citenamefont {Yang}, \citenamefont {Liu}, \citenamefont
  {Shen}, \citenamefont {Kimura}, \citenamefont {Qiao},\ and\ \citenamefont
  {Xie}}]{Ye_NComms_2015}%
  \BibitemOpen
  \bibfield  {author} {\bibinfo {author} {\bibfnamefont {M.}~\bibnamefont
  {Ye}}, \bibinfo {author} {\bibfnamefont {W.}~\bibnamefont {Li}}, \bibinfo
  {author} {\bibfnamefont {S.}~\bibnamefont {Zhu}}, \bibinfo {author}
  {\bibfnamefont {Y.}~\bibnamefont {Takeda}}, \bibinfo {author} {\bibfnamefont
  {Y.}~\bibnamefont {Saitoh}}, \bibinfo {author} {\bibfnamefont
  {J.}~\bibnamefont {Wang}}, \bibinfo {author} {\bibfnamefont {H.}~\bibnamefont
  {Pan}}, \bibinfo {author} {\bibfnamefont {M.}~\bibnamefont {Nurmamat}},
  \bibinfo {author} {\bibfnamefont {K.}~\bibnamefont {Sumida}}, \bibinfo
  {author} {\bibfnamefont {F.}~\bibnamefont {Ji}}, \bibinfo {author}
  {\bibfnamefont {Z.}~\bibnamefont {Liu}}, \bibinfo {author} {\bibfnamefont
  {H.}~\bibnamefont {Yang}}, \bibinfo {author} {\bibfnamefont {Z.}~\bibnamefont
  {Liu}}, \bibinfo {author} {\bibfnamefont {D.}~\bibnamefont {Shen}}, \bibinfo
  {author} {\bibfnamefont {A.}~\bibnamefont {Kimura}}, \bibinfo {author}
  {\bibfnamefont {S.}~\bibnamefont {Qiao}}, \ and\ \bibinfo {author}
  {\bibfnamefont {X.}~\bibnamefont {Xie}},\ }\href {\doibase
  10.1038/ncomms9913} {\bibfield  {journal} {\bibinfo  {journal} {Nat.
  Commun.}\ }\textbf {\bibinfo {volume} {6}},\ \bibinfo {pages} {8913}
  (\bibinfo {year} {2015})}\BibitemShut {NoStop}%
\bibitem [{\citenamefont {Steinke}\ \emph {et~al.}(2020)\citenamefont
  {Steinke}, \citenamefont {Zhang}, \citenamefont {Duffy}, \citenamefont
  {Kronast}, \citenamefont {Salman}, \citenamefont {Prokscha}, \citenamefont
  {Suter}, \citenamefont {Langridge}, \citenamefont {van~der Laan},\ and\
  \citenamefont {Hesjedal}}]{crpuddles}%
  \BibitemOpen
  \bibfield  {author} {\bibinfo {author} {\bibfnamefont {N.-J.}\ \bibnamefont
  {Steinke}}, \bibinfo {author} {\bibfnamefont {S.~L.}\ \bibnamefont {Zhang}},
  \bibinfo {author} {\bibfnamefont {L.~B.}\ \bibnamefont {Duffy}}, \bibinfo
  {author} {\bibfnamefont {F.}~\bibnamefont {Kronast}}, \bibinfo {author}
  {\bibfnamefont {Z.}~\bibnamefont {Salman}}, \bibinfo {author} {\bibfnamefont
  {T.}~\bibnamefont {Prokscha}}, \bibinfo {author} {\bibfnamefont
  {A.}~\bibnamefont {Suter}}, \bibinfo {author} {\bibfnamefont
  {S.}~\bibnamefont {Langridge}}, \bibinfo {author} {\bibfnamefont
  {G.}~\bibnamefont {van~der Laan}}, \ and\ \bibinfo {author} {\bibfnamefont
  {T.}~\bibnamefont {Hesjedal}},\ }\href@noop {} {\bibfield  {journal}
  {\bibinfo  {journal} {to be submitted}\ } (\bibinfo {year}
  {2020})}\BibitemShut {NoStop}%
\bibitem [{\citenamefont {Lachman}\ \emph {et~al.}(2015)\citenamefont
  {Lachman}, \citenamefont {Young}, \citenamefont {Richardella}, \citenamefont
  {Cuppens}, \citenamefont {Naren}, \citenamefont {Anahory}, \citenamefont
  {Meltzer}, \citenamefont {Kandala}, \citenamefont {Kempinger}, \citenamefont
  {Myasoedov}, \citenamefont {Huber}, \citenamefont {Samarth},\ and\
  \citenamefont {Zeldov}}]{Lachman2015}%
  \BibitemOpen
  \bibfield  {author} {\bibinfo {author} {\bibfnamefont {E.~O.}\ \bibnamefont
  {Lachman}}, \bibinfo {author} {\bibfnamefont {A.~F.}\ \bibnamefont {Young}},
  \bibinfo {author} {\bibfnamefont {A.}~\bibnamefont {Richardella}}, \bibinfo
  {author} {\bibfnamefont {J.}~\bibnamefont {Cuppens}}, \bibinfo {author}
  {\bibfnamefont {H.~R.}\ \bibnamefont {Naren}}, \bibinfo {author}
  {\bibfnamefont {Y.}~\bibnamefont {Anahory}}, \bibinfo {author} {\bibfnamefont
  {A.~Y.}\ \bibnamefont {Meltzer}}, \bibinfo {author} {\bibfnamefont
  {A.}~\bibnamefont {Kandala}}, \bibinfo {author} {\bibfnamefont
  {S.}~\bibnamefont {Kempinger}}, \bibinfo {author} {\bibfnamefont
  {Y.}~\bibnamefont {Myasoedov}}, \bibinfo {author} {\bibfnamefont {M.~E.}\
  \bibnamefont {Huber}}, \bibinfo {author} {\bibfnamefont {N.}~\bibnamefont
  {Samarth}}, \ and\ \bibinfo {author} {\bibfnamefont {E.}~\bibnamefont
  {Zeldov}},\ }\href {\doibase 10.1126/sciadv.1500740} {\bibfield  {journal}
  {\bibinfo  {journal} {Sci. Adv.}\ }\textbf {\bibinfo {volume} {1}},\ \bibinfo
  {pages} {e1500740} (\bibinfo {year} {2015})}\BibitemShut {NoStop}%
\bibitem [{\citenamefont {Liu}\ \emph {et~al.}(2019)\citenamefont {Liu},
  \citenamefont {Xu}, \citenamefont {He}, \citenamefont {van~der Laan},
  \citenamefont {Zhang},\ and\ \citenamefont {Wang}}]{Liu2019}%
  \BibitemOpen
  \bibfield  {author} {\bibinfo {author} {\bibfnamefont {W.}~\bibnamefont
  {Liu}}, \bibinfo {author} {\bibfnamefont {Y.}~\bibnamefont {Xu}}, \bibinfo
  {author} {\bibfnamefont {L.}~\bibnamefont {He}}, \bibinfo {author}
  {\bibfnamefont {G.}~\bibnamefont {van~der Laan}}, \bibinfo {author}
  {\bibfnamefont {R.}~\bibnamefont {Zhang}}, \ and\ \bibinfo {author}
  {\bibfnamefont {K.}~\bibnamefont {Wang}},\ }\href {\doibase
  10.1126/sciadv.aav2088} {\bibfield  {journal} {\bibinfo  {journal} {Sci.
  Adv.}\ }\textbf {\bibinfo {volume} {5}},\ \bibinfo {pages} {eaav2088}
  (\bibinfo {year} {2019})}\BibitemShut {NoStop}%
\bibitem [{\citenamefont {{Collins-McIntyre}}\ \emph
  {et~al.}(2016)\citenamefont {{Collins-McIntyre}}, \citenamefont {Duffy},
  \citenamefont {Singh}, \citenamefont {Steinke}, \citenamefont {Kinane},
  \citenamefont {Charlton}, \citenamefont {Pushp}, \citenamefont {Kellock},
  \citenamefont {Parkin}, \citenamefont {Holmes}, \citenamefont {Barnes},
  \citenamefont {van~der Laan}, \citenamefont {Langridge},\ and\ \citenamefont
  {Hesjedal}}]{collins2016structural}%
  \BibitemOpen
  \bibfield  {author} {\bibinfo {author} {\bibfnamefont {L.~J.}\ \bibnamefont
  {{Collins-McIntyre}}}, \bibinfo {author} {\bibfnamefont {L.~B.}\ \bibnamefont
  {Duffy}}, \bibinfo {author} {\bibfnamefont {A.}~\bibnamefont {Singh}},
  \bibinfo {author} {\bibfnamefont {N.-J.}\ \bibnamefont {Steinke}}, \bibinfo
  {author} {\bibfnamefont {C.~J.}\ \bibnamefont {Kinane}}, \bibinfo {author}
  {\bibfnamefont {T.~R.}\ \bibnamefont {Charlton}}, \bibinfo {author}
  {\bibfnamefont {A.}~\bibnamefont {Pushp}}, \bibinfo {author} {\bibfnamefont
  {A.~J.}\ \bibnamefont {Kellock}}, \bibinfo {author} {\bibfnamefont
  {S.~S.~P.}\ \bibnamefont {Parkin}}, \bibinfo {author} {\bibfnamefont {S.~N.}\
  \bibnamefont {Holmes}}, \bibinfo {author} {\bibfnamefont {C.~H.~W.}\
  \bibnamefont {Barnes}}, \bibinfo {author} {\bibfnamefont {G.}~\bibnamefont
  {van~der Laan}}, \bibinfo {author} {\bibfnamefont {S.}~\bibnamefont
  {Langridge}}, \ and\ \bibinfo {author} {\bibfnamefont {T.}~\bibnamefont
  {Hesjedal}},\ }\href@noop {} {\bibfield  {journal} {\bibinfo  {journal} {EPL
  (Europhys. Lett.)}\ }\textbf {\bibinfo {volume} {115}},\ \bibinfo {pages}
  {27006} (\bibinfo {year} {2016})}\BibitemShut {NoStop}%
\bibitem [{\citenamefont {Tcakaev}\ \emph {et~al.}(2020)\citenamefont
  {Tcakaev}, \citenamefont {Zabolotnyy}, \citenamefont {Green}, \citenamefont
  {Peixoto}, \citenamefont {Stier}, \citenamefont {Dettbarn}, \citenamefont
  {Schreyeck}, \citenamefont {Winnerlein}, \citenamefont {Vidal}, \citenamefont
  {Schatz}, \citenamefont {Vasili}, \citenamefont {Valvidares}, \citenamefont
  {Brunner}, \citenamefont {Gould}, \citenamefont {Bentmann}, \citenamefont
  {Reinert}, \citenamefont {Molenkamp},\ and\ \citenamefont
  {Hinkov}}]{Tcakaev2020}%
  \BibitemOpen
  \bibfield  {author} {\bibinfo {author} {\bibfnamefont {A.}~\bibnamefont
  {Tcakaev}}, \bibinfo {author} {\bibfnamefont {V.~B.}\ \bibnamefont
  {Zabolotnyy}}, \bibinfo {author} {\bibfnamefont {R.~J.}\ \bibnamefont
  {Green}}, \bibinfo {author} {\bibfnamefont {T.~R.}\ \bibnamefont {Peixoto}},
  \bibinfo {author} {\bibfnamefont {F.}~\bibnamefont {Stier}}, \bibinfo
  {author} {\bibfnamefont {M.}~\bibnamefont {Dettbarn}}, \bibinfo {author}
  {\bibfnamefont {S.}~\bibnamefont {Schreyeck}}, \bibinfo {author}
  {\bibfnamefont {M.}~\bibnamefont {Winnerlein}}, \bibinfo {author}
  {\bibfnamefont {R.~C.}\ \bibnamefont {Vidal}}, \bibinfo {author}
  {\bibfnamefont {S.}~\bibnamefont {Schatz}}, \bibinfo {author} {\bibfnamefont
  {H.~B.}\ \bibnamefont {Vasili}}, \bibinfo {author} {\bibfnamefont
  {M.}~\bibnamefont {Valvidares}}, \bibinfo {author} {\bibfnamefont
  {K.}~\bibnamefont {Brunner}}, \bibinfo {author} {\bibfnamefont
  {C.}~\bibnamefont {Gould}}, \bibinfo {author} {\bibfnamefont
  {H.}~\bibnamefont {Bentmann}}, \bibinfo {author} {\bibfnamefont
  {F.}~\bibnamefont {Reinert}}, \bibinfo {author} {\bibfnamefont {L.~W.}\
  \bibnamefont {Molenkamp}}, \ and\ \bibinfo {author} {\bibfnamefont
  {V.}~\bibnamefont {Hinkov}},\ }\href {\doibase 10.1103/PhysRevB.101.045127}
  {\bibfield  {journal} {\bibinfo  {journal} {Phys. Rev. B}\ }\textbf {\bibinfo
  {volume} {101}},\ \bibinfo {pages} {45127} (\bibinfo {year}
  {2020})}\BibitemShut {NoStop}%
\bibitem [{\citenamefont {Zhang}\ \emph {et~al.}(2017)\citenamefont {Zhang},
  \citenamefont {Zhao}, \citenamefont {Zang}, \citenamefont {Yuan},
  \citenamefont {Jiang}, \citenamefont {Liao}, \citenamefont {Zhang},
  \citenamefont {He}, \citenamefont {Ma},\ and\ \citenamefont
  {Xue}}]{Zhang2017}%
  \BibitemOpen
  \bibfield  {author} {\bibinfo {author} {\bibfnamefont {L.}~\bibnamefont
  {Zhang}}, \bibinfo {author} {\bibfnamefont {D.}~\bibnamefont {Zhao}},
  \bibinfo {author} {\bibfnamefont {Y.}~\bibnamefont {Zang}}, \bibinfo {author}
  {\bibfnamefont {Y.}~\bibnamefont {Yuan}}, \bibinfo {author} {\bibfnamefont
  {G.}~\bibnamefont {Jiang}}, \bibinfo {author} {\bibfnamefont
  {M.}~\bibnamefont {Liao}}, \bibinfo {author} {\bibfnamefont {D.}~\bibnamefont
  {Zhang}}, \bibinfo {author} {\bibfnamefont {K.}~\bibnamefont {He}}, \bibinfo
  {author} {\bibfnamefont {X.}~\bibnamefont {Ma}}, \ and\ \bibinfo {author}
  {\bibfnamefont {Q.}~\bibnamefont {Xue}},\ }\href {\doibase 10.1063/1.4990548}
  {\bibfield  {journal} {\bibinfo  {journal} {APL Mater.}\ }\textbf {\bibinfo
  {volume} {5}},\ \bibinfo {pages} {076106} (\bibinfo {year}
  {2017})}\BibitemShut {NoStop}%
\bibitem [{\citenamefont {Lee}\ \emph {et~al.}(2014)\citenamefont {Lee},
  \citenamefont {Richardella}, \citenamefont {Rench}, \citenamefont {Fraleigh},
  \citenamefont {Flanagan}, \citenamefont {Borchers}, \citenamefont {Tao},\
  and\ \citenamefont {Samarth}}]{Lee_PRB2014}%
  \BibitemOpen
  \bibfield  {author} {\bibinfo {author} {\bibfnamefont {J.~S.}\ \bibnamefont
  {Lee}}, \bibinfo {author} {\bibfnamefont {A.}~\bibnamefont {Richardella}},
  \bibinfo {author} {\bibfnamefont {D.~W.}\ \bibnamefont {Rench}}, \bibinfo
  {author} {\bibfnamefont {R.~D.}\ \bibnamefont {Fraleigh}}, \bibinfo {author}
  {\bibfnamefont {T.~C.}\ \bibnamefont {Flanagan}}, \bibinfo {author}
  {\bibfnamefont {J.~A.}\ \bibnamefont {Borchers}}, \bibinfo {author}
  {\bibfnamefont {J.}~\bibnamefont {Tao}}, \ and\ \bibinfo {author}
  {\bibfnamefont {N.}~\bibnamefont {Samarth}},\ }\href {\doibase
  10.1103/PhysRevB.89.174425} {\bibfield  {journal} {\bibinfo  {journal} {Phys.
  Rev. B}\ }\textbf {\bibinfo {volume} {89}},\ \bibinfo {pages} {174425}
  (\bibinfo {year} {2014})}\BibitemShut {NoStop}%
\bibitem [{\citenamefont {Figueroa}\ \emph {et~al.}(2015)\citenamefont
  {Figueroa}, \citenamefont {van~der Laan}, \citenamefont {Collins-McIntyre},
  \citenamefont {Cibin}, \citenamefont {Dent},\ and\ \citenamefont
  {Hesjedal}}]{Figueroa:2015aa}%
  \BibitemOpen
  \bibfield  {author} {\bibinfo {author} {\bibfnamefont {A.~I.}\ \bibnamefont
  {Figueroa}}, \bibinfo {author} {\bibfnamefont {G.}~\bibnamefont {van~der
  Laan}}, \bibinfo {author} {\bibfnamefont {L.~J.}\ \bibnamefont
  {Collins-McIntyre}}, \bibinfo {author} {\bibfnamefont {G.}~\bibnamefont
  {Cibin}}, \bibinfo {author} {\bibfnamefont {A.~J.}\ \bibnamefont {Dent}}, \
  and\ \bibinfo {author} {\bibfnamefont {T.}~\bibnamefont {Hesjedal}},\ }\href
  {\doibase 10.1021/jp511713s} {\bibfield  {journal} {\bibinfo  {journal} {J.
  Phys. Chem. C}\ }\textbf {\bibinfo {volume} {119}},\ \bibinfo {pages} {17344}
  (\bibinfo {year} {2015})}\BibitemShut {NoStop}%
\bibitem [{\citenamefont {R{\r{u}}{\v{z}}i{\v{c}}ka}\ \emph
  {et~al.}(2015)\citenamefont {R{\r{u}}{\v{z}}i{\v{c}}ka}, \citenamefont
  {Caha}, \citenamefont {Hol{\'{y}}}, \citenamefont {Steiner}, \citenamefont
  {Volobuiev}, \citenamefont {Ney}, \citenamefont {Bauer}, \citenamefont
  {Ducho{\v{n}}}, \citenamefont {Veltrusk{\'{a}}}, \citenamefont {Khalakhan},
  \citenamefont {Matol{\'{\i}}n}, \citenamefont {Schwier}, \citenamefont
  {Iwasawa}, \citenamefont {Shimada},\ and\ \citenamefont
  {Springholz}}]{Ruzika2015}%
  \BibitemOpen
  \bibfield  {author} {\bibinfo {author} {\bibfnamefont {J.}~\bibnamefont
  {R{\r{u}}{\v{z}}i{\v{c}}ka}}, \bibinfo {author} {\bibfnamefont
  {O.}~\bibnamefont {Caha}}, \bibinfo {author} {\bibfnamefont {V.}~\bibnamefont
  {Hol{\'{y}}}}, \bibinfo {author} {\bibfnamefont {H.}~\bibnamefont {Steiner}},
  \bibinfo {author} {\bibfnamefont {V.}~\bibnamefont {Volobuiev}}, \bibinfo
  {author} {\bibfnamefont {A.}~\bibnamefont {Ney}}, \bibinfo {author}
  {\bibfnamefont {G.}~\bibnamefont {Bauer}}, \bibinfo {author} {\bibfnamefont
  {T.}~\bibnamefont {Ducho{\v{n}}}}, \bibinfo {author} {\bibfnamefont
  {K.}~\bibnamefont {Veltrusk{\'{a}}}}, \bibinfo {author} {\bibfnamefont
  {I.}~\bibnamefont {Khalakhan}}, \bibinfo {author} {\bibfnamefont
  {V.}~\bibnamefont {Matol{\'{\i}}n}}, \bibinfo {author} {\bibfnamefont
  {E.~F.}\ \bibnamefont {Schwier}}, \bibinfo {author} {\bibfnamefont
  {H.}~\bibnamefont {Iwasawa}}, \bibinfo {author} {\bibfnamefont
  {K.}~\bibnamefont {Shimada}}, \ and\ \bibinfo {author} {\bibfnamefont
  {G.}~\bibnamefont {Springholz}},\ }\href {\doibase
  10.1088/1367-2630/17/1/013028} {\bibfield  {journal} {\bibinfo  {journal}
  {New J. Phys.}\ }\textbf {\bibinfo {volume} {17}},\ \bibinfo {pages} {013028}
  (\bibinfo {year} {2015})}\BibitemShut {NoStop}%
\bibitem [{\citenamefont {Figueroa}\ \emph
  {et~al.}(2016{\natexlab{a}})\citenamefont {Figueroa}, \citenamefont {{van der
  Laan}}, \citenamefont {Harrison}, \citenamefont {Cibin},\ and\ \citenamefont
  {Hesjedal}}]{Figueroa-oxEXAFS}%
  \BibitemOpen
  \bibfield  {author} {\bibinfo {author} {\bibfnamefont {A.~I.}\ \bibnamefont
  {Figueroa}}, \bibinfo {author} {\bibfnamefont {G.}~\bibnamefont {{van der
  Laan}}}, \bibinfo {author} {\bibfnamefont {S.~E.}\ \bibnamefont {Harrison}},
  \bibinfo {author} {\bibfnamefont {G.}~\bibnamefont {Cibin}}, \ and\ \bibinfo
  {author} {\bibfnamefont {T.}~\bibnamefont {Hesjedal}},\ }\href {\doibase
  10.1038/srep22935} {\bibfield  {journal} {\bibinfo  {journal} {Sci. Rep.}\
  }\textbf {\bibinfo {volume} {6}},\ \bibinfo {pages} {22935} (\bibinfo {year}
  {2016}{\natexlab{a}})}\BibitemShut {NoStop}%
\bibitem [{\citenamefont {Li}\ \emph {et~al.}(2010{\natexlab{b}})\citenamefont
  {Li}, \citenamefont {Wang}, \citenamefont {Kan}, \citenamefont {Guo},
  \citenamefont {He}, \citenamefont {Wang}, \citenamefont {Wang}, \citenamefont
  {Wong}, \citenamefont {Wang},\ and\ \citenamefont {Xie}}]{Li2010-MBE}%
  \BibitemOpen
  \bibfield  {author} {\bibinfo {author} {\bibfnamefont {H.~D.}\ \bibnamefont
  {Li}}, \bibinfo {author} {\bibfnamefont {Z.~Y.}\ \bibnamefont {Wang}},
  \bibinfo {author} {\bibfnamefont {X.}~\bibnamefont {Kan}}, \bibinfo {author}
  {\bibfnamefont {X.}~\bibnamefont {Guo}}, \bibinfo {author} {\bibfnamefont
  {H.~T.}\ \bibnamefont {He}}, \bibinfo {author} {\bibfnamefont
  {Z.}~\bibnamefont {Wang}}, \bibinfo {author} {\bibfnamefont {J.~N.}\
  \bibnamefont {Wang}}, \bibinfo {author} {\bibfnamefont {T.~L.}\ \bibnamefont
  {Wong}}, \bibinfo {author} {\bibfnamefont {N.}~\bibnamefont {Wang}}, \ and\
  \bibinfo {author} {\bibfnamefont {M.~H.}\ \bibnamefont {Xie}},\ }\href@noop
  {} {\bibfield  {journal} {\bibinfo  {journal} {{New J. Phys.}}\ }\textbf
  {\bibinfo {volume} {{12}}},\ \bibinfo {pages} {{103038}} (\bibinfo {year}
  {{2010}}{\natexlab{b}})}\BibitemShut {NoStop}%
\bibitem [{\citenamefont {Fornari}\ \emph {et~al.}(2020)\citenamefont
  {Fornari}, \citenamefont {Bentmann}, \citenamefont {Morelhao}, \citenamefont
  {Peixoto}, \citenamefont {Rappl}, \citenamefont {Tcakaev}, \citenamefont
  {Zabolotnyy}, \citenamefont {Kamp}, \citenamefont {Lee}, \citenamefont {Min},
  \citenamefont {Kagerer}, \citenamefont {Vidal}, \citenamefont {Isaeva},
  \citenamefont {Ruck}, \citenamefont {Hinkov}, \citenamefont {Reinert},\ and\
  \citenamefont {Abramof}}]{Fornari2020}%
  \BibitemOpen
  \bibfield  {author} {\bibinfo {author} {\bibfnamefont {C.~I.}\ \bibnamefont
  {Fornari}}, \bibinfo {author} {\bibfnamefont {H.}~\bibnamefont {Bentmann}},
  \bibinfo {author} {\bibfnamefont {S.~L.}\ \bibnamefont {Morelhao}}, \bibinfo
  {author} {\bibfnamefont {T.~R.~F.}\ \bibnamefont {Peixoto}}, \bibinfo
  {author} {\bibfnamefont {P.~H.~O.}\ \bibnamefont {Rappl}}, \bibinfo {author}
  {\bibfnamefont {A.-V.}\ \bibnamefont {Tcakaev}}, \bibinfo {author}
  {\bibfnamefont {V.}~\bibnamefont {Zabolotnyy}}, \bibinfo {author}
  {\bibfnamefont {M.}~\bibnamefont {Kamp}}, \bibinfo {author} {\bibfnamefont
  {T.-L.}\ \bibnamefont {Lee}}, \bibinfo {author} {\bibfnamefont {C.-H.}\
  \bibnamefont {Min}}, \bibinfo {author} {\bibfnamefont {P.}~\bibnamefont
  {Kagerer}}, \bibinfo {author} {\bibfnamefont {R.~C.}\ \bibnamefont {Vidal}},
  \bibinfo {author} {\bibfnamefont {A.}~\bibnamefont {Isaeva}}, \bibinfo
  {author} {\bibfnamefont {M.}~\bibnamefont {Ruck}}, \bibinfo {author}
  {\bibfnamefont {V.}~\bibnamefont {Hinkov}}, \bibinfo {author} {\bibfnamefont
  {F.}~\bibnamefont {Reinert}}, \ and\ \bibinfo {author} {\bibfnamefont
  {E.}~\bibnamefont {Abramof}},\ }\href {\doibase 10.1021/acs.jpcc.0c05077}
  {\bibfield  {journal} {\bibinfo  {journal} {J. Phys. Chem. C}\ }\textbf
  {\bibinfo {volume} {124}},\ \bibinfo {pages} {16048} (\bibinfo {year}
  {2020})}\BibitemShut {NoStop}%
\bibitem [{\citenamefont {Aronzon}\ \emph {et~al.}(2018)\citenamefont
  {Aronzon}, \citenamefont {Oveshnikov}, \citenamefont {Prudkoglyad},
  \citenamefont {Selivanov}, \citenamefont {Chizhevskii}, \citenamefont
  {Kugel}, \citenamefont {Karateev}, \citenamefont {Vasiliev},\ and\
  \citenamefont {L\"ahderanta}}]{Aronzon2018}%
  \BibitemOpen
  \bibfield  {author} {\bibinfo {author} {\bibfnamefont {B.~A.}\ \bibnamefont
  {Aronzon}}, \bibinfo {author} {\bibfnamefont {L.~N.}\ \bibnamefont
  {Oveshnikov}}, \bibinfo {author} {\bibfnamefont {V.~A.}\ \bibnamefont
  {Prudkoglyad}}, \bibinfo {author} {\bibfnamefont {Y.~G.}\ \bibnamefont
  {Selivanov}}, \bibinfo {author} {\bibfnamefont {E.~G.}\ \bibnamefont
  {Chizhevskii}}, \bibinfo {author} {\bibfnamefont {K.~I.}\ \bibnamefont
  {Kugel}}, \bibinfo {author} {\bibfnamefont {I.~A.}\ \bibnamefont {Karateev}},
  \bibinfo {author} {\bibfnamefont {A.~L.}\ \bibnamefont {Vasiliev}}, \ and\
  \bibinfo {author} {\bibfnamefont {E.}~\bibnamefont {L\"ahderanta}},\ }\href
  {\doibase 10.1016/j.jmmm.2017.09.058} {\bibfield  {journal} {\bibinfo
  {journal} {J. Magn. Magn.Mater.}\ }\textbf {\bibinfo {volume} {459}},\
  \bibinfo {pages} {331} (\bibinfo {year} {2018})}\BibitemShut {NoStop}%
\bibitem [{\citenamefont {Yu}\ \emph {et~al.}(2010)\citenamefont {Yu},
  \citenamefont {Zhang}, \citenamefont {Zhang}, \citenamefont {Zhang},
  \citenamefont {Dai},\ and\ \citenamefont {Fang}}]{Yu2010}%
  \BibitemOpen
  \bibfield  {author} {\bibinfo {author} {\bibfnamefont {R.}~\bibnamefont
  {Yu}}, \bibinfo {author} {\bibfnamefont {W.}~\bibnamefont {Zhang}}, \bibinfo
  {author} {\bibfnamefont {H.-J.}\ \bibnamefont {Zhang}}, \bibinfo {author}
  {\bibfnamefont {S.-C.}\ \bibnamefont {Zhang}}, \bibinfo {author}
  {\bibfnamefont {X.}~\bibnamefont {Dai}}, \ and\ \bibinfo {author}
  {\bibfnamefont {Z.}~\bibnamefont {Fang}},\ }\href@noop {} {\bibfield
  {journal} {\bibinfo  {journal} {{Science}}\ }\textbf {\bibinfo {volume}
  {{329}}},\ \bibinfo {pages} {{61}} (\bibinfo {year} {{2010}})}\BibitemShut
  {NoStop}%
\bibitem [{\citenamefont {Li}\ \emph {et~al.}(2015)\citenamefont {Li},
  \citenamefont {Chang}, \citenamefont {Wu}, \citenamefont {Tao}, \citenamefont
  {Zhao}, \citenamefont {Chan}, \citenamefont {Moodera}, \citenamefont {Li},\
  and\ \citenamefont {Zhu}}]{Li2015}%
  \BibitemOpen
  \bibfield  {author} {\bibinfo {author} {\bibfnamefont {M.}~\bibnamefont
  {Li}}, \bibinfo {author} {\bibfnamefont {C.~Z.}\ \bibnamefont {Chang}},
  \bibinfo {author} {\bibfnamefont {L.}~\bibnamefont {Wu}}, \bibinfo {author}
  {\bibfnamefont {J.}~\bibnamefont {Tao}}, \bibinfo {author} {\bibfnamefont
  {W.}~\bibnamefont {Zhao}}, \bibinfo {author} {\bibfnamefont {M.~H.}\
  \bibnamefont {Chan}}, \bibinfo {author} {\bibfnamefont {J.~S.}\ \bibnamefont
  {Moodera}}, \bibinfo {author} {\bibfnamefont {J.}~\bibnamefont {Li}}, \ and\
  \bibinfo {author} {\bibfnamefont {Y.}~\bibnamefont {Zhu}},\ }\href@noop {}
  {\bibfield  {journal} {\bibinfo  {journal} {Phys. Rev. Lett.}\ }\textbf
  {\bibinfo {volume} {114}},\ \bibinfo {pages} {146802} (\bibinfo {year}
  {2015})}\BibitemShut {NoStop}%
\bibitem [{\citenamefont {{Van Vleck}}(1953)}]{VanVleck1953}%
  \BibitemOpen
  \bibfield  {author} {\bibinfo {author} {\bibfnamefont {J.~H.}\ \bibnamefont
  {{Van Vleck}}},\ }\href {\doibase 10.1103/RevModPhys.25.220} {\bibfield
  {journal} {\bibinfo  {journal} {Rev. Mod. Phys.}\ }\textbf {\bibinfo {volume}
  {25}},\ \bibinfo {pages} {220} (\bibinfo {year} {1953})}\BibitemShut
  {NoStop}%
\bibitem [{\citenamefont {Islam}\ \emph {et~al.}(2018)\citenamefont {Islam},
  \citenamefont {Canali}, \citenamefont {Pertsova}, \citenamefont {Balatsky},
  \citenamefont {Mahatha}, \citenamefont {Carbone}, \citenamefont {Barla},
  \citenamefont {Kokh}, \citenamefont {Tereshchenko}, \citenamefont
  {Jim{\'{e}}nez}, \citenamefont {Brookes}, \citenamefont {Gargiani},
  \citenamefont {Valvidares}, \citenamefont {Schatz}, \citenamefont {Peixoto},
  \citenamefont {Bentmann}, \citenamefont {Reinert}, \citenamefont {Jung},
  \citenamefont {Bathon}, \citenamefont {Fauth}, \citenamefont {Bode},\ and\
  \citenamefont {Sessi}}]{Islam_PRB_2018}%
  \BibitemOpen
  \bibfield  {author} {\bibinfo {author} {\bibfnamefont {M.~F.}\ \bibnamefont
  {Islam}}, \bibinfo {author} {\bibfnamefont {C.~M.}\ \bibnamefont {Canali}},
  \bibinfo {author} {\bibfnamefont {A.}~\bibnamefont {Pertsova}}, \bibinfo
  {author} {\bibfnamefont {A.}~\bibnamefont {Balatsky}}, \bibinfo {author}
  {\bibfnamefont {S.~K.}\ \bibnamefont {Mahatha}}, \bibinfo {author}
  {\bibfnamefont {C.}~\bibnamefont {Carbone}}, \bibinfo {author} {\bibfnamefont
  {A.}~\bibnamefont {Barla}}, \bibinfo {author} {\bibfnamefont {K.~A.}\
  \bibnamefont {Kokh}}, \bibinfo {author} {\bibfnamefont {O.~E.}\ \bibnamefont
  {Tereshchenko}}, \bibinfo {author} {\bibfnamefont {E.}~\bibnamefont
  {Jim{\'{e}}nez}}, \bibinfo {author} {\bibfnamefont {N.~B.}\ \bibnamefont
  {Brookes}}, \bibinfo {author} {\bibfnamefont {P.}~\bibnamefont {Gargiani}},
  \bibinfo {author} {\bibfnamefont {M.}~\bibnamefont {Valvidares}}, \bibinfo
  {author} {\bibfnamefont {S.}~\bibnamefont {Schatz}}, \bibinfo {author}
  {\bibfnamefont {T.~R.}\ \bibnamefont {Peixoto}}, \bibinfo {author}
  {\bibfnamefont {H.}~\bibnamefont {Bentmann}}, \bibinfo {author}
  {\bibfnamefont {F.}~\bibnamefont {Reinert}}, \bibinfo {author} {\bibfnamefont
  {J.}~\bibnamefont {Jung}}, \bibinfo {author} {\bibfnamefont {T.}~\bibnamefont
  {Bathon}}, \bibinfo {author} {\bibfnamefont {K.}~\bibnamefont {Fauth}},
  \bibinfo {author} {\bibfnamefont {M.}~\bibnamefont {Bode}}, \ and\ \bibinfo
  {author} {\bibfnamefont {P.}~\bibnamefont {Sessi}},\ }\href {\doibase
  10.1103/PhysRevB.97.155429} {\bibfield  {journal} {\bibinfo  {journal} {Phys.
  Rev. B}\ }\textbf {\bibinfo {volume} {97}},\ \bibinfo {pages} {155429}
  (\bibinfo {year} {2018})},\ \Eprint {http://arxiv.org/abs/1802.05302}
  {1802.05302} \BibitemShut {NoStop}%
\bibitem [{\citenamefont {Vergniory}\ \emph {et~al.}(2014)\citenamefont
  {Vergniory}, \citenamefont {Otrokov}, \citenamefont {Thonig}, \citenamefont
  {Hoffmann}, \citenamefont {Maznichenko}, \citenamefont {Geilhufe},
  \citenamefont {Zubizarreta}, \citenamefont {Ostanin}, \citenamefont
  {Marmodoro}, \citenamefont {Henk}, \citenamefont {Hergert}, \citenamefont
  {Mertig}, \citenamefont {Chulkov},\ and\ \citenamefont
  {Ernst}}]{Vergniory2014}%
  \BibitemOpen
  \bibfield  {author} {\bibinfo {author} {\bibfnamefont {M.~G.}\ \bibnamefont
  {Vergniory}}, \bibinfo {author} {\bibfnamefont {M.~M.}\ \bibnamefont
  {Otrokov}}, \bibinfo {author} {\bibfnamefont {D.}~\bibnamefont {Thonig}},
  \bibinfo {author} {\bibfnamefont {M.}~\bibnamefont {Hoffmann}}, \bibinfo
  {author} {\bibfnamefont {I.~V.}\ \bibnamefont {Maznichenko}}, \bibinfo
  {author} {\bibfnamefont {M.}~\bibnamefont {Geilhufe}}, \bibinfo {author}
  {\bibfnamefont {X.}~\bibnamefont {Zubizarreta}}, \bibinfo {author}
  {\bibfnamefont {S.}~\bibnamefont {Ostanin}}, \bibinfo {author} {\bibfnamefont
  {A.}~\bibnamefont {Marmodoro}}, \bibinfo {author} {\bibfnamefont
  {J.}~\bibnamefont {Henk}}, \bibinfo {author} {\bibfnamefont {W.}~\bibnamefont
  {Hergert}}, \bibinfo {author} {\bibfnamefont {I.}~\bibnamefont {Mertig}},
  \bibinfo {author} {\bibfnamefont {E.~V.}\ \bibnamefont {Chulkov}}, \ and\
  \bibinfo {author} {\bibfnamefont {A.}~\bibnamefont {Ernst}},\ }\href
  {\doibase 10.1103/PhysRevB.89.165202} {\bibfield  {journal} {\bibinfo
  {journal} {Phys. Rev.}\ }\textbf {\bibinfo {volume} {89}},\ \bibinfo {pages}
  {165202} (\bibinfo {year} {2014})}\BibitemShut {NoStop}%
\bibitem [{\citenamefont {Ye}\ \emph {et~al.}(2019)\citenamefont {Ye},
  \citenamefont {Xu}, \citenamefont {Li}, \citenamefont {Qiao}, \citenamefont
  {Takeda}, \citenamefont {Saitoh}, \citenamefont {Zhu}, \citenamefont
  {Nurmamat}, \citenamefont {Sumida}, \citenamefont {Ishida}, \citenamefont
  {Shin},\ and\ \citenamefont {Kimura}}]{Ye2019}%
  \BibitemOpen
  \bibfield  {author} {\bibinfo {author} {\bibfnamefont {M.}~\bibnamefont
  {Ye}}, \bibinfo {author} {\bibfnamefont {T.}~\bibnamefont {Xu}}, \bibinfo
  {author} {\bibfnamefont {G.}~\bibnamefont {Li}}, \bibinfo {author}
  {\bibfnamefont {S.}~\bibnamefont {Qiao}}, \bibinfo {author} {\bibfnamefont
  {Y.}~\bibnamefont {Takeda}}, \bibinfo {author} {\bibfnamefont
  {Y.}~\bibnamefont {Saitoh}}, \bibinfo {author} {\bibfnamefont {S.~Y.}\
  \bibnamefont {Zhu}}, \bibinfo {author} {\bibfnamefont {M.}~\bibnamefont
  {Nurmamat}}, \bibinfo {author} {\bibfnamefont {K.}~\bibnamefont {Sumida}},
  \bibinfo {author} {\bibfnamefont {Y.}~\bibnamefont {Ishida}}, \bibinfo
  {author} {\bibfnamefont {S.}~\bibnamefont {Shin}}, \ and\ \bibinfo {author}
  {\bibfnamefont {A.}~\bibnamefont {Kimura}},\ }\href {\doibase
  10.1103/PhysRevB.99.144413} {\bibfield  {journal} {\bibinfo  {journal} {Phys.
  Rev. B}\ }\textbf {\bibinfo {volume} {99}},\ \bibinfo {pages} {144413}
  (\bibinfo {year} {2019})}\BibitemShut {NoStop}%
\bibitem [{\citenamefont {Dietl}\ \emph {et~al.}(2000)\citenamefont {Dietl},
  \citenamefont {Ohno}, \citenamefont {Matsukura}, \citenamefont {Cibert},\
  and\ \citenamefont {Ferrand}}]{Dietl2000}%
  \BibitemOpen
  \bibfield  {author} {\bibinfo {author} {\bibfnamefont {T.}~\bibnamefont
  {Dietl}}, \bibinfo {author} {\bibfnamefont {H.}~\bibnamefont {Ohno}},
  \bibinfo {author} {\bibfnamefont {F.}~\bibnamefont {Matsukura}}, \bibinfo
  {author} {\bibfnamefont {J.}~\bibnamefont {Cibert}}, \ and\ \bibinfo {author}
  {\bibfnamefont {D.}~\bibnamefont {Ferrand}},\ }\href {\doibase
  10.1126/science.287.5455.1019} {\bibfield  {journal} {\bibinfo  {journal}
  {Science}\ }\textbf {\bibinfo {volume} {287}},\ \bibinfo {pages} {1019}
  (\bibinfo {year} {2000})}\BibitemShut {NoStop}%
\bibitem [{\citenamefont {Storchak}\ \emph {et~al.}(2008)\citenamefont
  {Storchak}, \citenamefont {Eshchenko}, \citenamefont {Morenzoni},
  \citenamefont {Prokscha}, \citenamefont {Suter}, \citenamefont {Liu},\ and\
  \citenamefont {Furdyna}}]{Storchak2008}%
  \BibitemOpen
  \bibfield  {author} {\bibinfo {author} {\bibfnamefont {V.~G.}\ \bibnamefont
  {Storchak}}, \bibinfo {author} {\bibfnamefont {D.~G.}\ \bibnamefont
  {Eshchenko}}, \bibinfo {author} {\bibfnamefont {E.}~\bibnamefont
  {Morenzoni}}, \bibinfo {author} {\bibfnamefont {T.}~\bibnamefont {Prokscha}},
  \bibinfo {author} {\bibfnamefont {A.}~\bibnamefont {Suter}}, \bibinfo
  {author} {\bibfnamefont {X.}~\bibnamefont {Liu}}, \ and\ \bibinfo {author}
  {\bibfnamefont {J.~K.}\ \bibnamefont {Furdyna}},\ }\href {\doibase
  10.1103/PhysRevLett.101.027202} {\bibfield  {journal} {\bibinfo  {journal}
  {Phys. Rev. Lett.}\ }\textbf {\bibinfo {volume} {101}},\ \bibinfo {pages}
  {027202} (\bibinfo {year} {2008})}\BibitemShut {NoStop}%
\bibitem [{\citenamefont {Fang}\ \emph {et~al.}(2012)\citenamefont {Fang},
  \citenamefont {Jia}, \citenamefont {Miller}, \citenamefont {Latimer},
  \citenamefont {Xiao}, \citenamefont {Welp}, \citenamefont {Crabtree},\ and\
  \citenamefont {Kwok}}]{Fang2012}%
  \BibitemOpen
  \bibfield  {author} {\bibinfo {author} {\bibfnamefont {L.}~\bibnamefont
  {Fang}}, \bibinfo {author} {\bibfnamefont {Y.}~\bibnamefont {Jia}}, \bibinfo
  {author} {\bibfnamefont {D.~J.}\ \bibnamefont {Miller}}, \bibinfo {author}
  {\bibfnamefont {M.~L.}\ \bibnamefont {Latimer}}, \bibinfo {author}
  {\bibfnamefont {Z.~L.}\ \bibnamefont {Xiao}}, \bibinfo {author}
  {\bibfnamefont {U.}~\bibnamefont {Welp}}, \bibinfo {author} {\bibfnamefont
  {G.~W.}\ \bibnamefont {Crabtree}}, \ and\ \bibinfo {author} {\bibfnamefont
  {W.-K.}\ \bibnamefont {Kwok}},\ }\href@noop {} {\bibfield  {journal}
  {\bibinfo  {journal} {Nano Lett.}\ }\textbf {\bibinfo {volume} {12}},\
  \bibinfo {pages} {6164} (\bibinfo {year} {2012})}\BibitemShut {NoStop}%
\bibitem [{\citenamefont {Krieger}\ \emph {et~al.}(2019)\citenamefont
  {Krieger}, \citenamefont {Ou}, \citenamefont {Caputo}, \citenamefont
  {Chikina}, \citenamefont {D\"obeli}, \citenamefont {Husanu}, \citenamefont
  {Keren}, \citenamefont {Prokscha}, \citenamefont {Suter}, \citenamefont
  {Chang}, \citenamefont {Moodera}, \citenamefont {Strocov},\ and\
  \citenamefont {Salman}}]{Krieger_PRB2019}%
  \BibitemOpen
  \bibfield  {author} {\bibinfo {author} {\bibfnamefont {J.~A.}\ \bibnamefont
  {Krieger}}, \bibinfo {author} {\bibfnamefont {Y.}~\bibnamefont {Ou}},
  \bibinfo {author} {\bibfnamefont {M.}~\bibnamefont {Caputo}}, \bibinfo
  {author} {\bibfnamefont {A.}~\bibnamefont {Chikina}}, \bibinfo {author}
  {\bibfnamefont {M.}~\bibnamefont {D\"obeli}}, \bibinfo {author}
  {\bibfnamefont {M.-A.}\ \bibnamefont {Husanu}}, \bibinfo {author}
  {\bibfnamefont {I.}~\bibnamefont {Keren}}, \bibinfo {author} {\bibfnamefont
  {T.}~\bibnamefont {Prokscha}}, \bibinfo {author} {\bibfnamefont
  {A.}~\bibnamefont {Suter}}, \bibinfo {author} {\bibfnamefont {C.-Z.}\
  \bibnamefont {Chang}}, \bibinfo {author} {\bibfnamefont {J.~S.}\ \bibnamefont
  {Moodera}}, \bibinfo {author} {\bibfnamefont {V.~N.}\ \bibnamefont
  {Strocov}}, \ and\ \bibinfo {author} {\bibfnamefont {Z.}~\bibnamefont
  {Salman}},\ }\href {\doibase 10.1103/PhysRevB.99.064423} {\bibfield
  {journal} {\bibinfo  {journal} {Phys. Rev. B}\ }\textbf {\bibinfo {volume}
  {99}},\ \bibinfo {pages} {064423} (\bibinfo {year} {2019})}\BibitemShut
  {NoStop}%
\bibitem [{\citenamefont {Fujii}\ \emph {et~al.}(2011)\citenamefont {Fujii},
  \citenamefont {Sperl}, \citenamefont {Ueda}, \citenamefont {Kobayashi},
  \citenamefont {Yamashita}, \citenamefont {Kobata}, \citenamefont {Torelli},
  \citenamefont {Borgatti}, \citenamefont {Utz}, \citenamefont {Fadley},
  \citenamefont {Gray}, \citenamefont {Monaco}, \citenamefont {Back},
  \citenamefont {van~der Laan},\ and\ \citenamefont {Panaccione}}]{Fujii2011}%
  \BibitemOpen
  \bibfield  {author} {\bibinfo {author} {\bibfnamefont {J.}~\bibnamefont
  {Fujii}}, \bibinfo {author} {\bibfnamefont {M.}~\bibnamefont {Sperl}},
  \bibinfo {author} {\bibfnamefont {S.}~\bibnamefont {Ueda}}, \bibinfo {author}
  {\bibfnamefont {K.}~\bibnamefont {Kobayashi}}, \bibinfo {author}
  {\bibfnamefont {Y.}~\bibnamefont {Yamashita}}, \bibinfo {author}
  {\bibfnamefont {M.}~\bibnamefont {Kobata}}, \bibinfo {author} {\bibfnamefont
  {P.}~\bibnamefont {Torelli}}, \bibinfo {author} {\bibfnamefont
  {F.}~\bibnamefont {Borgatti}}, \bibinfo {author} {\bibfnamefont
  {M.}~\bibnamefont {Utz}}, \bibinfo {author} {\bibfnamefont {C.~S.}\
  \bibnamefont {Fadley}}, \bibinfo {author} {\bibfnamefont {A.~X.}\
  \bibnamefont {Gray}}, \bibinfo {author} {\bibfnamefont {G.}~\bibnamefont
  {Monaco}}, \bibinfo {author} {\bibfnamefont {C.~H.}\ \bibnamefont {Back}},
  \bibinfo {author} {\bibfnamefont {G.}~\bibnamefont {van~der Laan}}, \ and\
  \bibinfo {author} {\bibfnamefont {G.}~\bibnamefont {Panaccione}},\
  }\href@noop {} {\bibfield  {journal} {\bibinfo  {journal} {{Phys. Rev.
  Lett.}}\ }\textbf {\bibinfo {volume} {107}},\ \bibinfo {pages} {187203}
  (\bibinfo {year} {2011})}\BibitemShut {NoStop}%
\bibitem [{\citenamefont {Priour}\ and\ \citenamefont {{Das
  Sarma}}(2006)}]{Priour2006}%
  \BibitemOpen
  \bibfield  {author} {\bibinfo {author} {\bibfnamefont {D.~J.}\ \bibnamefont
  {Priour}}\ and\ \bibinfo {author} {\bibfnamefont {S.}~\bibnamefont {{Das
  Sarma}}},\ }\href {\doibase 10.1103/PhysRevB.73.165203} {\bibfield  {journal}
  {\bibinfo  {journal} {Phys. Rev. B}\ }\textbf {\bibinfo {volume} {73}},\
  \bibinfo {pages} {165203} (\bibinfo {year} {2006})}\BibitemShut {NoStop}%
\bibitem [{\citenamefont {Figueroa}\ \emph {et~al.}(2017)\citenamefont
  {Figueroa}, \citenamefont {Baker}, \citenamefont {Harrison}, \citenamefont
  {Kummer}, \citenamefont {{van der Laan}},\ and\ \citenamefont
  {Hesjedal}}]{Figueroa_2017}%
  \BibitemOpen
  \bibfield  {author} {\bibinfo {author} {\bibfnamefont {A.~I.}\ \bibnamefont
  {Figueroa}}, \bibinfo {author} {\bibfnamefont {A.~A.}\ \bibnamefont {Baker}},
  \bibinfo {author} {\bibfnamefont {S.}~\bibnamefont {Harrison}}, \bibinfo
  {author} {\bibfnamefont {K.}~\bibnamefont {Kummer}}, \bibinfo {author}
  {\bibfnamefont {G.}~\bibnamefont {{van der Laan}}}, \ and\ \bibinfo {author}
  {\bibfnamefont {T.}~\bibnamefont {Hesjedal}},\ }\href {\doibase
  10.1016/j.jmmm.2016.08.063} {\bibfield  {journal} {\bibinfo  {journal} {{J.
  Magn. Magn. Mater.}}\ }\textbf {\bibinfo {volume} {422}},\ \bibinfo {pages}
  {93} (\bibinfo {year} {2017})}\BibitemShut {NoStop}%
\bibitem [{\citenamefont {Duffy}\ \emph
  {et~al.}(2018{\natexlab{a}})\citenamefont {Duffy}, \citenamefont {Steinke},
  \citenamefont {Krieger}, \citenamefont {Figueroa}, \citenamefont {Kummer},
  \citenamefont {Lancaster}, \citenamefont {Giblin}, \citenamefont {Pratt},
  \citenamefont {Blundell}, \citenamefont {Prokscha}, \citenamefont {Suter},
  \citenamefont {Langridge}, \citenamefont {Strocov}, \citenamefont {Salman},
  \citenamefont {van~der Laan},\ and\ \citenamefont {Hesjedal}}]{Duffy2018_Dy}%
  \BibitemOpen
  \bibfield  {author} {\bibinfo {author} {\bibfnamefont {L.~B.}\ \bibnamefont
  {Duffy}}, \bibinfo {author} {\bibfnamefont {N.-J.}\ \bibnamefont {Steinke}},
  \bibinfo {author} {\bibfnamefont {J.~A.}\ \bibnamefont {Krieger}}, \bibinfo
  {author} {\bibfnamefont {A.~I.}\ \bibnamefont {Figueroa}}, \bibinfo {author}
  {\bibfnamefont {K.}~\bibnamefont {Kummer}}, \bibinfo {author} {\bibfnamefont
  {T.}~\bibnamefont {Lancaster}}, \bibinfo {author} {\bibfnamefont {S.~R.}\
  \bibnamefont {Giblin}}, \bibinfo {author} {\bibfnamefont {F.~L.}\
  \bibnamefont {Pratt}}, \bibinfo {author} {\bibfnamefont {S.~J.}\ \bibnamefont
  {Blundell}}, \bibinfo {author} {\bibfnamefont {T.}~\bibnamefont {Prokscha}},
  \bibinfo {author} {\bibfnamefont {A.}~\bibnamefont {Suter}}, \bibinfo
  {author} {\bibfnamefont {S.}~\bibnamefont {Langridge}}, \bibinfo {author}
  {\bibfnamefont {V.~N.}\ \bibnamefont {Strocov}}, \bibinfo {author}
  {\bibfnamefont {Z.}~\bibnamefont {Salman}}, \bibinfo {author} {\bibfnamefont
  {G.}~\bibnamefont {van~der Laan}}, \ and\ \bibinfo {author} {\bibfnamefont
  {T.}~\bibnamefont {Hesjedal}},\ }\href {\doibase 10.1103/PhysRevB.97.174427}
  {\bibfield  {journal} {\bibinfo  {journal} {Phys. Rev. B}\ }\textbf {\bibinfo
  {volume} {97}},\ \bibinfo {pages} {174427} (\bibinfo {year}
  {2018}{\natexlab{a}})}\BibitemShut {NoStop}%
\bibitem [{\citenamefont {Duffy}\ \emph
  {et~al.}(2018{\natexlab{b}})\citenamefont {Duffy}, \citenamefont {Frisk},
  \citenamefont {Burn}, \citenamefont {Steinke}, \citenamefont
  {Herrero-Martin}, \citenamefont {Ernst}, \citenamefont {van~der Laan},\ and\
  \citenamefont {Hesjedal}}]{Duffy2018}%
  \BibitemOpen
  \bibfield  {author} {\bibinfo {author} {\bibfnamefont {L.~B.}\ \bibnamefont
  {Duffy}}, \bibinfo {author} {\bibfnamefont {A.}~\bibnamefont {Frisk}},
  \bibinfo {author} {\bibfnamefont {D.~M.}\ \bibnamefont {Burn}}, \bibinfo
  {author} {\bibfnamefont {N.-J.}\ \bibnamefont {Steinke}}, \bibinfo {author}
  {\bibfnamefont {J.}~\bibnamefont {Herrero-Martin}}, \bibinfo {author}
  {\bibfnamefont {A.}~\bibnamefont {Ernst}}, \bibinfo {author} {\bibfnamefont
  {G.}~\bibnamefont {van~der Laan}}, \ and\ \bibinfo {author} {\bibfnamefont
  {T.}~\bibnamefont {Hesjedal}},\ }\href {\doibase
  10.1103/PhysRevMaterials.2.054201} {\bibfield  {journal} {\bibinfo  {journal}
  {Phys. Rev. Materials}\ }\textbf {\bibinfo {volume} {2}},\ \bibinfo {pages}
  {054201} (\bibinfo {year} {2018}{\natexlab{b}})}\BibitemShut {NoStop}%
\bibitem [{\citenamefont {Duffy}\ \emph {et~al.}(2019)\citenamefont {Duffy},
  \citenamefont {Steinke}, \citenamefont {Burn}, \citenamefont {Frisk},
  \citenamefont {Lari}, \citenamefont {Kuerbanjiang}, \citenamefont {Lazarov},
  \citenamefont {{van der Laan}}, \citenamefont {Langridge},\ and\
  \citenamefont {Hesjedal}}]{Duffy2019}%
  \BibitemOpen
  \bibfield  {author} {\bibinfo {author} {\bibfnamefont {L.~B.}\ \bibnamefont
  {Duffy}}, \bibinfo {author} {\bibfnamefont {N.~J.}\ \bibnamefont {Steinke}},
  \bibinfo {author} {\bibfnamefont {D.~M.}\ \bibnamefont {Burn}}, \bibinfo
  {author} {\bibfnamefont {A.}~\bibnamefont {Frisk}}, \bibinfo {author}
  {\bibfnamefont {L.}~\bibnamefont {Lari}}, \bibinfo {author} {\bibfnamefont
  {B.}~\bibnamefont {Kuerbanjiang}}, \bibinfo {author} {\bibfnamefont {V.~K.}\
  \bibnamefont {Lazarov}}, \bibinfo {author} {\bibfnamefont {G.}~\bibnamefont
  {{van der Laan}}}, \bibinfo {author} {\bibfnamefont {S.}~\bibnamefont
  {Langridge}}, \ and\ \bibinfo {author} {\bibfnamefont {T.}~\bibnamefont
  {Hesjedal}},\ }\href {\doibase 10.1103/PhysRevB.100.054402} {\bibfield
  {journal} {\bibinfo  {journal} {Phys. Rev. B}\ }\textbf {\bibinfo {volume}
  {100}},\ \bibinfo {pages} {54402} (\bibinfo {year} {2019})}\BibitemShut
  {NoStop}%
\bibitem [{\citenamefont {Mogi}\ \emph {et~al.}(2015)\citenamefont {Mogi},
  \citenamefont {Yoshimi}, \citenamefont {Tsukazaki}, \citenamefont {Yasuda},
  \citenamefont {Kozuka}, \citenamefont {Takahashi}, \citenamefont {Kawasaki},\
  and\ \citenamefont {Tokura}}]{Mogi2015}%
  \BibitemOpen
  \bibfield  {author} {\bibinfo {author} {\bibfnamefont {M.}~\bibnamefont
  {Mogi}}, \bibinfo {author} {\bibfnamefont {R.}~\bibnamefont {Yoshimi}},
  \bibinfo {author} {\bibfnamefont {A.}~\bibnamefont {Tsukazaki}}, \bibinfo
  {author} {\bibfnamefont {K.}~\bibnamefont {Yasuda}}, \bibinfo {author}
  {\bibfnamefont {Y.}~\bibnamefont {Kozuka}}, \bibinfo {author} {\bibfnamefont
  {K.~S.}\ \bibnamefont {Takahashi}}, \bibinfo {author} {\bibfnamefont
  {M.}~\bibnamefont {Kawasaki}}, \ and\ \bibinfo {author} {\bibfnamefont
  {Y.}~\bibnamefont {Tokura}},\ }\href {\doibase 10.1063/1.4935075} {\bibfield
  {journal} {\bibinfo  {journal} {Appl. Phys. Lett.}\ }\textbf {\bibinfo
  {volume} {107}},\ \bibinfo {pages} {182401} (\bibinfo {year}
  {2015})}\BibitemShut {NoStop}%
\bibitem [{\citenamefont {Mogi}\ \emph {et~al.}(2017)\citenamefont {Mogi},
  \citenamefont {Kawamura}, \citenamefont {Yoshimi}, \citenamefont {Tsukazaki},
  \citenamefont {Kozuka}, \citenamefont {Shirakawa}, \citenamefont {Takahashi},
  \citenamefont {Kawasaki},\ and\ \citenamefont {Tokura}}]{Mogi2017}%
  \BibitemOpen
  \bibfield  {author} {\bibinfo {author} {\bibfnamefont {M.}~\bibnamefont
  {Mogi}}, \bibinfo {author} {\bibfnamefont {M.}~\bibnamefont {Kawamura}},
  \bibinfo {author} {\bibfnamefont {R.}~\bibnamefont {Yoshimi}}, \bibinfo
  {author} {\bibfnamefont {A.}~\bibnamefont {Tsukazaki}}, \bibinfo {author}
  {\bibfnamefont {Y.}~\bibnamefont {Kozuka}}, \bibinfo {author} {\bibfnamefont
  {N.}~\bibnamefont {Shirakawa}}, \bibinfo {author} {\bibfnamefont {K.~S.}\
  \bibnamefont {Takahashi}}, \bibinfo {author} {\bibfnamefont {M.}~\bibnamefont
  {Kawasaki}}, \ and\ \bibinfo {author} {\bibfnamefont {Y.}~\bibnamefont
  {Tokura}},\ }\href {\doibase 10.1038/nmat4855} {\bibfield  {journal}
  {\bibinfo  {journal} {Nat. Mater.}\ }\textbf {\bibinfo {volume} {16}},\
  \bibinfo {pages} {516} (\bibinfo {year} {2017})}\BibitemShut {NoStop}%
\bibitem [{\citenamefont {Liu}\ \emph {et~al.}(2015)\citenamefont {Liu},
  \citenamefont {He}, \citenamefont {Xu}, \citenamefont {Murata}, \citenamefont
  {Onbasli}, \citenamefont {Lang}, \citenamefont {Maltby}, \citenamefont {Li},
  \citenamefont {Wang}, \citenamefont {Ross}, \citenamefont {Bencok},
  \citenamefont {van~der Laan}, \citenamefont {Zhang},\ and\ \citenamefont
  {Wang}}]{Liu2015}%
  \BibitemOpen
  \bibfield  {author} {\bibinfo {author} {\bibfnamefont {W.}~\bibnamefont
  {Liu}}, \bibinfo {author} {\bibfnamefont {L.}~\bibnamefont {He}}, \bibinfo
  {author} {\bibfnamefont {Y.}~\bibnamefont {Xu}}, \bibinfo {author}
  {\bibfnamefont {K.}~\bibnamefont {Murata}}, \bibinfo {author} {\bibfnamefont
  {M.~C.}\ \bibnamefont {Onbasli}}, \bibinfo {author} {\bibfnamefont
  {M.}~\bibnamefont {Lang}}, \bibinfo {author} {\bibfnamefont {N.~J.}\
  \bibnamefont {Maltby}}, \bibinfo {author} {\bibfnamefont {S.}~\bibnamefont
  {Li}}, \bibinfo {author} {\bibfnamefont {X.}~\bibnamefont {Wang}}, \bibinfo
  {author} {\bibfnamefont {C.~A.}\ \bibnamefont {Ross}}, \bibinfo {author}
  {\bibfnamefont {P.}~\bibnamefont {Bencok}}, \bibinfo {author} {\bibfnamefont
  {G.}~\bibnamefont {van~der Laan}}, \bibinfo {author} {\bibfnamefont
  {R.}~\bibnamefont {Zhang}}, \ and\ \bibinfo {author} {\bibfnamefont {K.~L.}\
  \bibnamefont {Wang}},\ }\href@noop {} {\bibfield  {journal} {\bibinfo
  {journal} {{Nano Lett.}}\ }\textbf {\bibinfo {volume} {15}},\ \bibinfo
  {pages} {764} (\bibinfo {year} {2015})}\BibitemShut {NoStop}%
\bibitem [{\citenamefont {Vobornik}\ \emph {et~al.}(2011)\citenamefont
  {Vobornik}, \citenamefont {Manju}, \citenamefont {Fujii}, \citenamefont
  {Borgatti}, \citenamefont {Torelli}, \citenamefont {Krizmancic},
  \citenamefont {Hor}, \citenamefont {Cava},\ and\ \citenamefont
  {Panaccione}}]{Vobornik2011}%
  \BibitemOpen
  \bibfield  {author} {\bibinfo {author} {\bibfnamefont {I.}~\bibnamefont
  {Vobornik}}, \bibinfo {author} {\bibfnamefont {U.}~\bibnamefont {Manju}},
  \bibinfo {author} {\bibfnamefont {J.}~\bibnamefont {Fujii}}, \bibinfo
  {author} {\bibfnamefont {F.}~\bibnamefont {Borgatti}}, \bibinfo {author}
  {\bibfnamefont {P.}~\bibnamefont {Torelli}}, \bibinfo {author} {\bibfnamefont
  {D.}~\bibnamefont {Krizmancic}}, \bibinfo {author} {\bibfnamefont {Y.~S.}\
  \bibnamefont {Hor}}, \bibinfo {author} {\bibfnamefont {R.~J.}\ \bibnamefont
  {Cava}}, \ and\ \bibinfo {author} {\bibfnamefont {G.}~\bibnamefont
  {Panaccione}},\ }\href@noop {} {\bibfield  {journal} {\bibinfo  {journal}
  {Nano Lett.}\ }\textbf {\bibinfo {volume} {11}},\ \bibinfo {pages} {4079}
  (\bibinfo {year} {2011})}\BibitemShut {NoStop}%
\bibitem [{\citenamefont {Baker}\ \emph
  {et~al.}(2015{\natexlab{a}})\citenamefont {Baker}, \citenamefont {Figueroa},
  \citenamefont {Kummer}, \citenamefont {{Collins-McIntyre}}, \citenamefont
  {Hesjedal},\ and\ \citenamefont {{van der Laan}}}]{baker2015magnetic}%
  \BibitemOpen
  \bibfield  {author} {\bibinfo {author} {\bibfnamefont {A.~A.}\ \bibnamefont
  {Baker}}, \bibinfo {author} {\bibfnamefont {A.~I.}\ \bibnamefont {Figueroa}},
  \bibinfo {author} {\bibfnamefont {K.}~\bibnamefont {Kummer}}, \bibinfo
  {author} {\bibfnamefont {L.~J.}\ \bibnamefont {{Collins-McIntyre}}}, \bibinfo
  {author} {\bibfnamefont {T.}~\bibnamefont {Hesjedal}}, \ and\ \bibinfo
  {author} {\bibfnamefont {G.}~\bibnamefont {{van der Laan}}},\ }\href@noop {}
  {\bibfield  {journal} {\bibinfo  {journal} {Phys. Rev. B}\ }\textbf {\bibinfo
  {volume} {92}},\ \bibinfo {pages} {094420} (\bibinfo {year}
  {2015}{\natexlab{a}})}\BibitemShut {NoStop}%
\bibitem [{\citenamefont {Hesjedal}\ and\ \citenamefont
  {Chen}(2017)}]{Hesjedal_hetero:2017}%
  \BibitemOpen
  \bibfield  {author} {\bibinfo {author} {\bibfnamefont {T.}~\bibnamefont
  {Hesjedal}}\ and\ \bibinfo {author} {\bibfnamefont {Y.}~\bibnamefont
  {Chen}},\ }\href {\doibase 10.1038/nmat4835} {\bibfield  {journal} {\bibinfo
  {journal} {Nat. Mater.}\ }\textbf {\bibinfo {volume} {16}},\ \bibinfo {pages}
  {3} (\bibinfo {year} {2017})}\BibitemShut {NoStop}%
\bibitem [{\citenamefont {He}\ \emph {et~al.}(2017)\citenamefont {He},
  \citenamefont {Kou}, \citenamefont {Grutter}, \citenamefont {Yin},
  \citenamefont {Pan}, \citenamefont {Che}, \citenamefont {Liu}, \citenamefont
  {Nie}, \citenamefont {Zhang}, \citenamefont {Disseler}, \citenamefont
  {Kirby}, \citenamefont {Ratcliff~II}, \citenamefont {Shao}, \citenamefont
  {Murata}, \citenamefont {Zhu}, \citenamefont {Yu}, \citenamefont {Fan},
  \citenamefont {Montazeri}, \citenamefont {Han}, \citenamefont {Borchers},\
  and\ \citenamefont {Wang}}]{He_hetero_2017}%
  \BibitemOpen
  \bibfield  {author} {\bibinfo {author} {\bibfnamefont {Q.~L.}\ \bibnamefont
  {He}}, \bibinfo {author} {\bibfnamefont {X.}~\bibnamefont {Kou}}, \bibinfo
  {author} {\bibfnamefont {A.~J.}\ \bibnamefont {Grutter}}, \bibinfo {author}
  {\bibfnamefont {G.}~\bibnamefont {Yin}}, \bibinfo {author} {\bibfnamefont
  {L.}~\bibnamefont {Pan}}, \bibinfo {author} {\bibfnamefont {X.}~\bibnamefont
  {Che}}, \bibinfo {author} {\bibfnamefont {Y.}~\bibnamefont {Liu}}, \bibinfo
  {author} {\bibfnamefont {T.}~\bibnamefont {Nie}}, \bibinfo {author}
  {\bibfnamefont {B.}~\bibnamefont {Zhang}}, \bibinfo {author} {\bibfnamefont
  {S.~M.}\ \bibnamefont {Disseler}}, \bibinfo {author} {\bibfnamefont {B.~J.}\
  \bibnamefont {Kirby}}, \bibinfo {author} {\bibfnamefont {W.}~\bibnamefont
  {Ratcliff~II}}, \bibinfo {author} {\bibfnamefont {Q.}~\bibnamefont {Shao}},
  \bibinfo {author} {\bibfnamefont {K.}~\bibnamefont {Murata}}, \bibinfo
  {author} {\bibfnamefont {X.}~\bibnamefont {Zhu}}, \bibinfo {author}
  {\bibfnamefont {G.}~\bibnamefont {Yu}}, \bibinfo {author} {\bibfnamefont
  {Y.}~\bibnamefont {Fan}}, \bibinfo {author} {\bibfnamefont {M.}~\bibnamefont
  {Montazeri}}, \bibinfo {author} {\bibfnamefont {X.}~\bibnamefont {Han}},
  \bibinfo {author} {\bibfnamefont {J.~A.}\ \bibnamefont {Borchers}}, \ and\
  \bibinfo {author} {\bibfnamefont {K.~L.}\ \bibnamefont {Wang}},\ }\href
  {\doibase 10.1038/nmat4783} {\bibfield  {journal} {\bibinfo  {journal} {Nat.
  Mater.}\ }\textbf {\bibinfo {volume} {16}},\ \bibinfo {pages} {94} (\bibinfo
  {year} {2017})}\BibitemShut {NoStop}%
\bibitem [{\citenamefont {Otrokov}\ \emph
  {et~al.}(2017{\natexlab{a}})\citenamefont {Otrokov}, \citenamefont
  {Menshchikova}, \citenamefont {Rusinov}, \citenamefont {Vergniory},
  \citenamefont {Kuznetsov},\ and\ \citenamefont {Chulkov}}]{Otrokov_JETP2017}%
  \BibitemOpen
  \bibfield  {author} {\bibinfo {author} {\bibfnamefont {M.~M.}\ \bibnamefont
  {Otrokov}}, \bibinfo {author} {\bibfnamefont {T.~V.}\ \bibnamefont
  {Menshchikova}}, \bibinfo {author} {\bibfnamefont {I.~P.}\ \bibnamefont
  {Rusinov}}, \bibinfo {author} {\bibfnamefont {M.~G.}\ \bibnamefont
  {Vergniory}}, \bibinfo {author} {\bibfnamefont {V.~M.}\ \bibnamefont
  {Kuznetsov}}, \ and\ \bibinfo {author} {\bibfnamefont {E.~V.}\ \bibnamefont
  {Chulkov}},\ }\href {\doibase 10.1088/2053-1583/aa6bec} {\bibfield  {journal}
  {\bibinfo  {journal} {JETP Letters}\ }\textbf {\bibinfo {volume} {105}},\
  \bibinfo {pages} {297} (\bibinfo {year} {2017}{\natexlab{a}})}\BibitemShut
  {NoStop}%
\bibitem [{\citenamefont {Otrokov}\ \emph
  {et~al.}(2017{\natexlab{b}})\citenamefont {Otrokov}, \citenamefont
  {Menshchikova}, \citenamefont {Vergniory}, \citenamefont {Rusinov},
  \citenamefont {Vyazovskaya}, \citenamefont {Koroteev}, \citenamefont
  {Bihlmayer}, \citenamefont {Ernst}, \citenamefont {Echenique}, \citenamefont
  {Arnau},\ and\ \citenamefont {Chulkov}}]{Otrokov_2DMat2017}%
  \BibitemOpen
  \bibfield  {author} {\bibinfo {author} {\bibfnamefont {M.~M.}\ \bibnamefont
  {Otrokov}}, \bibinfo {author} {\bibfnamefont {T.~V.}\ \bibnamefont
  {Menshchikova}}, \bibinfo {author} {\bibfnamefont {M.~G.}\ \bibnamefont
  {Vergniory}}, \bibinfo {author} {\bibfnamefont {I.~P.}\ \bibnamefont
  {Rusinov}}, \bibinfo {author} {\bibfnamefont {A.~Y.}\ \bibnamefont
  {Vyazovskaya}}, \bibinfo {author} {\bibfnamefont {Y.~M.}\ \bibnamefont
  {Koroteev}}, \bibinfo {author} {\bibfnamefont {G.}~\bibnamefont {Bihlmayer}},
  \bibinfo {author} {\bibfnamefont {A.}~\bibnamefont {Ernst}}, \bibinfo
  {author} {\bibfnamefont {P.~M.}\ \bibnamefont {Echenique}}, \bibinfo {author}
  {\bibfnamefont {A.}~\bibnamefont {Arnau}}, \ and\ \bibinfo {author}
  {\bibfnamefont {E.~V.}\ \bibnamefont {Chulkov}},\ }\href {\doibase
  10.1088/2053-1583/aa6bec} {\bibfield  {journal} {\bibinfo  {journal} {2D
  Mater.}\ }\textbf {\bibinfo {volume} {4}},\ \bibinfo {pages} {025082}
  (\bibinfo {year} {2017}{\natexlab{b}})}\BibitemShut {NoStop}%
\bibitem [{\citenamefont {Hirahara}\ \emph {et~al.}(2017)\citenamefont
  {Hirahara}, \citenamefont {Eremeev}, \citenamefont {Shirasawa}, \citenamefont
  {Okuyama}, \citenamefont {Kubo}, \citenamefont {Nakanishi}, \citenamefont
  {Akiyama}, \citenamefont {Takayama}, \citenamefont {Hajiri}, \citenamefont
  {Ideta}, \citenamefont {Matsunami}, \citenamefont {Sumida}, \citenamefont
  {Miyamoto}, \citenamefont {Takagi}, \citenamefont {Tanaka}, \citenamefont
  {Okuda}, \citenamefont {Yokoyama}, \citenamefont {Kimura}, \citenamefont
  {Hasegawa},\ and\ \citenamefont {Chulkov}}]{Hirahara_NanoLett2017}%
  \BibitemOpen
  \bibfield  {author} {\bibinfo {author} {\bibfnamefont {T.}~\bibnamefont
  {Hirahara}}, \bibinfo {author} {\bibfnamefont {S.~V.}\ \bibnamefont
  {Eremeev}}, \bibinfo {author} {\bibfnamefont {T.}~\bibnamefont {Shirasawa}},
  \bibinfo {author} {\bibfnamefont {Y.}~\bibnamefont {Okuyama}}, \bibinfo
  {author} {\bibfnamefont {T.}~\bibnamefont {Kubo}}, \bibinfo {author}
  {\bibfnamefont {R.}~\bibnamefont {Nakanishi}}, \bibinfo {author}
  {\bibfnamefont {R.}~\bibnamefont {Akiyama}}, \bibinfo {author} {\bibfnamefont
  {A.}~\bibnamefont {Takayama}}, \bibinfo {author} {\bibfnamefont
  {T.}~\bibnamefont {Hajiri}}, \bibinfo {author} {\bibfnamefont {S.-i.}\
  \bibnamefont {Ideta}}, \bibinfo {author} {\bibfnamefont {M.}~\bibnamefont
  {Matsunami}}, \bibinfo {author} {\bibfnamefont {K.}~\bibnamefont {Sumida}},
  \bibinfo {author} {\bibfnamefont {K.}~\bibnamefont {Miyamoto}}, \bibinfo
  {author} {\bibfnamefont {Y.}~\bibnamefont {Takagi}}, \bibinfo {author}
  {\bibfnamefont {K.}~\bibnamefont {Tanaka}}, \bibinfo {author} {\bibfnamefont
  {T.}~\bibnamefont {Okuda}}, \bibinfo {author} {\bibfnamefont
  {T.}~\bibnamefont {Yokoyama}}, \bibinfo {author} {\bibfnamefont {S.-i.}\
  \bibnamefont {Kimura}}, \bibinfo {author} {\bibfnamefont {S.}~\bibnamefont
  {Hasegawa}}, \ and\ \bibinfo {author} {\bibfnamefont {E.~V.}\ \bibnamefont
  {Chulkov}},\ }\href {\doibase 10.1021/acs.nanolett.7b00560} {\bibfield
  {journal} {\bibinfo  {journal} {Nano Lett.}\ }\textbf {\bibinfo {volume}
  {17}},\ \bibinfo {pages} {3493} (\bibinfo {year} {2017})}\BibitemShut
  {NoStop}%
\bibitem [{\citenamefont {Eremeev}, \citenamefont {Otrokov},\ and\
  \citenamefont {Chulkov}(2018)}]{EremeevNanolett2018}%
  \BibitemOpen
  \bibfield  {author} {\bibinfo {author} {\bibfnamefont {S.~V.}\ \bibnamefont
  {Eremeev}}, \bibinfo {author} {\bibfnamefont {M.~M.}\ \bibnamefont
  {Otrokov}}, \ and\ \bibinfo {author} {\bibfnamefont {E.~V.}\ \bibnamefont
  {Chulkov}},\ }\href {\doibase 10.1021/acs.nanolett.8b03057} {\bibfield
  {journal} {\bibinfo  {journal} {Nano Lett.}\ }\textbf {\bibinfo {volume}
  {18}},\ \bibinfo {pages} {6521} (\bibinfo {year} {2018})}\BibitemShut
  {NoStop}%
\bibitem [{\citenamefont {Klimovskikh}\ \emph {et~al.}(2020)\citenamefont
  {Klimovskikh}, \citenamefont {Otrokov}, \citenamefont {Estyunin},
  \citenamefont {Eremeev}, \citenamefont {Filnov}, \citenamefont {Koroleva},
  \citenamefont {Shevchenko}, \citenamefont {Voroshnin}, \citenamefont
  {Rybkin}, \citenamefont {Rusinov}, \citenamefont {Blanco-Rey}, \citenamefont
  {Hoffmann}, \citenamefont {Aliev}, \citenamefont {Babanly}, \citenamefont
  {Amiraslanov}, \citenamefont {Abdullayev}, \citenamefont {Zverev},
  \citenamefont {Kimura}, \citenamefont {Tereshchenko}, \citenamefont {Kokh},
  \citenamefont {Petaccia}, \citenamefont {Di~Santo}, \citenamefont {Ernst},
  \citenamefont {Echenique}, \citenamefont {Mamedov}, \citenamefont {Shikin},\
  and\ \citenamefont {Chulkov}}]{Klimovskikh2020}%
  \BibitemOpen
  \bibfield  {author} {\bibinfo {author} {\bibfnamefont {I.~I.}\ \bibnamefont
  {Klimovskikh}}, \bibinfo {author} {\bibfnamefont {M.~M.}\ \bibnamefont
  {Otrokov}}, \bibinfo {author} {\bibfnamefont {D.}~\bibnamefont {Estyunin}},
  \bibinfo {author} {\bibfnamefont {S.~V.}\ \bibnamefont {Eremeev}}, \bibinfo
  {author} {\bibfnamefont {S.~O.}\ \bibnamefont {Filnov}}, \bibinfo {author}
  {\bibfnamefont {A.}~\bibnamefont {Koroleva}}, \bibinfo {author}
  {\bibfnamefont {E.}~\bibnamefont {Shevchenko}}, \bibinfo {author}
  {\bibfnamefont {V.}~\bibnamefont {Voroshnin}}, \bibinfo {author}
  {\bibfnamefont {A.~G.}\ \bibnamefont {Rybkin}}, \bibinfo {author}
  {\bibfnamefont {I.~P.}\ \bibnamefont {Rusinov}}, \bibinfo {author}
  {\bibfnamefont {M.}~\bibnamefont {Blanco-Rey}}, \bibinfo {author}
  {\bibfnamefont {M.}~\bibnamefont {Hoffmann}}, \bibinfo {author}
  {\bibfnamefont {Z.~S.}\ \bibnamefont {Aliev}}, \bibinfo {author}
  {\bibfnamefont {M.~B.}\ \bibnamefont {Babanly}}, \bibinfo {author}
  {\bibfnamefont {I.~R.}\ \bibnamefont {Amiraslanov}}, \bibinfo {author}
  {\bibfnamefont {N.~A.}\ \bibnamefont {Abdullayev}}, \bibinfo {author}
  {\bibfnamefont {V.~N.}\ \bibnamefont {Zverev}}, \bibinfo {author}
  {\bibfnamefont {A.}~\bibnamefont {Kimura}}, \bibinfo {author} {\bibfnamefont
  {O.~E.}\ \bibnamefont {Tereshchenko}}, \bibinfo {author} {\bibfnamefont
  {K.~A.}\ \bibnamefont {Kokh}}, \bibinfo {author} {\bibfnamefont
  {L.}~\bibnamefont {Petaccia}}, \bibinfo {author} {\bibfnamefont
  {G.}~\bibnamefont {Di~Santo}}, \bibinfo {author} {\bibfnamefont
  {A.}~\bibnamefont {Ernst}}, \bibinfo {author} {\bibfnamefont {P.~M.}\
  \bibnamefont {Echenique}}, \bibinfo {author} {\bibfnamefont {N.~T.}\
  \bibnamefont {Mamedov}}, \bibinfo {author} {\bibfnamefont {A.~M.}\
  \bibnamefont {Shikin}}, \ and\ \bibinfo {author} {\bibfnamefont {E.~V.}\
  \bibnamefont {Chulkov}},\ }\href {\doibase 10.1038/s41535-020-00255-9}
  {\bibfield  {journal} {\bibinfo  {journal} {npj Quantum Mater.}\ }\textbf
  {\bibinfo {volume} {5}},\ \bibinfo {pages} {54} (\bibinfo {year}
  {2020})}\BibitemShut {NoStop}%
\bibitem [{\citenamefont {Wei}\ \emph {et~al.}(2013)\citenamefont {Wei},
  \citenamefont {Katmis}, \citenamefont {Assaf}, \citenamefont {Steinberg},
  \citenamefont {Jarillo-Herrero}, \citenamefont {Heiman},\ and\ \citenamefont
  {Moodera}}]{WeiPRL2013}%
  \BibitemOpen
  \bibfield  {author} {\bibinfo {author} {\bibfnamefont {P.}~\bibnamefont
  {Wei}}, \bibinfo {author} {\bibfnamefont {F.}~\bibnamefont {Katmis}},
  \bibinfo {author} {\bibfnamefont {B.~A.}\ \bibnamefont {Assaf}}, \bibinfo
  {author} {\bibfnamefont {H.}~\bibnamefont {Steinberg}}, \bibinfo {author}
  {\bibfnamefont {P.}~\bibnamefont {Jarillo-Herrero}}, \bibinfo {author}
  {\bibfnamefont {D.}~\bibnamefont {Heiman}}, \ and\ \bibinfo {author}
  {\bibfnamefont {J.~S.}\ \bibnamefont {Moodera}},\ }\href {\doibase
  10.1103/PhysRevLett.110.186807} {\bibfield  {journal} {\bibinfo  {journal}
  {Phys. Rev. Lett.}\ }\textbf {\bibinfo {volume} {110}},\ \bibinfo {pages}
  {186807} (\bibinfo {year} {2013})}\BibitemShut {NoStop}%
\bibitem [{\citenamefont {Katmis}\ \emph {et~al.}(2016)\citenamefont {Katmis},
  \citenamefont {Lauter}, \citenamefont {Nogueira}, \citenamefont {Assaf},
  \citenamefont {Jamer}, \citenamefont {Wei}, \citenamefont {Satpati},
  \citenamefont {Freeland}, \citenamefont {Eremin}, \citenamefont {Heiman},
  \citenamefont {Jarillo-Herrero},\ and\ \citenamefont
  {Moodera}}]{KatmisNat2016}%
  \BibitemOpen
  \bibfield  {author} {\bibinfo {author} {\bibfnamefont {F.}~\bibnamefont
  {Katmis}}, \bibinfo {author} {\bibfnamefont {V.}~\bibnamefont {Lauter}},
  \bibinfo {author} {\bibfnamefont {F.~S.}\ \bibnamefont {Nogueira}}, \bibinfo
  {author} {\bibfnamefont {B.~A.}\ \bibnamefont {Assaf}}, \bibinfo {author}
  {\bibfnamefont {M.~E.}\ \bibnamefont {Jamer}}, \bibinfo {author}
  {\bibfnamefont {P.}~\bibnamefont {Wei}}, \bibinfo {author} {\bibfnamefont
  {B.}~\bibnamefont {Satpati}}, \bibinfo {author} {\bibfnamefont {J.~W.}\
  \bibnamefont {Freeland}}, \bibinfo {author} {\bibfnamefont {I.}~\bibnamefont
  {Eremin}}, \bibinfo {author} {\bibfnamefont {D.}~\bibnamefont {Heiman}},
  \bibinfo {author} {\bibfnamefont {P.}~\bibnamefont {Jarillo-Herrero}}, \ and\
  \bibinfo {author} {\bibfnamefont {J.~S.}\ \bibnamefont {Moodera}},\
  }\href@noop {} {\bibfield  {journal} {\bibinfo  {journal} {Nature}\ }\textbf
  {\bibinfo {volume} {533}},\ \bibinfo {pages} {513} (\bibinfo {year}
  {2016})}\BibitemShut {NoStop}%
\bibitem [{\citenamefont {Lee}\ \emph {et~al.}(2016)\citenamefont {Lee},
  \citenamefont {Katmis}, \citenamefont {Jarillo-Herrero}, \citenamefont
  {Moodera},\ and\ \citenamefont {Gedik}}]{Lee2016-prox}%
  \BibitemOpen
  \bibfield  {author} {\bibinfo {author} {\bibfnamefont {C.}~\bibnamefont
  {Lee}}, \bibinfo {author} {\bibfnamefont {F.}~\bibnamefont {Katmis}},
  \bibinfo {author} {\bibfnamefont {P.}~\bibnamefont {Jarillo-Herrero}},
  \bibinfo {author} {\bibfnamefont {J.~S.}\ \bibnamefont {Moodera}}, \ and\
  \bibinfo {author} {\bibfnamefont {N.}~\bibnamefont {Gedik}},\ }\href
  {\doibase 10.1038/ncomms12014} {\bibfield  {journal} {\bibinfo  {journal}
  {Nat. Commun.}\ }\textbf {\bibinfo {volume} {7}},\ \bibinfo {pages} {12014}
  (\bibinfo {year} {2016})}\BibitemShut {NoStop}%
\bibitem [{\citenamefont {Jiang}\ \emph {et~al.}(2015)\citenamefont {Jiang},
  \citenamefont {Chang}, \citenamefont {Tang}, \citenamefont {Wei},
  \citenamefont {Moodera},\ and\ \citenamefont {Shi}}]{JiangNanoLett2015}%
  \BibitemOpen
  \bibfield  {author} {\bibinfo {author} {\bibfnamefont {Z.}~\bibnamefont
  {Jiang}}, \bibinfo {author} {\bibfnamefont {C.-Z.}\ \bibnamefont {Chang}},
  \bibinfo {author} {\bibfnamefont {C.}~\bibnamefont {Tang}}, \bibinfo {author}
  {\bibfnamefont {P.}~\bibnamefont {Wei}}, \bibinfo {author} {\bibfnamefont
  {J.~S.}\ \bibnamefont {Moodera}}, \ and\ \bibinfo {author} {\bibfnamefont
  {J.}~\bibnamefont {Shi}},\ }\href {\doibase 10.1021/acs.nanolett.5b01905}
  {\bibfield  {journal} {\bibinfo  {journal} {Nano Lett.}\ }\textbf {\bibinfo
  {volume} {15}},\ \bibinfo {pages} {5835} (\bibinfo {year}
  {2015})}\BibitemShut {NoStop}%
\bibitem [{\citenamefont {Li}\ \emph {et~al.}(2017)\citenamefont {Li},
  \citenamefont {Song}, \citenamefont {Zhao}, \citenamefont {Garlow},
  \citenamefont {Liu}, \citenamefont {Wu}, \citenamefont {Zhu}, \citenamefont
  {Moodera}, \citenamefont {Chan}, \citenamefont {Chen},\ and\ \citenamefont
  {Chang}}]{LiPRB2017}%
  \BibitemOpen
  \bibfield  {author} {\bibinfo {author} {\bibfnamefont {M.}~\bibnamefont
  {Li}}, \bibinfo {author} {\bibfnamefont {Q.}~\bibnamefont {Song}}, \bibinfo
  {author} {\bibfnamefont {W.}~\bibnamefont {Zhao}}, \bibinfo {author}
  {\bibfnamefont {J.~A.}\ \bibnamefont {Garlow}}, \bibinfo {author}
  {\bibfnamefont {T.-H.}\ \bibnamefont {Liu}}, \bibinfo {author} {\bibfnamefont
  {L.}~\bibnamefont {Wu}}, \bibinfo {author} {\bibfnamefont {Y.}~\bibnamefont
  {Zhu}}, \bibinfo {author} {\bibfnamefont {J.~S.}\ \bibnamefont {Moodera}},
  \bibinfo {author} {\bibfnamefont {M.~H.~W.}\ \bibnamefont {Chan}}, \bibinfo
  {author} {\bibfnamefont {G.}~\bibnamefont {Chen}}, \ and\ \bibinfo {author}
  {\bibfnamefont {C.-Z.}\ \bibnamefont {Chang}},\ }\href {\doibase
  10.1103/PhysRevB.96.201301} {\bibfield  {journal} {\bibinfo  {journal} {Phys.
  Rev. B}\ }\textbf {\bibinfo {volume} {96}},\ \bibinfo {pages} {201301}
  (\bibinfo {year} {2017})}\BibitemShut {NoStop}%
\bibitem [{\citenamefont {Tang}\ \emph {et~al.}(2017)\citenamefont {Tang},
  \citenamefont {Chang}, \citenamefont {Zhao}, \citenamefont {Liu},
  \citenamefont {Jiang}, \citenamefont {Liu}, \citenamefont {McCartney},
  \citenamefont {Smith}, \citenamefont {Chen}, \citenamefont {Moodera},\ and\
  \citenamefont {Shi}}]{TangSciAdv2017}%
  \BibitemOpen
  \bibfield  {author} {\bibinfo {author} {\bibfnamefont {C.}~\bibnamefont
  {Tang}}, \bibinfo {author} {\bibfnamefont {C.-Z.}\ \bibnamefont {Chang}},
  \bibinfo {author} {\bibfnamefont {G.}~\bibnamefont {Zhao}}, \bibinfo {author}
  {\bibfnamefont {Y.}~\bibnamefont {Liu}}, \bibinfo {author} {\bibfnamefont
  {Z.}~\bibnamefont {Jiang}}, \bibinfo {author} {\bibfnamefont {C.-X.}\
  \bibnamefont {Liu}}, \bibinfo {author} {\bibfnamefont {M.~R.}\ \bibnamefont
  {McCartney}}, \bibinfo {author} {\bibfnamefont {D.~J.}\ \bibnamefont
  {Smith}}, \bibinfo {author} {\bibfnamefont {T.}~\bibnamefont {Chen}},
  \bibinfo {author} {\bibfnamefont {J.~S.}\ \bibnamefont {Moodera}}, \ and\
  \bibinfo {author} {\bibfnamefont {J.}~\bibnamefont {Shi}},\ }\href {\doibase
  10.1126/sciadv.1700307} {\bibfield  {journal} {\bibinfo  {journal} {Sci.
  Adv.}\ }\textbf {\bibinfo {volume} {3}},\ \bibinfo {pages} {e1700307}
  (\bibinfo {year} {2017})}\BibitemShut {NoStop}%
\bibitem [{\citenamefont {Mogi}\ \emph {et~al.}(2019)\citenamefont {Mogi},
  \citenamefont {Nakajima}, \citenamefont {Ukleev}, \citenamefont {Tsukazaki},
  \citenamefont {Yoshimi}, \citenamefont {Kawamura}, \citenamefont {Takahashi},
  \citenamefont {Hanashima}, \citenamefont {Kakurai}, \citenamefont {Arima},
  \citenamefont {Kawasaki},\ and\ \citenamefont {Tokura}}]{Mogi2019}%
  \BibitemOpen
  \bibfield  {author} {\bibinfo {author} {\bibfnamefont {M.}~\bibnamefont
  {Mogi}}, \bibinfo {author} {\bibfnamefont {T.}~\bibnamefont {Nakajima}},
  \bibinfo {author} {\bibfnamefont {V.}~\bibnamefont {Ukleev}}, \bibinfo
  {author} {\bibfnamefont {A.}~\bibnamefont {Tsukazaki}}, \bibinfo {author}
  {\bibfnamefont {R.}~\bibnamefont {Yoshimi}}, \bibinfo {author} {\bibfnamefont
  {M.}~\bibnamefont {Kawamura}}, \bibinfo {author} {\bibfnamefont {K.~S.}\
  \bibnamefont {Takahashi}}, \bibinfo {author} {\bibfnamefont {T.}~\bibnamefont
  {Hanashima}}, \bibinfo {author} {\bibfnamefont {K.}~\bibnamefont {Kakurai}},
  \bibinfo {author} {\bibfnamefont {T.-h.}\ \bibnamefont {Arima}}, \bibinfo
  {author} {\bibfnamefont {M.}~\bibnamefont {Kawasaki}}, \ and\ \bibinfo
  {author} {\bibfnamefont {Y.}~\bibnamefont {Tokura}},\ }\href {\doibase
  10.1103/PhysRevLett.123.016804} {\bibfield  {journal} {\bibinfo  {journal}
  {Phys. Rev. Lett.}\ }\textbf {\bibinfo {volume} {123}},\ \bibinfo {pages}
  {016804} (\bibinfo {year} {2019})}\BibitemShut {NoStop}%
\bibitem [{\citenamefont {Watanabe}\ \emph {et~al.}(2019)\citenamefont
  {Watanabe}, \citenamefont {Yoshimi}, \citenamefont {Kawamura}, \citenamefont
  {Mogi}, \citenamefont {Tsukazaki}, \citenamefont {Yu}, \citenamefont
  {Nakajima}, \citenamefont {Takahashi}, \citenamefont {Kawasaki},\ and\
  \citenamefont {Tokura}}]{Watanabe2019}%
  \BibitemOpen
  \bibfield  {author} {\bibinfo {author} {\bibfnamefont {R.}~\bibnamefont
  {Watanabe}}, \bibinfo {author} {\bibfnamefont {R.}~\bibnamefont {Yoshimi}},
  \bibinfo {author} {\bibfnamefont {M.}~\bibnamefont {Kawamura}}, \bibinfo
  {author} {\bibfnamefont {M.}~\bibnamefont {Mogi}}, \bibinfo {author}
  {\bibfnamefont {A.}~\bibnamefont {Tsukazaki}}, \bibinfo {author}
  {\bibfnamefont {X.~Z.}\ \bibnamefont {Yu}}, \bibinfo {author} {\bibfnamefont
  {K.}~\bibnamefont {Nakajima}}, \bibinfo {author} {\bibfnamefont {K.~S.}\
  \bibnamefont {Takahashi}}, \bibinfo {author} {\bibfnamefont {M.}~\bibnamefont
  {Kawasaki}}, \ and\ \bibinfo {author} {\bibfnamefont {Y.}~\bibnamefont
  {Tokura}},\ }\href {\doibase 10.1063/1.5111891} {\bibfield  {journal}
  {\bibinfo  {journal} {Appl. Phys. Lett.}\ }\textbf {\bibinfo {volume}
  {115}},\ \bibinfo {pages} {102403} (\bibinfo {year} {2019})}\BibitemShut
  {NoStop}%
\bibitem [{\citenamefont {Pereira}\ \emph {et~al.}(2020)\citenamefont
  {Pereira}, \citenamefont {Altendorf}, \citenamefont {Liu}, \citenamefont
  {Liao}, \citenamefont {Komarek}, \citenamefont {Guo}, \citenamefont {Lin},
  \citenamefont {Chen}, \citenamefont {Hong}, \citenamefont {Kwo},
  \citenamefont {Tjeng},\ and\ \citenamefont {Wu}}]{Pereira_PRM2020}%
  \BibitemOpen
  \bibfield  {author} {\bibinfo {author} {\bibfnamefont {V.~M.}\ \bibnamefont
  {Pereira}}, \bibinfo {author} {\bibfnamefont {S.~G.}\ \bibnamefont
  {Altendorf}}, \bibinfo {author} {\bibfnamefont {C.~E.}\ \bibnamefont {Liu}},
  \bibinfo {author} {\bibfnamefont {S.~C.}\ \bibnamefont {Liao}}, \bibinfo
  {author} {\bibfnamefont {A.~C.}\ \bibnamefont {Komarek}}, \bibinfo {author}
  {\bibfnamefont {M.}~\bibnamefont {Guo}}, \bibinfo {author} {\bibfnamefont
  {H.-J.}\ \bibnamefont {Lin}}, \bibinfo {author} {\bibfnamefont {C.~T.}\
  \bibnamefont {Chen}}, \bibinfo {author} {\bibfnamefont {M.}~\bibnamefont
  {Hong}}, \bibinfo {author} {\bibfnamefont {J.}~\bibnamefont {Kwo}}, \bibinfo
  {author} {\bibfnamefont {L.~H.}\ \bibnamefont {Tjeng}}, \ and\ \bibinfo
  {author} {\bibfnamefont {C.~N.}\ \bibnamefont {Wu}},\ }\href {\doibase
  10.1103/PhysRevMaterials.4.064202} {\bibfield  {journal} {\bibinfo  {journal}
  {Phys. Rev. Materials}\ }\textbf {\bibinfo {volume} {4}},\ \bibinfo {pages}
  {064202} (\bibinfo {year} {2020})}\BibitemShut {NoStop}%
\bibitem [{\citenamefont {Shiomi}\ \emph {et~al.}(2014)\citenamefont {Shiomi},
  \citenamefont {Nomura}, \citenamefont {Kajiwara}, \citenamefont {Eto},
  \citenamefont {Novak}, \citenamefont {Segawa}, \citenamefont {Ando},\ and\
  \citenamefont {Saitoh}}]{Shiomi_PRL2014}%
  \BibitemOpen
  \bibfield  {author} {\bibinfo {author} {\bibfnamefont {Y.}~\bibnamefont
  {Shiomi}}, \bibinfo {author} {\bibfnamefont {K.}~\bibnamefont {Nomura}},
  \bibinfo {author} {\bibfnamefont {Y.}~\bibnamefont {Kajiwara}}, \bibinfo
  {author} {\bibfnamefont {K.}~\bibnamefont {Eto}}, \bibinfo {author}
  {\bibfnamefont {M.}~\bibnamefont {Novak}}, \bibinfo {author} {\bibfnamefont
  {K.}~\bibnamefont {Segawa}}, \bibinfo {author} {\bibfnamefont
  {Y.}~\bibnamefont {Ando}}, \ and\ \bibinfo {author} {\bibfnamefont
  {E.}~\bibnamefont {Saitoh}},\ }\href {\doibase
  10.1103/PhysRevLett.113.196601} {\bibfield  {journal} {\bibinfo  {journal}
  {Phys. Rev. Lett.}\ }\textbf {\bibinfo {volume} {113}},\ \bibinfo {pages}
  {196601} (\bibinfo {year} {2014})}\BibitemShut {NoStop}%
\bibitem [{\citenamefont {Baker}\ \emph
  {et~al.}(2015{\natexlab{b}})\citenamefont {Baker}, \citenamefont {Figueroa},
  \citenamefont {Collins-McIntyre}, \citenamefont {{van der Laan}},\ and\
  \citenamefont {Hesjedal}}]{Baker_XFMR2015}%
  \BibitemOpen
  \bibfield  {author} {\bibinfo {author} {\bibfnamefont {A.~A.}\ \bibnamefont
  {Baker}}, \bibinfo {author} {\bibfnamefont {A.~I.}\ \bibnamefont {Figueroa}},
  \bibinfo {author} {\bibfnamefont {L.~J.}\ \bibnamefont {Collins-McIntyre}},
  \bibinfo {author} {\bibfnamefont {G.}~\bibnamefont {{van der Laan}}}, \ and\
  \bibinfo {author} {\bibfnamefont {T.}~\bibnamefont {Hesjedal}},\ }\href
  {\doibase 10.1038/srep07907} {\bibfield  {journal} {\bibinfo  {journal} {Sci.
  Rep.}\ }\textbf {\bibinfo {volume} {5}},\ \bibinfo {pages} {7907} (\bibinfo
  {year} {2015}{\natexlab{b}})}\BibitemShut {NoStop}%
\bibitem [{\citenamefont {Jamali}\ \emph {et~al.}(2015)\citenamefont {Jamali},
  \citenamefont {Lee}, \citenamefont {Jeong}, \citenamefont {Mahfouzi},
  \citenamefont {Lv}, \citenamefont {Zhao}, \citenamefont {Nikolic},
  \citenamefont {Mkhoyan}, \citenamefont {Samarth},\ and\ \citenamefont
  {Wang}}]{Jamali_2015}%
  \BibitemOpen
  \bibfield  {author} {\bibinfo {author} {\bibfnamefont {M.}~\bibnamefont
  {Jamali}}, \bibinfo {author} {\bibfnamefont {J.~S.}\ \bibnamefont {Lee}},
  \bibinfo {author} {\bibfnamefont {J.~S.}\ \bibnamefont {Jeong}}, \bibinfo
  {author} {\bibfnamefont {F.}~\bibnamefont {Mahfouzi}}, \bibinfo {author}
  {\bibfnamefont {Y.}~\bibnamefont {Lv}}, \bibinfo {author} {\bibfnamefont
  {Z.}~\bibnamefont {Zhao}}, \bibinfo {author} {\bibfnamefont {B.~K.}\
  \bibnamefont {Nikolic}}, \bibinfo {author} {\bibfnamefont {K.~A.}\
  \bibnamefont {Mkhoyan}}, \bibinfo {author} {\bibfnamefont {N.}~\bibnamefont
  {Samarth}}, \ and\ \bibinfo {author} {\bibfnamefont {J.-P.}\ \bibnamefont
  {Wang}},\ }\href {\doibase 10.1021/acs.nanolett.5b03274} {\bibfield
  {journal} {\bibinfo  {journal} {Nano Lett.}\ }\textbf {\bibinfo {volume}
  {15}},\ \bibinfo {pages} {7126} (\bibinfo {year} {2015})}\BibitemShut
  {NoStop}%
\bibitem [{\citenamefont {Figueroa}\ \emph
  {et~al.}(2016{\natexlab{b}})\citenamefont {Figueroa}, \citenamefont {Baker},
  \citenamefont {Collins-McIntyre}, \citenamefont {Hesjedal},\ and\
  \citenamefont {{van der Laan}}}]{Figueroa_XFMR2016}%
  \BibitemOpen
  \bibfield  {author} {\bibinfo {author} {\bibfnamefont {A.~I.}\ \bibnamefont
  {Figueroa}}, \bibinfo {author} {\bibfnamefont {A.~A.}\ \bibnamefont {Baker}},
  \bibinfo {author} {\bibfnamefont {L.~J.}\ \bibnamefont {Collins-McIntyre}},
  \bibinfo {author} {\bibfnamefont {T.}~\bibnamefont {Hesjedal}}, \ and\
  \bibinfo {author} {\bibfnamefont {G.}~\bibnamefont {{van der Laan}}},\ }\href
  {\doibase 10.1016/j.jmmm.2015.07.013} {\bibfield  {journal} {\bibinfo
  {journal} {{J. Magn. Magn. Mater.}}\ }\textbf {\bibinfo {volume} {400}},\
  \bibinfo {pages} {178} (\bibinfo {year} {2016}{\natexlab{b}})}\BibitemShut
  {NoStop}%
\bibitem [{\citenamefont {Baker}\ \emph {et~al.}(2019)\citenamefont {Baker},
  \citenamefont {Figueroa}, \citenamefont {Hesjedal},\ and\ \citenamefont {{van
  der Laan}}}]{Baker_XFMR2019}%
  \BibitemOpen
  \bibfield  {author} {\bibinfo {author} {\bibfnamefont {A.~A.}\ \bibnamefont
  {Baker}}, \bibinfo {author} {\bibfnamefont {A.~I.}\ \bibnamefont {Figueroa}},
  \bibinfo {author} {\bibfnamefont {T.}~\bibnamefont {Hesjedal}}, \ and\
  \bibinfo {author} {\bibfnamefont {G.}~\bibnamefont {{van der Laan}}},\ }\href
  {\doibase 10.1016/j.jmmm.2018.10.109} {\bibfield  {journal} {\bibinfo
  {journal} {{J. Magn. Magn. Mater.}}\ }\textbf {\bibinfo {volume} {473}},\
  \bibinfo {pages} {470} (\bibinfo {year} {2019})}\BibitemShut {NoStop}%
\bibitem [{\citenamefont {Mellnik}\ \emph {et~al.}(2014)\citenamefont
  {Mellnik}, \citenamefont {Lee}, \citenamefont {Richardella}, \citenamefont
  {Grab}, \citenamefont {Mintun}, \citenamefont {Fischer}, \citenamefont
  {Vaezi}, \citenamefont {Manchon}, \citenamefont {Kim}, \citenamefont
  {Samarth},\ and\ \citenamefont {Ralph}}]{Mellnik_Nature2014}%
  \BibitemOpen
  \bibfield  {author} {\bibinfo {author} {\bibfnamefont {A.~R.}\ \bibnamefont
  {Mellnik}}, \bibinfo {author} {\bibfnamefont {J.~S.}\ \bibnamefont {Lee}},
  \bibinfo {author} {\bibfnamefont {A.}~\bibnamefont {Richardella}}, \bibinfo
  {author} {\bibfnamefont {J.~L.}\ \bibnamefont {Grab}}, \bibinfo {author}
  {\bibfnamefont {P.~J.}\ \bibnamefont {Mintun}}, \bibinfo {author}
  {\bibfnamefont {M.~H.}\ \bibnamefont {Fischer}}, \bibinfo {author}
  {\bibfnamefont {A.}~\bibnamefont {Vaezi}}, \bibinfo {author} {\bibfnamefont
  {A.}~\bibnamefont {Manchon}}, \bibinfo {author} {\bibfnamefont {E.-A.}\
  \bibnamefont {Kim}}, \bibinfo {author} {\bibfnamefont {N.}~\bibnamefont
  {Samarth}}, \ and\ \bibinfo {author} {\bibfnamefont {D.}~\bibnamefont
  {Ralph}},\ }\href@noop {} {\bibfield  {journal} {\bibinfo  {journal}
  {Nature}\ }\textbf {\bibinfo {volume} {511}},\ \bibinfo {pages} {449}
  (\bibinfo {year} {2014})}\BibitemShut {NoStop}%
\bibitem [{\citenamefont {Wang}\ \emph {et~al.}(2015)\citenamefont {Wang},
  \citenamefont {Deorani}, \citenamefont {Banerjee}, \citenamefont {Koirala},
  \citenamefont {Brahlek}, \citenamefont {Oh},\ and\ \citenamefont
  {Yang}}]{Wang_PRL2015}%
  \BibitemOpen
  \bibfield  {author} {\bibinfo {author} {\bibfnamefont {Y.}~\bibnamefont
  {Wang}}, \bibinfo {author} {\bibfnamefont {P.}~\bibnamefont {Deorani}},
  \bibinfo {author} {\bibfnamefont {K.}~\bibnamefont {Banerjee}}, \bibinfo
  {author} {\bibfnamefont {N.}~\bibnamefont {Koirala}}, \bibinfo {author}
  {\bibfnamefont {M.}~\bibnamefont {Brahlek}}, \bibinfo {author} {\bibfnamefont
  {S.}~\bibnamefont {Oh}}, \ and\ \bibinfo {author} {\bibfnamefont
  {H.}~\bibnamefont {Yang}},\ }\href {\doibase 10.1103/PhysRevLett.114.257202}
  {\bibfield  {journal} {\bibinfo  {journal} {Phys. Rev. Lett.}\ }\textbf
  {\bibinfo {volume} {114}},\ \bibinfo {pages} {257202} (\bibinfo {year}
  {2015})}\BibitemShut {NoStop}%
\bibitem [{\citenamefont {Rojas-S\'anchez}\ \emph {et~al.}(2016)\citenamefont
  {Rojas-S\'anchez}, \citenamefont {Oyarz\'un}, \citenamefont {Fu},
  \citenamefont {Marty}, \citenamefont {Vergnaud}, \citenamefont {Gambarelli},
  \citenamefont {Vila}, \citenamefont {Jamet}, \citenamefont {Ohtsubo},
  \citenamefont {Taleb-Ibrahimi}, \citenamefont {Le~F\`evre}, \citenamefont
  {Bertran}, \citenamefont {Reyren}, \citenamefont {George},\ and\
  \citenamefont {Fert}}]{Rojas_PRL2016}%
  \BibitemOpen
  \bibfield  {author} {\bibinfo {author} {\bibfnamefont {J.-C.}\ \bibnamefont
  {Rojas-S\'anchez}}, \bibinfo {author} {\bibfnamefont {S.}~\bibnamefont
  {Oyarz\'un}}, \bibinfo {author} {\bibfnamefont {Y.}~\bibnamefont {Fu}},
  \bibinfo {author} {\bibfnamefont {A.}~\bibnamefont {Marty}}, \bibinfo
  {author} {\bibfnamefont {C.}~\bibnamefont {Vergnaud}}, \bibinfo {author}
  {\bibfnamefont {S.}~\bibnamefont {Gambarelli}}, \bibinfo {author}
  {\bibfnamefont {L.}~\bibnamefont {Vila}}, \bibinfo {author} {\bibfnamefont
  {M.}~\bibnamefont {Jamet}}, \bibinfo {author} {\bibfnamefont
  {Y.}~\bibnamefont {Ohtsubo}}, \bibinfo {author} {\bibfnamefont
  {A.}~\bibnamefont {Taleb-Ibrahimi}}, \bibinfo {author} {\bibfnamefont
  {P.}~\bibnamefont {Le~F\`evre}}, \bibinfo {author} {\bibfnamefont
  {F.}~\bibnamefont {Bertran}}, \bibinfo {author} {\bibfnamefont
  {N.}~\bibnamefont {Reyren}}, \bibinfo {author} {\bibfnamefont {J.-M.}\
  \bibnamefont {George}}, \ and\ \bibinfo {author} {\bibfnamefont
  {A.}~\bibnamefont {Fert}},\ }\href {\doibase 10.1103/PhysRevLett.116.096602}
  {\bibfield  {journal} {\bibinfo  {journal} {Phys. Rev. Lett.}\ }\textbf
  {\bibinfo {volume} {116}},\ \bibinfo {pages} {096602} (\bibinfo {year}
  {2016})}\BibitemShut {NoStop}%
\bibitem [{\citenamefont {Wang}\ \emph {et~al.}(2017)\citenamefont {Wang},
  \citenamefont {Zhu}, \citenamefont {Wu}, \citenamefont {Yang}, \citenamefont
  {Yu}, \citenamefont {Ramaswamy}, \citenamefont {Mishra}, \citenamefont {Shi},
  \citenamefont {Elyasi}, \citenamefont {Teo}, \citenamefont {Wu},\ and\
  \citenamefont {Yang}}]{Wang_NatCom2017}%
  \BibitemOpen
  \bibfield  {author} {\bibinfo {author} {\bibfnamefont {Y.}~\bibnamefont
  {Wang}}, \bibinfo {author} {\bibfnamefont {D.}~\bibnamefont {Zhu}}, \bibinfo
  {author} {\bibfnamefont {Y.}~\bibnamefont {Wu}}, \bibinfo {author}
  {\bibfnamefont {Y.}~\bibnamefont {Yang}}, \bibinfo {author} {\bibfnamefont
  {J.}~\bibnamefont {Yu}}, \bibinfo {author} {\bibfnamefont {R.}~\bibnamefont
  {Ramaswamy}}, \bibinfo {author} {\bibfnamefont {R.}~\bibnamefont {Mishra}},
  \bibinfo {author} {\bibfnamefont {S.}~\bibnamefont {Shi}}, \bibinfo {author}
  {\bibfnamefont {M.}~\bibnamefont {Elyasi}}, \bibinfo {author} {\bibfnamefont
  {K.-L.}\ \bibnamefont {Teo}}, \bibinfo {author} {\bibfnamefont
  {Y.}~\bibnamefont {Wu}}, \ and\ \bibinfo {author} {\bibfnamefont
  {H.}~\bibnamefont {Yang}},\ }\href {\doibase {10.1038/s41467-017-01583-4}}
  {\bibfield  {journal} {\bibinfo  {journal} {{Nat. Commun.}}\ }\textbf
  {\bibinfo {volume} {{8}}},\ \bibinfo {pages} {{1364}} (\bibinfo {year}
  {{2017}})}\BibitemShut {NoStop}%
\bibitem [{\citenamefont {Wu}\ \emph {et~al.}(2019)\citenamefont {Wu},
  \citenamefont {Zhang}, \citenamefont {Deng}, \citenamefont {Lan},
  \citenamefont {Pan}, \citenamefont {Razavi}, \citenamefont {Che},
  \citenamefont {Huang}, \citenamefont {Dai}, \citenamefont {Wong},
  \citenamefont {Han},\ and\ \citenamefont {Wang}}]{Wu_PRL2019}%
  \BibitemOpen
  \bibfield  {author} {\bibinfo {author} {\bibfnamefont {H.}~\bibnamefont
  {Wu}}, \bibinfo {author} {\bibfnamefont {P.}~\bibnamefont {Zhang}}, \bibinfo
  {author} {\bibfnamefont {P.}~\bibnamefont {Deng}}, \bibinfo {author}
  {\bibfnamefont {Q.}~\bibnamefont {Lan}}, \bibinfo {author} {\bibfnamefont
  {Q.}~\bibnamefont {Pan}}, \bibinfo {author} {\bibfnamefont {S.~A.}\
  \bibnamefont {Razavi}}, \bibinfo {author} {\bibfnamefont {X.}~\bibnamefont
  {Che}}, \bibinfo {author} {\bibfnamefont {L.}~\bibnamefont {Huang}}, \bibinfo
  {author} {\bibfnamefont {B.}~\bibnamefont {Dai}}, \bibinfo {author}
  {\bibfnamefont {K.}~\bibnamefont {Wong}}, \bibinfo {author} {\bibfnamefont
  {X.}~\bibnamefont {Han}}, \ and\ \bibinfo {author} {\bibfnamefont {K.~L.}\
  \bibnamefont {Wang}},\ }\href {\doibase 10.1103/PhysRevLett.123.207205}
  {\bibfield  {journal} {\bibinfo  {journal} {Phys. Rev. Lett.}\ }\textbf
  {\bibinfo {volume} {123}},\ \bibinfo {pages} {207205} (\bibinfo {year}
  {2019})}\BibitemShut {NoStop}%
\bibitem [{\citenamefont {Han}\ \emph {et~al.}(2017)\citenamefont {Han},
  \citenamefont {Richardella}, \citenamefont {Siddiqui}, \citenamefont
  {Finley}, \citenamefont {Samarth},\ and\ \citenamefont {Liu}}]{Han_PRL2017}%
  \BibitemOpen
  \bibfield  {author} {\bibinfo {author} {\bibfnamefont {J.}~\bibnamefont
  {Han}}, \bibinfo {author} {\bibfnamefont {A.}~\bibnamefont {Richardella}},
  \bibinfo {author} {\bibfnamefont {S.~A.}\ \bibnamefont {Siddiqui}}, \bibinfo
  {author} {\bibfnamefont {J.}~\bibnamefont {Finley}}, \bibinfo {author}
  {\bibfnamefont {N.}~\bibnamefont {Samarth}}, \ and\ \bibinfo {author}
  {\bibfnamefont {L.}~\bibnamefont {Liu}},\ }\href {\doibase
  10.1103/PhysRevLett.119.077702} {\bibfield  {journal} {\bibinfo  {journal}
  {Phys. Rev. Lett.}\ }\textbf {\bibinfo {volume} {119}},\ \bibinfo {pages}
  {077702} (\bibinfo {year} {2017})}\BibitemShut {NoStop}%
\bibitem [{\citenamefont {Bonell}\ \emph {et~al.}(2020)\citenamefont {Bonell},
  \citenamefont {Goto}, \citenamefont {Sauthier}, \citenamefont {Sierra},
  \citenamefont {Figueroa}, \citenamefont {Costache}, \citenamefont {Miwa},
  \citenamefont {Suzuki},\ and\ \citenamefont
  {Valenzuela}}]{Bonell_NanoLett2020}%
  \BibitemOpen
  \bibfield  {author} {\bibinfo {author} {\bibfnamefont {F.}~\bibnamefont
  {Bonell}}, \bibinfo {author} {\bibfnamefont {M.}~\bibnamefont {Goto}},
  \bibinfo {author} {\bibfnamefont {G.}~\bibnamefont {Sauthier}}, \bibinfo
  {author} {\bibfnamefont {J.~F.}\ \bibnamefont {Sierra}}, \bibinfo {author}
  {\bibfnamefont {A.~I.}\ \bibnamefont {Figueroa}}, \bibinfo {author}
  {\bibfnamefont {M.~V.}\ \bibnamefont {Costache}}, \bibinfo {author}
  {\bibfnamefont {S.}~\bibnamefont {Miwa}}, \bibinfo {author} {\bibfnamefont
  {Y.}~\bibnamefont {Suzuki}}, \ and\ \bibinfo {author} {\bibfnamefont {S.~O.}\
  \bibnamefont {Valenzuela}},\ }\href {\doibase 10.1021/acs.nanolett.0c01850}
  {\bibfield  {journal} {\bibinfo  {journal} {Nano Lett.}\ }\textbf {\bibinfo
  {volume} {20}},\ \bibinfo {pages} {5893} (\bibinfo {year}
  {2020})}\BibitemShut {NoStop}%
\end{thebibliography}
%

\end{document}